\documentclass{article}

\usepackage{amsmath}
\usepackage{amssymb}
\usepackage{booktabs}
\usepackage{epubtk}    
\usepackage{graphicx}
\usepackage{makeidx}

\makeindex

\def\beq{\begin{equation}}
\def\eeq{\end{equation}}
\def\beqn{\begin{eqnarray}}
\def\eeqn{\end{eqnarray}}

\begin{document}

\title{Minimal Length Scale Scenarios for Quantum Gravity}

\author{%
\epubtkAuthorData{Sabine Hossenfelder}{%
Nordita \\
Roslagstullsbacken 23\\
106 91 Stockholm\\
Sweden}{%
hossi@nordita.org}{}
}

\date{}
\maketitle

\begin{abstract}
We review the question of whether the fundamental laws of nature limit 
our ability to probe arbitrarily short distances. First, we examine what insights
can be gained from thought experiments for probes of shortest distances, and
summarize what can be
learned from different approaches to a theory of quantum gravity. Then we discuss 
some models that have been developed to implement a minimal length scale in quantum mechanics
and quantum field theory. These models have entered the literature as the generalized uncertainty principle or the modified dispersion relation, and have allowed 
the study of the effects of a minimal length scale in quantum mechanics, quantum electrodynamics, 
thermodynamics,
black-hole physics and cosmology. Finally, we touch upon the question of ways to circumvent
the manifestation of a minimal length scale in short-distance physics.
\end{abstract}



\epubtkKeywords{minimal length, quantum gravity, generalized
  uncertainty principle}

\newpage
\tableofcontents

\newpage

\section{Introduction}

In the 5th century B.C., Democritus postulated the existence of
smallest objects that all matter is built from and called them
`atoms'. In Greek, the prefix `a' means `not' and the word `tomos'
means `cut'. Thus, atomos or atom means uncuttable or
indivisible. According to Democritus' theory of atomism, ``Nothing
exists except atoms and empty space, everything else is opinion.''
Though variable in shape, Democritus' atoms were the hypothetical
fundamental constituents of matter, the elementary building blocks of
all that exists, the smallest possible entities. They were conjectured
to be of finite size, but homogeneous and without substructure. They
were the first envisioned end of reductionism.
 
2500 years later, we know that Democritus was right in that solids and
liquids are composed of smaller entities with universal properties
that are called atoms in his honor. But these atoms turned out to be
divisible. And stripped of its electrons, the atomic nucleus too was
found to be a composite of smaller particles, neutrons and
protons. Looking closer still, we have found that even neutrons and
protons have a substructure of quarks and gluons. At present, the
standard model of particle physics with three generations of quarks
and fermions and the vector fields associated to the gauge groups are
the most fundamental constituents of matter that we know.

Like a Russian doll, reality has so far revealed one after another
layer on smaller and smaller scales. This begs the question: Will we
continue to look closer into the structure of matter, and possibly
find more layers? Or is there a fundamental limit to this search, a
limit beyond which we cannot go? And if so, is this a limit in
principle or one in practice?

Any answer to this question has to include not only the structure of
matter, but the structure of space and time itself, and therefore it
has to include gravity. For one, this is because Democritus' search
for the most fundamental constituents carries over to space and time
too. Are space and time fundamental, or are they just good
approximations that emerge from a more fundamental concept in the
limits that we have tested so far? Is spacetime made of something
else? Are there `atoms' of space? And second, testing short distances
requires focusing large energies in small volumes, and when energy
densities increase, one finally cannot neglect anymore the curvature
of the background.

In this review we will study this old question of whether there is a
fundamental limit to the resolution of structures beyond which we
cannot discover anything more. In Section~\ref{motivations}, we will
summarize different approaches to this question, and
how they connect with our search for a theory of quantum gravity. We
will see that almost all such approaches lead us to find that the
possible resolution of structures is finite or, more graphically, that
nature features a minimal length scale -- though we will also see that
the expression `minimal length scale' can have different
interpretations. While we will not go into many of the details of the
presently pursued candidate theories for quantum gravity, we will
learn what some of them have to say about the question. After the
motivations, we will in Section~\ref{models} briefly review some
approaches that investigate the consequences of a minimal length scale
in quantum mechanics and quantum field theory, models that have
flourished into one of the best motivated and best developed areas  of
the phenomenology of quantum gravity.

In the following, we use the unit convention $c=\hbar=1$, so that the
Planck length $l_{\mathrm{Pl}}$ is the inverse of the Planck mass
$m_{\mathrm{Pl}} = 1/ l_{\mathrm{Pl}}$, and Newton's constant $G =
l_{\mathrm{Pl}}^2 = 1/m_{\mathrm{Pl}}^2$. The signature of the metric is
$(1,-1,-1,-1)$. Small Greek indices run from 0 to 3,  large Latin
indices from 0 to 4, and small Latin indices from 1 to 3, except for
Section~\ref{string}, where small Greek indices run from 0 to $D$, and
small Latin indices run from 2 to $D$. An arrow denotes the spatial
component of a vector, for example $\vec a = (a_1, a_2, a_3)$. Bold-faced quantities are tensors in an index-free notation that will be
used in the text for better readability, for example $\mathbf{p} = (p_0,
p_1, p_2, p_3)$. Acronyms and abbreviations can be found in the index.

We begin with a brief historical background.

\newpage

\section{A Minimal History}

Special relativity and quantum mechanics are characterized by two universal
constants, the speed of light, $c$, and \index{Planck's constant} Planck's constant, $\hbar$. Yet, from these
constants alone one cannot construct either a constant of dimension length or mass. Though, if one
had either, they could be converted into each other by use of $\hbar$ and $c$. But in 1899, Max Planck
pointed out that adding Newton's constant $G$ to the universal constants $c$ and $\hbar$ allows
one to construct units of mass, length and time~\cite{Planck}:
\beqn
t_{\mathrm{Pl}} &\approx& 10^{-43}\mathrm{\ s} \nonumber \\
l_{\mathrm{Pl}} &\approx& 10^{-33}\mathrm{\ cm} \nonumber \\
m_{\mathrm{Pl}} &\approx& 1.2 \times 10^{19}\mathrm{\ GeV} \,.
\eeqn
Today these are known as the Planck time, Planck length and Planck mass\index{Planck scale}, respectively. As we will
see later, they mark the scale at which quantum effects of the gravitational interaction are expected
to become important. But back in Planck's days their relevance was their universality,
because they are constructed entirely from fundamental constants. 

The idea of a minimal length was predated by that of the ``chronon,'' a smallest unit of time,
proposed by Robert L\'evi~\cite{Levi} in 1927 in his ``Hyphoth\`ese de l'atome de temps'' (hypothesis of time atoms), \index{Time atoms}
that was further developed by Pokrowski in the years following  L\'evi's proposal~\cite{Pokrowski}.
But that there might be limits to the divisibility
of space and time remained a far-fetched speculation on the fringes of a community rapidly
pushing forward the development of general relativity and quantum mechanics. It was not until
special relativity and quantum mechanics were joined in the framework of quantum field theory
that the possible existence of a minimal length scale rose to the awareness of the community. 

With the advent of quantum 
field theory in the 1930s, it was widely believed that a fundamental length was necessary to cure 
troublesome divergences. The most commonly
used regularization was a cut-off or some other dimensionful quantity to render integrals finite.
It seemed natural to think of this pragmatic cut-off as having fundamental significance, an interpretation
that however inevitably caused problems with Lorentz invariance, since the cut-off would not be
independent of the frame of reference. Heisenberg was among the first to consider a
fundamentally-discrete spacetime that would yield a cut-off, laid out in his letters to Bohr and Pauli. The idea of a fundamentally finite 
length or a maximum frequency was in these years studied by many, including Flint~\cite{Flint}, March~\cite{March}, 
M\"oglich~\cite{Moglich}
and Goudsmit~\cite{Goudsmit}, just to mention a few. They all had in common that they considered the
fundamental length to be in the realm of subatomic physics on the
order of the femtometer ($10^{-15}\mathrm{\ m}$). 
\index{Lorentz-invariance}

The one exception was a young Russian, Matvei Bronstein. Today recognized as
the first to comprehend the problem of quantizing gravity~\cite{Bronstein}, Bronstein was decades 
ahead of his time. Already in 1936, he argued that gravity is in one important way fundamentally
different from electrodynamics: Gravity does not allow an arbitrarily high
 concentration of charge in a small region of spacetime, since the gravitational `charge' is 
energy and, if concentrated too much, will collapse to a black hole. Using the weak field approximation
of gravity, he concluded that this leads to an inevitable limit to the precision of which one
can measure the strength of the gravitational field (in terms of the Christoffel symbols). 

In his 1936 article ``Quantentheorie schwacher Gravitationsfelder''
(Quantum theory of weak gravitational fields), Bronstein
wrote~\cite{Bronstein,Bronstein2}:
\begin{quote}
``[T]he gravitational radius of the test-body $(G\rho V/c^2)$ used
  for the measurements should by no means be larger than its linear
  dimensions $(V^{1/3})$; from this one obtains an upper bound for its
  density $(\rho \lesssim c^2/G V^{2/3})$. Thus, the possibilities for
  measurements in this region are even more restricted than one
  concludes from the quantum-mechanical commutation relations. Without
  a profound change of the classical notions it therefore seems hardly
  possible to extend the quantum theory of gravitation to this
  region.''%
\epubtkFootnote[orgQuoteDe]{``[D]er Gravitationsradius des zur
    Messung dienenden Probek\"orpers ($G\rho V/c^2$) soll keineswegs
    gr\"o{\ss}er als seine linearen Abmessungen ($V^{1/3}$) sein;
    daraus entsteht eine obere Grenze f\"ur seine Dichte ($\rho
    \lesssim c^2/G V^{2/3}$). Die Messungsm\"oglichkeiten sind also in
    dem Gebiet noch mehr beschr\"ankt als es sich aus den
    quantenmechanischen Vertauschungsrelationen schliessen
    l\"a{\ss}t. Ohne eine tiefgreifende \"Anderung der klassischen
    Begriffe, scheint es daher kaum m\"oglich, die Quantentheorie der
    Gravitation auch auf dieses Gebiet auszudehnen.''}
  (\cite{Bronstein2}, p.~150)%
\epubtkFootnote{Translations from German to English: SH.}
\end{quote}
Few people took note of Bronstein's argument and, unfortunately, the history of this promising young physicist ended in a
Leningrad prison in February 1938, where Matvei Bronstein was executed at the age of 31. 

Heisenberg meanwhile continued in his attempt to make sense of the notion of a fundamental minimal length of nuclear dimensions. 
In 1938, Heisenberg wrote ``{\"U}ber die in der Theorie der Elementarteilchen auftretende universelle
L{\"a}nge'' (On the universal length appearing in the theory of elementary particles)~\cite{Heisenberg1}, in which he 
argued that this fundamental length, which he denoted $r_0$, should appear somewhere not too far beyond the classical 
electron radius (of the order 100~fm).

This idea seems curious today, and has to be put into perspective. Heisenberg was
very worried about the non-renormalizability \index{Renormalizability} of Fermi's theory of $\beta$-decay. He had previously
shown~\cite{Heisenberg2} that applying Fermi's theory to the high center-of-mass energies of some hundred GeV 
lead to an `explosion,'
by which he referred to events of very high multiplicity. Heisenberg argued this would explain the observed
cosmic ray showers, whose large number of secondary particles we know today are created by cascades (a possibility that was discussed already at the time of Heisenberg's writing, but
not agreed upon). We also know today that what Heisenberg actually discovered is that Fermi's theory
breaks down at such high energies, and the four-fermion coupling has to be replaced by the exchange
of a gauge boson in the electroweak interaction. But in the 1930s neither the strong nor the
electroweak force was known. Heisenberg then connected the problem of regularization with the
breakdown of the perturbation expansion of Fermi's theory, and argued that the presence
of the alleged explosions would prohibit the resolution of finer structures:
\begin{quote}
``If the explosions actually exist and represent the processes
  characteristic for the constant $r_0$, then they maybe convey a
  first, still unclear, understanding of the obscure properties
  connected with the constant $r_0$. These should certainly express
  themselves in difficulties of measurements with a precision better
  than $r_0$\ldots The explosions would have the effect\ldots that
  measurements of positions are not possible to a precision better
  than $r_0$.''%
\epubtkFootnote[orgQuoteDe]{``Wenn die Explosionen tats\"achlich
    existieren und die f\"ur die Konstante $r_0$ eigentlich
    charakeristischen Prozesse darstellen, so vermitteln sie
    vielleicht ein erstes, noch unklares Verst\"andnis der
    unanschaulichen Z\"uge, die mit der Konstanten $r_0$ verbunden
    sind. Diese sollten sich ja wohl zun\"achst darin \"au{\ss}ern,
    da{\ss} die Messung einer den Wert $r_0$ unterschreitenden
    Genauigkeit zu Schwierigkeiten f\"uhrt\ldots [D]ie Explosionen
    [w\"urden] daf\"ur sorgen\ldots, da{\ss} Ortsmessungen mit einer
    $r_0$ unterschreitenden Genauigkeit unm\"oglich sind.''} 
  (\cite{Heisenberg1}, p.~31)
\end{quote}

In hindsight we know that Heisenberg was,
correctly, arguing that the theory of elementary particles known in the 1930s was
incomplete. The strong interaction was missing and Fermi's theory indeed non-renormalizable,
but not fundamental. Today we also know that the standard model of particle physics
is renormalizable and know techniques to deal with divergent integrals that do not
necessitate cut-offs, such as dimensional regularization. But lacking that knowledge, it
is understandable that Heisenberg argued that taking into account gravity was irrelevant
for the existence of a fundamental length:
\begin{quote}
``The fact that [the Planck length] is significantly smaller than
  $r_0$ makes it valid to leave aside the obscure properties of the
  description of nature due to gravity, since they -- at least in
  atomic physics -- are totally negligible relative to the much
  coarser obscure properties that go back to the universal constant
  $r_0$. For this reason, it seems hardly possible to integrate
  electric and gravitational phenomena into the rest of physics until
  the problems connected to the length $r_0$ are solved.''%
\epubtkFootnote[orgQuoteDe]{``Der Umstand, da{\ss} [die
      Planckl\"ange] wesentlich kleiner ist als $r_0$, gibt uns das
    Recht, von den durch die Gravitation bedingten unanschaulichen
    Z\"ugen der Naturbeschreibung zun\"achst abzusehen, da sie --
    wenigstens in der Atomphysik -- v\"ollig untergehen in den viel
    gr\"oberen unanschaulichen Z\"ugen, die von der universellen
    Konstanten $r_0$ herr\"uhren. Es d\"urfte aus diesen Gr\"unden
    wohl kaum m\"oglich sein, die elektrischen und die
    Gravitationserscheinungen in die \"ubrige Physik einzuordnen,
    bevor die mit der L\"ange $r_0$ zusammenh\"angenden Probleme
    gel\"ost sind.''}
  (\cite{Heisenberg1}, p.~26)
\end{quote}
Heisenberg apparently put great hope in the notion of a fundamental length to move forward the
understanding of elementary matter. In 1939 he expressed his belief that a quantum theory with a
minimal length scale would be able to account for the discrete mass spectrum of the (then known)
elementary particles~\cite{Heisenberg39}.
However, the theory of quantum electrodynamics was developed to maturity, the `explosions' were satisfactorily
explained and, without being hindered by the appearance of any fundamentally finite resolution, experiments
probed shorter and shorter scales. The divergences in quantum field theory became better
understood and discrete approaches to space and time remained unappealing due to their problems with
Lorentz invariance.

In a 1947 letter to Heisenberg, Pauli commented on
the idea of a smallest length that Heisenberg still held dearly and explained his reservations, concluding
``Extremely put, I would not be surprised if your `universal' length turned out to be a mere
figment of imagination.''~\cite{Pauliletter}. (For more about Heisenberg's historical 
involvement with the universal length, the interested reader is referred to Kragh's very recommendable
article~\cite{Kragh}.)

In 1930, in a letter to his student Rudolf Peierls~\cite{Peierls},
Heisenberg mentioned that he was trying to make sense of a minimal
length by letting the position operators be non-commuting $[\hat
  x^\nu, \hat x^\mu] \neq 0$. He expressed his hope that Peierls ask
Pauli how to proceed with this idea: 
\begin{quote}
``So far, I have not been able to make mathematical sense of such
  commutation relations\ldots Do you or Pauli have anything to say
  about the mathematical meaning of such commutation relations?''%
\epubtkFootnote[orgQuoteDe]{``Mir ist es bisher nicht gelungen,
    solchen Vertauschungs-Relationen einen vern\"unftigen
    mathematischen Sinn zuzuordnen\dots F\"allt Ihnen oder Pauli nicht
    vielleicht etwas \"uber den mathematischen Sinn solcher
    Vertauschungs-Relationen ein?''} 
  (\cite{Peierls}, p.~16)
\end{quote} 

But it took 17 years until Snyder,\index{Non-commutative geometry}\index{Commutation relations} in 1947, made mathematical sense of Heisenberg's idea.%
\epubtkFootnote{The story has been told~\cite{Wess} that Peierls asked
  Pauli, Pauli passed the question on to his colleague Oppenheimer,
  who asked his student Hartland Snyder. However, in a 1946 letter to
  Pauli~\cite{Snyderletter}, Snyder encloses his paper without any
  mention of it being an answer to a question posed to him by
  others.}
Snyder, who felt that that the use of a cut-off \index{Cut-off} in momentum space was a ``distasteful arbitrary procedure''~\cite{Snyder}, worked out a modification of the canonical commutation relations of position and momentum operators. 
In that way, spacetime became
Lorentz-covariantly non-commutative, but the modification of commutation relations increased the Heisenberg uncertainty,
such that a smallest possible resolution of structures was introduced (a consequence Snyder did not explicitly mention in his paper). Though Snyder's approach was criticized for the difficulties of inclusion of translations~\cite{Yang}, it has received
a lot of attention as the first to show that a minimal length scale need not
be in conflict with Lorentz invariance. 

In 1960, Peres and Rosen~\cite{Peres:1960zz} studied uncertainties in
the measurement of the average values of Christoffel symbols due to
the impossibility of concentrating a mass to a region smaller than its
Schwarzschild radius, and came to the same conclusion as Bronstein
already had, in 1936,
\begin{quote}
``The existence of these quantum uncertainties in the gravitational
  field is a strong argument for the necessity of quantizing it. It is
  very likely that a quantum theory of gravitation would then
  generalize these uncertainty relations to all other Christoffel
  symbols.'' (\cite{Peres:1960zz}, p. 336)
\end{quote}
While they considered the limitations for measuring the gravitational field itself, they did not study the limitations
these uncertainties induce on the ability to measure distances in general. 

It was not until 1964, that Mead pointed out the peculiar role that gravity plays in our
attempts to test physics at short distances~\cite{Mead,Mead1}. He showed, in a series of thought
experiments that we will discuss in Section~\ref{thought}, that this influence does
have the effect of amplifying Heisenberg's measurement uncertainty, making it impossible
to measure distances to a precision better than Planck's length. And, since gravity couples universally, this
is, though usually negligible, an inescapable influence on all our experiments. 

Mead's work did not originally attain a lot of attention. Decades
later, he submitted his recollection~\cite{Mead2} that ``Planck's
proposal that the Planck mass, length, and time should form a
fundamental system of units\ldots was still considered heretical well
into the 1960s,'' and that his argument for the fundamental relevance
of the Planck length met strong resistance:
\begin{quote}
``At the time, I read many referee reports on my papers and discussed
  the matter with every theoretical physicist who was willing to
  listen; nobody that I contacted recognized the connection with the
  Planck proposal, and few took seriously the idea of [the Planck
    length] as a possible fundamental length. The view was nearly
  unanimous, not just that I had failed to prove my result, but that
  the Planck length could never play a fundamental role in physics. A
  minority held that there could be no fundamental length at all, but
  most were then convinced that a [different] fundamental
  length\ldots, of the order of the proton Compton wavelength, was the
  wave of the future. Moreover, the people I contacted seemed to treat
  this much longer fundamental length as established fact, not
  speculation, despite the lack of actual evidence for it.''
  (\cite{Mead2}, p.~15)
\end{quote}

But then in the mid 1970s then Hawking's \index{Hawking-radiation} calculation of a black hole's \index{Black hole} 
thermodynamical properties~\cite{Hawking} introduced the `transplanckian problem.' Due to the, in principle infinite, blue shift of
photons approaching a black-hole horizon, modes with energies exceeding the Planck scale had to be taken into account
to calculate the emission rate. A great many physicists have significantly advanced our understanding of black-hole physics and the Planck scale, too many to be named here. However, the prominent
role played by John Wheeler, whose contributions, though not directly on the topic of a minimal length, has
connected black-hole physics with spacetime foam and the Planckian limit, and by this inspired much of
what followed.

Unruh suggested in 1995~\cite{Unruh} that one use a
modified dispersion relation to deal with the difficulty of transplanckian modes, so that a smallest possible wavelength takes care of the contributions beyond the Planck scale. A similar problem exists in inflationary
cosmology~\cite{Martin:2000xs} since tracing back in time small frequencies increases the frequency till it eventually might
surpass the Planck scale at which point we no longer know how to make sense of general relativity.  Thus, this issue of transplanckian modes in cosmology brought up another reason to reconsider the possibility
of a minimal length or a maximal frequency, but this time the maximal frequency was
at the Planck scale rather than at the nuclear scale. Therefore, it was proposed~\cite{Kempf:2000ac,Hassan:2002qk} 
that this problem too might be cured by implementing a minimum length
uncertainty principle into inflationary cosmology.  \index{Transplanckian problem}

Almost at the same time, Majid and Ruegg~\cite{Majid:1994cy} proposed a modification for
the commutators of spacetime coordinates, similar to that of Snyder, following from a generalization of the 
Poincar\'e algebra to a Hopf algebra\index{Hopf algebra}, which became known as $\kappa$-Poincar\'e\index{$\kappa$-Poincar\'e}. Kempf et al.~\cite{Kempf:1993bq,Kempf:1994qp,Kempf:1994su,Kempf:1996ss}
developed the mathematical basis of quantum mechanics that took into account a minimal length scale
and ventured towards quantum field theory. There are by now many variants of models employing modifications
of the canonical commutation relations in order to accommodate a minimal length scale, 
not all of which make use of the complete $\kappa$-Poincar\'e framework, as will be
discussed later in Sections~\ref{mcr} and \ref{dsr}. Some of these approaches were shown to give rise to a 
modification of the dispersion relation, though the physical interpretation and relevance, as well
as the phenomenological consequences of this relation are still under debate. 

In parallel to this, developments in string theory revealed the impossibility of resolving arbitrarily
small structures with an object of finite extension. It had already been shown in the late 1980s~\cite{Gross:1987ar,Amati:1987uf,Amati:1987wq,Amati:1990xe,Veneziano:1989fc}
that string scattering in the super-Planckian regime would result in a generalized uncertainty principle,
preventing a localization to better than the string scale (more on this in Section~\ref{string}). In 1996, John Schwarz gave a 
talk at {\sc SLAC} about the generalized uncertainty principles resulting from string theory and thereby inspired the 1999
work by Adler and Santiago~\cite{Adler} who almost exactly reproduced Mead's earlier argument, apparently
without being aware of Mead's work. This picture was later refined when it became understood that string theory
not only contains strings but also higher dimensional objects, known as branes, which will be discussed
in Section~\ref{string}.

In the following years, a generalized uncertainty principle and
quantum mechanics with the Planck length as a minimal length received an increasing amount of attention
as potential cures for the transplanckian problem, a natural UV-regulator, and as possible manifestations of a fundamental
property of quantum spacetime. In the late 1990s, it was also noted that it is compatible with string
theory to have large or warped extra dimensions that can effectively lower the Planck scale into the
TeV range. With this, the fundamental length scale also moved into the reach of collider physics,
resulting in a flurry of activity.%
\epubtkFootnote{Though the hope of a lowered Planck scale pushing
  quantum gravitational effects into the reach of the Large Hadron
  Collider seems, at the time of writing, to not have been
  fulfilled.}\index{Planck scale!lowered}

Today, how to resolve the
apparent disagreements between the quantum field theories of the standard model and general relativity is one of the big open questions in theoretical physics. 
It is not that we cannot quantize gravity, but that the attempt to do so
leads to a perturbatively non-renormalizable and thus fundamentally nonsensical theory. The basic reason is
 that the coupling constant of gravity, Newton's constant, is dimensionful.
This leads to the necessity to introduce an infinite number of counter-terms, eventually
rendering the theory incapable of prediction.

But the same is true for Fermi's theory that Heisenberg was so worried about that
he argued for a finite resolution where the theory breaks down, and mistakenly so, 
since he was merely pushing an effective theory beyond its limits. 
So we have to ask then if we might be making the same mistake as Heisenberg, 
in that we falsely interpret the failure of general relativity to extend beyond the Planck scale
as the occurrence of a fundamentally finite resolution of structures, rather than just
the limit beyond which we have to look for a new theory that will allow us to resolve
smaller distances still? 

If it was only the extension of classical gravity, laid out in many thought experiments 
that will be discussed in Section~\ref{thought}, that had us believing the Planck length 
is of fundamental importance, then the above historical lesson should caution us we might 
be on the wrong track. Yet, the situation today is different from the one that Heisenberg 
faced. Rather than pushing a quantum theory beyond its limits, we are pushing a classical 
theory and conclude that its short-distance behavior is troublesome, which we hope to 
resolve with quantizing the theory. And, as we will see, several attempts at a 
UV-completion of gravity, discussed in Sections~\ref{string}\,--\,\ref{motmisc}, 
suggest that the role of the Planck length as a minimal length carries over into 
the quantum regime as a dimensionful regulator, though in very different ways, feeding our hopes
that we are working on unveiling the last and final Russian doll. 

For a more exhaustive coverage of the history of the minimal length, the interested
reader is referred to~\cite{Hagar:LM}.

\newpage

\section{Motivations}
\label{motivations}

\subsection{Thought experiments}
\label{thought}
\index{Thought experiment}

Thought experiments have played an important role in the history of physics as the poor theoretician's way to 
test the limits of a theory. This poverty might be an actual one of lacking experimental equipment, or it
might be one of practical impossibility. Luckily, technological advances sometimes turn thought experiments
into real experiments, as was the case with Einstein, Podolsky and Rosen's 1935 paradox. But even if
an experiment is not experimentally realizable in the near future, thought
experiments serve two important purposes. First, by allowing the thinker to test ranges of parameter space
that are inaccessible to experiment, they may reveal inconsistencies or paradoxes and thereby open doors to
an improvement in the fundamentals of the theory. The complete evaporation of a black hole and the question of
information loss in that process is a good example for this. Second, thought experiments tie the theory 
to reality by the necessity to investigate in detail what constitutes a measurable entity. The thought 
experiments discussed in the following are examples of this. 

\subsubsection{The Heisenberg microscope with Newtonian gravity}
\label{sec:heisenberg-microscope-newtonian}

Let us first recall Heisenberg's microscope, that lead to the uncertainty principle~\cite{Heisenberg3}. Consider a photon 
with frequency $\omega$ moving
in direction $x$, which scatters on a particle whose position on the $x$-axis we want to
measure. The scattered photons that reach the lens of the microscope have to lie within an
angle $\epsilon$ to produce an image from which we
want to infer the position of the particle (see Figure~\ref{1}). According to classical optics, the wavelength of the photon sets a limit
to the possible resolution $\Delta x$  
\beqn
\Delta x \gtrsim \frac{1}{2 \pi \omega \sin \epsilon} \,. \label{usualdelta}
\eeqn

But the photon used to measure the position of the particle has a recoil when it scatters 
and transfers a momentum to the particle. Since one does not know the direction of the
photon to better than $\epsilon$, this results in an uncertainty for the momentum of the
particle in direction $x$
\beqn
\Delta p_x \gtrsim \omega \sin \epsilon \,. \label{deltap}
\eeqn 
Taken together one obtains Heisenberg's uncertainty\index{Uncertainty principle} (up to a factor of order one)
\beqn
\Delta x \Delta p_x \gtrsim \frac{1}{2 \pi} \,.
\eeqn

\epubtkImage{microscope.png}{%
  \begin{figure}[ht]
    \centerline{\includegraphics[width=8.0cm]{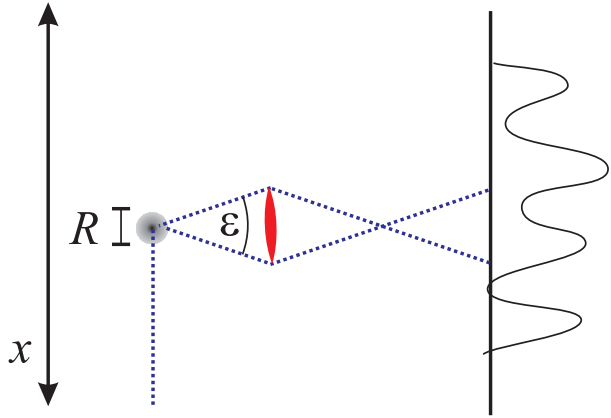}}
    \caption{Heisenberg's microscope. A photon moving along the
      $x$-axis scatters off a probe within an interaction region of
      radius $R$ and is detected by a microscope (indicated by a lens
      and screen) with opening angle $\epsilon$.}
    \label{1}
\end{figure}}

We know today that Heisenberg's uncertainty is not just a peculiarity of a measurement
method but much more than that -- it is a fundamental property of the quantum nature of matter. 
It does not, strictly speaking, even make sense to consider
the position and momentum of the particle at the same time. Consequently, instead of speaking
about the photon scattering off the particle as if that would happen in one particular
point, we should speak of the photon having a strong interaction with the particle in
some region of size $R$.

Now we will include gravity in the picture, following the treatment of Mead~\cite{Mead}. 
For any interaction to take place and subsequent measurement to be possible, the time
elapsed between the interaction and measurement has to be at least on the order of the
time, $\tau$, the photon needs to travel the distance $R$, so that $\tau \gtrsim R$.
The photon carries an energy that, though in general tiny, exerts a gravitational pull on
the particle whose position we wish to measure. The gravitational acceleration 
acting on the particle is
at least on the order of
\beqn
a \approx \frac{G \omega}{R^2} \,,
\eeqn
and, assuming that the particle is non-relativistic and much slower than the photon, 
the acceleration lasts about the duration the photon is in the region of strong interaction. From this,
the particle acquires a velocity of $v \approx a R$, or
\beqn
v \approx \frac{G \omega}{R} \,.
\eeqn
Thus, in the time $R$, the acquired velocity allows the particle to travel a distance of
\beqn
L \approx G \omega \,.
\eeqn
However, since the direction of the photon was unknown to within the angle $\epsilon$, the direction of the acceleration
and the motion of the particle is also unknown. Projection on the $x$-axis then yields the additional 
uncertainty of
\beqn
\Delta x \gtrsim G \omega \sin \epsilon. \label{extradelta}
\eeqn 
Combining (\ref{extradelta}) with (\ref{usualdelta}), one obtains
\beqn
\Delta x \gtrsim \sqrt{G} = l_{\mathrm{Pl}} \,. \label{minimaldelta}
\eeqn
One can refine this argument by taking into account that strictly speaking during the measurement, the
momentum of the photon, $\omega$, increases by $G m \omega / R$, where $m$ is the mass of the
particle. This increases the uncertainty in the particle's momentum (\ref{deltap}) to
\beqn
\Delta p_x \gtrsim \omega \left( 1 + \frac{Gm}{R} \right) \sin \epsilon \,,
\eeqn
and, for the time the photon is in the interaction region, translates into a position uncertainty $\Delta x \approx R \Delta p /m$
\beqn
\Delta x \gtrsim  \omega \left( \frac{R}{m} + G \right) \sin \epsilon \,,
\eeqn
which is larger than the previously found uncertainty (\ref{extradelta}) and thus (\ref{minimaldelta}) still follows. 

Adler and Santiago~\cite{Adler} offer pretty much the same argument, but add that
the particle's momentum uncertainty $\Delta p$ should be on the order of the photon's momentum $\omega$. Then one finds
\beqn 
\Delta x \gtrsim G \Delta p \,.
\eeqn
Assuming that the normal uncertainty and the gravitational uncertainties add linearly, one arrives at
\beqn
\Delta x \gtrsim \frac{1}{\Delta p} + G \Delta p \,. \label{ASgup}
\eeqn
Any uncertainty principle with a modification of this or similar form has become known in the literature as
`generalized uncertainty principle' ({\sc GUP}). \index{Generalized uncertainty principle} Adler and Santiago's
work was inspired by the appearance of such an uncertainty principle in string theory, which we will investigate in
Section~\ref{string}. Adler and Santiago make the interesting observation that the {\sc GUP} (\ref{ASgup}) is invariant under the replacement
\beqn
l_{\mathrm{Pl}}  \Delta p \leftrightarrow \frac{1}{l_{\mathrm{Pl}}  \Delta p} \,,
\eeqn
which relates long to short distances and high to low energies. 

These limitations, refinements of which we will discuss in the following Sections \ref{grhm}\,--\,\ref{volumes}, apply to the possible spatial
resolution in a microscope-like measurement. At the high energies necessary to reach the Planckian limit, the scattering
is unlikely to be elastic, but the same considerations apply to inelastic scattering events. Heisenberg's microscope
revealed a fundamental limit that is a consequence of the non-commutativity of position and momentum operators in
quantum mechanics. The question that the GUP then raises is what modification
of quantum mechanics would give rise to the generalized uncertainty, a question we will return to in Section~\ref{mcr}.

Another related argument has been put forward by Scardigli~\cite{Scardigli:1999jh}, who employs the idea that once
one arrives at energies of about the Planck mass and concentrates them to within a volume of radius of the Planck
length, one creates tiny black holes, which subsequently evaporate. This effects scales in the same
way as the one discussed here, and one arrives again at (\ref{ASgup}). 

\subsubsection{The general relativistic Heisenberg microscope}
\label{grhm}

The above result makes use of Newtonian gravity, and has to be refined when one takes into account
general relativity. Before we look into the details, let us start with a heuristic but instructive argument.
One of the most general features of general relativity is the formation of black holes under
certain circumstances, roughly speaking when the energy density in some region of spacetime
becomes too high. Once matter becomes very dense, its gravitational pull leads to a total
collapse that ends in the formation of a horizon.%
\epubtkFootnote{In the classical theory, inside the horizon lies a
  singularity. This singularity is expected to be avoided in quantum
  gravity, but how that works or doesn't work is not relevant in the
  following.}
It is usually assumed that the
Hoop conjecture holds~\cite{Hoop}: \index{Hoop conjecture} If an amount of energy $\omega$ is compacted at any time into
a region whose circumference in every direction is $R \leq 4 \pi G \omega$ , then the region will eventually 
develop into a black hole. The Hoop conjecture is unproven, but we know from both analytical and numerical
studies that it holds to very good precision~\cite{Eardley:2002re,Hsu:2002bd}.

Consider now that we have a particle of energy $\omega$. Its extension $R$ has to be larger than the Compton
wavelength associated to the energy, so $R \geq 1/\omega$. Thus, the larger the energy,
the better the particle can be focused. On the other hand, if the extension drops
below $4 \pi G E$, then a black hole is formed with radius $2 \omega G$. The important point to notice 
here is that the extension of the black hole grows linearly with the energy, and 
therefore one can achieve a minimal possible
extension, which is on the order of $R\sim \sqrt{G}$. 

For the more detailed argument, we follow Mead~\cite{Mead} with the general relativistic 
version of the Heisenberg microscope that was discussed in Section~\ref{sec:heisenberg-microscope-newtonian}. Again, we
have a particle whose position we want to measure by help of a test particle. The
test particle has a momentum vector $(\omega, \vec k)$, and for completeness we consider a 
particle with rest mass $\mu$, though we will see later that the tightest constraints
come from the limit $\mu \to 0$. 

The velocity $v$ of the test particle is
\beqn
v = \frac{k}{\sqrt{\mu^2 + k^2}}\,, 
\eeqn
where $k^2 = \omega^2 - \mu^2$, and $k=|\vec k|$. As before, the test particle moves in the $x$ direction. 
The task is now to compute the gravitational field of the test particle and the motion
it causes on the measured particle.

To obtain the metric that the test particle creates, we first change into the rest 
frame of the particle by boosting into $x$-direction. Denoting the new coordinates with
primes, the measured
particle moves towards the test particle in direction $-x'$, and the metric 
is a Schwarzschild metric. We will only need it on the
$x$-axis where we have $y=z=0$, and thus
\beqn
g'_{00} = 1 + 2 \phi' \,, \quad 
g'_{11} = - \frac{1}{g'_{00}} \,, \quad 
g'_{22}=g'_{33} = -1 \,,
\eeqn
where
\beqn
\phi' = \frac{G\mu}{| x' |}\,,
\eeqn
and the remaining components of the metric vanish. Using the transformation law for tensors
\beqn
g_{\mu\nu} = \frac{\partial (x')^\kappa}{\partial x^\mu} \frac{\partial (x')^\alpha}{\partial x^\nu} g'_{\kappa \alpha} \,,
\eeqn
with the notation $x^0 = t, x^1 =x, x^2 =y, x^3 = z$, and the same for the primed coordinates,
the Lorentz boost from the primed to unprimed coordinates yields in the rest frame of the measured particle
\beqn
g_{00} &=& \frac{1 + 2 \phi}{1+ 2\phi(1-v^2)} + 2 \phi \,,\qquad g_{11} = - \frac{-1 + 2 \phi v^2}{1+2 \phi (1-v^2)} + 2 v^2 \phi ,\\
g_{01} &=& g_{10} = - \frac{2v\phi}{1+ \phi (1-v^2)} - 2 v \phi \,,\qquad g'_{22}=g'_{33} = -1 \,, \label{gphi}
\eeqn
where 
\beqn
\phi = \frac{\phi'}{1-v^2} = - \frac{G\omega}{R} \,. \label{defphi}
\eeqn
Here, $R=vt -x$ is the mean distance between the test particle and the measured particle.
To avoid a horizon in the rest frame, we must have $2\phi' <1$, and thus from Eq.~(\ref{defphi}) 
\beqn
- 2 \phi' = 2 \frac{G\omega}{R} (1-v^2) < 1 \,. \label{ineqphi}
\eeqn
Because of Eq.~(\ref{usualdelta}), $\Delta x \geq 1/\omega$ but also $\Delta x \geq R$, which
is the area in which the particle may scatter, thus 
\beqn
\Delta x^2 \gtrsim \frac{R}{\omega} \gtrsim 2 G (1 -v^2) \,.
\eeqn
We see from this that, as long as $v^2 \ll 1$,  the previously found lower bound on the spatial resolution $\Delta x$ can already
be read off here, and we turn our attention towards the case where $1- v^2 \ll 1$. From (\ref{defphi}) we see that
this means we work in the limit where $-\phi \gg 1$. 

To proceed, we need to estimate now how much the measured particle moves due to the test particle's
vicinity. For this, we note that the world line of the measured particle must be 
timelike. We denote the velocity in the $x$-direction with $u$, then
we need
\beqn
ds^2 = \left( g_{00} + 2 g_{10} u + g_{11} u^2 \right) \mathrm{d}t^2 \geq 0 \,. \label{ds2}
\eeqn
Now we insert Eq.~(\ref{gphi}) and follow Mead~\cite{Mead} by introducing the abbreviation
\beqn
\alpha = 1 + 2\phi (1-v^2) \,.
\eeqn
Because of Eq.~(\ref{ineqphi}), $0 < \alpha < 1$. We simplify the requirement of Eq.~(\ref{ds2}) 
by leaving $u^2$ alone on the left side of the inequality, subtracting $1$ and dividing by $u-1$. Taking
into account that $\phi \leq 0$ and $v \leq 1$, one finds after some algebra
\beqn
u \geq \frac{1+2 \phi (1+\alpha)}{1-2\phi v^2(1+\alpha)} \,, 
\eeqn
and
\beqn
\frac{u}{1-u} \geq - \frac{1}{2} \left(1 + 2 \phi \right) . \label{uu1}
\eeqn
One arrives at this estimate with reduced effort if one makes it clear to oneself what we want to estimate. 
We want to know,
as previously, how much the particle, whose position we are trying to measure, will move due to the gravitational
attraction of the particle we are using for the measurement. The faster the particles pass by each other, the
shorter the interaction time and, all other things being equal, the less the particle we want to measure
will move. Thus, if we consider a photon with $v=1$, we are dealing with the case with the least influence, and
if we find a minimal length in this case, it should be there for all cases. Setting $v=1$, one obtains the
inequality Eq.~(\ref{uu1}) with greatly reduced work.

Now we can continue as before in the non-relativistic case. The time $\tau$ required for the test particle to move a distance
$R$ away from the measured particle is at least $\tau \gtrsim R/(1-u)$, and during this time the measured particle
moves a distance
\beqn
L = u \tau \gtrsim R \frac{u}{1-u} \gtrsim \frac{R}{2} \left(-1 - 2 \phi \right) .
\eeqn
Since we work in the limit $- \phi \gg 1$, this means
\beqn
L \gtrsim G \omega \,, \label{meadl}
\eeqn
and projection on the $x$-axis yields as before (compare to Eq.~(\ref{extradelta})) for the uncertainty added to the measured particle
because the photon's direction was known only to precision $\epsilon$
\beqn
\Delta x \gtrsim G \omega \sin \epsilon \,.
\eeqn
This combines with (\ref{usualdelta}), to again give
\beqn
\Delta x \gtrsim l_{\mathrm{Pl}} \,.
\eeqn

Adler and Santiago~\cite{Adler} found the same result by using the linear approximation of Einstein's field equation for
a cylindrical source with length $l$ and radius $\rho$ of comparable size, filled by a radiation field with 
total energy $\omega$, and moving in the $x$ direction. With cylindrical coordinates $x,r,\phi$, the line element takes
the form ~\cite{Adler} 
\beqn
\mathrm{d}s^2 = \mathrm{d}t^2 - \mathrm{d}x^2 - \mathrm{d}y^2 - \mathrm{d}z^2 + f(r,x,t) (\mathrm{dt} -\mathrm{d}x)^2  \,, \label{adlermetric}
\eeqn 
where the function $f$ is given by
\beqn
f(r,x,t) &=& \frac{4 G \omega}{l} g(r) \theta(x-t) \theta(t-x-l) \\
g(r) &=& \left\{ 
\begin{array}{lll} 
r^2/\rho^2 & {\mbox{for}} & r< \rho \\
1 + \ln( r^2/\rho^2) & {\mbox{for}} & r> \rho
\end{array} .
\right. 
\eeqn
In this background, one can then compute the motion of the measured particle by using the Newtonian limit of
the geodesic equation, provided the particle remains non-relativistic. In the longitudinal direction,
along the motion of the test particle one finds
\beqn
\frac{\mathrm{d}^2 x}{\mathrm{d}t^2} = \frac{1}{2} \frac{\partial f}{\partial x} \,.
\eeqn
The derivative of $f$ gives two delta-functions at the front and back of the cylinder with equal momentum transfer but of
opposite direction. The change in velocity to the measured particle is
\beqn
\Delta \dot x = 2 G \frac{\omega}{l} g(r) \,.
\eeqn
Near the cylinder $g(r)$ is of order one, and in the time of passage $\tau \sim l$, the particle thus moves approximately
\beqn
2 G \omega \,,
\eeqn
which is, up to a factor of 2, the same result as Mead's (\ref{meadl}). 
We note that Adler and Santiago's argument does not make use of the requirement that 
no black hole should be formed, but that the appropriateness of the 
non-relativistic and weak-field limit is
questionable.

\subsubsection{Limit to distance measurements}
\label{Wigner}

Wigner and Salecker~\cite{Salecker:1957be} proposed the following thought experiment to show that the precision of length measurements 
is limited. Consider that we try to measure a length by help of a clock that detects photons, which are reflected 
by a mirror at distance $D$ and return to the clock. Knowing the speed of light is universal, from the travel-time of the 
photon we can then extract the distance it has traveled. 
How precisely can we measure the distance in this way? 

Consider that at emission of the photon, we know the position of
the (non-relativistic) clock to precision $\Delta x$. This means, according to the Heisenberg uncertainty principle, 
we cannot know its velocity to better than
\beqn
\Delta v = \frac{1}{2 M \Delta x} \,,
\eeqn
where $M$ is the mass of the clock. During the time $T=2D$ that the photon needed to travel towards the mirror and 
back, the clock moves by $T\Delta v$, and
so acquires an
uncertainty in position of
\beqn
\Delta x + \frac{T}{2 M \Delta x} \,,
\eeqn
which bounds the accuracy by which we can determine the distance $D$. The minimal value that this uncertainty can
take is found by varying with respect to $\Delta x$ and reads
\beqn
\Delta x_{\min} = \sqrt{\frac{T}{2 M}} \,.
\eeqn
Taking into account that our measurement will not be causally connected to the rest of the world if it creates
a black hole, we require $D > 2 M G$ and thus
\beqn
\Delta x_{\min} \gtrsim l_{\mathrm{Pl}} \,.
\eeqn

\subsubsection{Limit to clock synchronization}

\index{Clock synchronization}

From Mead's~\cite{Mead} investigation of the limit for the precision 
of distance measurements due to the gravitational force also follows a limit on the precision by which clocks can be
synchronized. 

We will consider the clock synchronization to be performed by the passing
of light signals from some standard clock to the clock under question.  Since the emission
of a photon with energy spread $\Delta \omega$ by the usual Heisenberg uncertainty 
is uncertain by $\Delta T \sim 1/(2 \Delta \omega)$, we have to take into account the same uncertainty 
for the  synchronization.

The new ingredient comes again from the gravitational field of the photon, which interacts
with the clock in a region $R$ over a time $\tau \gtrsim R$. If the clock (or the part of the
clock that interacts with the photon) remains stationary, the (proper) time it records stands 
in relation to $\tau$ by $T = \tau \sqrt{g_{00}}$
with $g_{00}$ in the rest frame of the clock, given by Eq.~(\ref{gphi}), thus
\beqn
T = \tau \sqrt{1-\frac{4G\omega}{r}} \,.
\eeqn

Since the metric depends on the energy of the photon and this energy is not known precisely, the error on $\omega$ 
propagates into $T$ by 
\beqn
(\Delta T)^2 = \left(\frac{\partial T}{\partial \omega} \right)^2 (\Delta \omega)^2 \,,
\eeqn
thus
\beqn
\Delta T \sim  \frac{2 G \tau}{r \sqrt{1 - 4 G \omega/r}} \Delta \omega \,.
\eeqn
Since in the interaction region $\tau \gtrsim R \gtrsim r$, we can estimate
\beqn
\Delta T \gtrsim \frac{2 G}{\sqrt{1 - 4 G \omega/R}} \Delta \omega  \gtrsim 2 G \Delta \omega \,. \label{deltat}
\eeqn
Multiplication of (\ref{deltat}) with the normal uncertainty $\Delta T \gtrsim 1/(2 \Delta \omega)$ yields
\beqn
\Delta T \gtrsim l_{\mathrm{Pl}} \,. \label{deltatmead}
\eeqn
So we see that the precision by which clocks can be synchronized is also bound by the Planck scale.

However, strictly speaking the clock does not remain stationary during the interaction, since it
moves towards the photon due to the particles' mutual gravitational attraction. If the clock has a velocity $u$, then
the proper time it records is more generally given by
\beqn
T = \int \mathrm{d}s \sim \tau \sqrt{g_{00} + 2g_{01}u + g_{11}u^2} \quad .
\eeqn
Using (\ref{gphi}) and proceeding as before, one estimates the propagation of the error in the frequency by using $v=1$ and $u\leq 1$ 
\beqn
\Big| \frac{\mathrm{d}T}{\mathrm{d}\omega} \Big| \gtrsim \tau \frac{8G}{r} \frac{1}{\sqrt{1 + 4G\omega/r}} \,,
\eeqn
and so with $\tau \gtrsim R \gtrsim r$
\beqn
\Delta T \gtrsim \tau \frac{G}{R} \Delta \omega \gtrsim  G \Delta \omega \,.
\eeqn
Therefore, taking into account that the clock does not remain stationary, one still arrives at (\ref{deltatmead}).

\subsubsection{Limit to the measurement of the black-hole--horizon area}
\index{Black hole}

The above microscope experiment investigates how precisely one can measure the location of a
particle, and finds the precision bounded by the inevitable formation of a black hole. 
However, this position uncertainty
is for the location of the measured particle however and not for the size of the black hole or 
its radius.
There is a simple argument why one would expect there to also be a limit to the precision
by which the size of a black hole can be measured, first put forward in~\cite{Wilczek}. 
When the mass of a black-hole approaches the Planck mass, the horizon radius $R \sim G M$ associated
to the mass becomes
comparable to its Compton wavelength $\lambda = 1/M$. Then, quantum fluctuations in the position of the
black hole should affect the definition of the horizon.

A somewhat more elaborate argument has been studied by Maggiore~\cite{Maggiore} by a thought
experiment that makes use once again of Heisenberg's microscope. However, this time
one wants to measure not the position of a particle, but the area of a (non-rotating) charged black hole's
horizon. In Boyer--Lindquist coordinates, the horizon is located at the radius
\beqn
R_H = GM \left[ 1 + \left(1 - \frac{Q^2}{GM^2} \right)^\frac{1}{2}\right] , \label{reissner}
\eeqn
where $Q$ is the charge and $M$ is the mass of the black hole.

To deduce the area of the black hole, we detect the black hole's Hawking radiation and aim
at tracing it back to the emission point with the best possible accuracy. \index{Black hole}
For the case of an extremal black hole ($Q^2 = GM^2$) the temperature is zero and we
perturb the black hole by sending in photons from asymptotic infinity and wait for
re-emission. 

If the microscope detects a photon of some frequency $\omega$, it is subject to the
usual uncertainty (\ref{usualdelta}) arising from the photon's finite wavelength that
limits our knowledge about the photon's origin. However, in
addition, during the
process of emission the mass of the black hole changes from $M+\omega$ to $M$, and
the horizon radius, which we want to measure, has to change accordingly. If the
energy of the photon is known only up to an uncertainty $\Delta p$, then the error
propagates into the precision by which we can deduce the radius of the black hole
\beqn
\Delta R_H \sim \Big| \frac{\partial R_H}{\partial M}\Big| \Delta p \,.
\eeqn
With use of (\ref{reissner}) and assuming that no naked singularities exist in nature $M^2 G \leq Q^2$ 
one always finds that
\beqn
\Delta R_H \gtrsim 2 G \Delta p \,.
\label{hawkdelta}
\eeqn
In an argument similar to that of Adler and Santiago discussed in Section~\ref{grhm}, Maggiore 
then suggests that the two uncertainties, the usual one inversely proportional to
the photon's energy and the additional one (\ref{hawkdelta}), should be linearly added to
\beqn
\Delta R_H \gtrsim \frac{1}{\Delta p} + \alpha G \Delta p \,,
\eeqn
where the constant $\alpha$ would have to be fixed by using a specific theory. Minimizing
the possible position uncertainty, one thus finds again a minimum error of 
$\approx \alpha l_{\mathrm{Pl}}$. 

It is clear that the uncertainty
Maggiore considered is of a different kind than the one considered by Mead, though both
have the same origin. Maggiore's uncertainty is due
to the impossibility of directly measuring a black hole without it emitting a particle
that carries energy and thereby changing the black-hole--horizon area. The smaller the wavelength
of the emitted particle, the larger the so-caused distortion. Mead's uncertainty is due to the formation
of black holes if one uses probes of too high an energy, which limits the possible precision. But both
uncertainties go back to the relation between a black hole's area and its mass.

\subsubsection{A device-independent limit for non-relativistic particles}
\label{xavier}

Even though the Heisenberg microscope is a very general instrument and the above considerations
carry over to many other experiments, one may wonder if there is not some possibility to
overcome the limitation of the Planck length by use of massive test particles
that have smaller Compton wavelengths, or interferometers that allow one to improve
on the limitations on measurement precisions set by the test
particles' wavelengths. To fill in this gap, Calmet, Graesser and
Hsu~\cite{Calmet:2004mp,Calmet:2005mh} put forward an elegant
device-independent argument. They first consider a discrete spacetime
with a sub-Planckian spacing and then show that no experiment is able
to rule out this possibility. The point of the argument is not the
particular spacetime discreteness they consider, but that it cannot be
ruled out in principle. 

The setting is a position operator $\hat x$ with discrete eigenvalues
$\{x_i\}$ that have a separation of order $l_{\mathrm{Pl}}$ or smaller. To exclude the model, one
would have to measure position eigenvalues $x$ and $x'$, for example, of some
test particle of mass $M$, with $|x-x'| \leq l_{\mathrm{Pl}}$. Assuming the non-relativistic
Schr{\"o}dinger equation without potential, the time-evolution of the position operator is 
given by $\mathrm{d}\hat x(t)/\mathrm{d}t = i [ \hat H, \hat x(t)] = \hat p/M$, and thus
\beqn
\hat x(t) = \hat x(0) + \hat p(0) \frac{t}{M} \,. \label{0t}
\eeqn
We want to measure the expectation value of position at two subsequent times in order
to attempt to measure a spacing smaller than the Planck length. The spectra of any two 
Hermitian operators have to fulfill the inequality 
\beqn
\Delta A \Delta B \geq \frac{1}{2 i} \langle [ \hat A, \hat B] \rangle \,,
\eeqn
where $\Delta$ denotes, as usual, the variance and $\langle \cdot \rangle$ the expectation value of
the operator. From~(\ref{0t}) one has
\beqn
[\hat x(0), \hat x(t) ] = i\frac{t}{M} \,,
\eeqn
and thus
\beqn
\Delta x(0) \Delta x(t) \geq \frac{t}{2M} \,.
\eeqn
Since one needs to measure two positions to determine a distance, the minimal uncertainty
to the distance measurement is 
\beqn
\Delta x \geq \sqrt{\frac{t}{2M}} \label{deltasqrt} \,.
\eeqn

This is the same bound as previously discussed in Section~\ref{Wigner} for the measurement of distances by
help of a clock, yet we arrived here at this bound without making assumptions about exactly
what is measured and how. If we take into account gravity, the argument can be completed similar to
Wigner's and still without making assumptions about the type of measurement, as follows.  

We use an apparatus 
of size $R$. To get the spacing as precise as possible, 
we would use a test particle of high mass. But then we will run into the, by now familiar,
problem of black-hole formation when the mass becomes too large, so we have to require
\beqn 
M < 2 \frac{R}{G} \,.
\eeqn
Thus, we cannot make the detector arbitrarily small. However, we also cannot make it arbitrarily
large, since the components of the detector have to at least be in causal
contact with the position we want to measure, and so $t > R$. Taken together, one
finds
\beqn
\Delta x \geq \sqrt{\frac{t}{2M}} \geq \sqrt{\frac{R}{2M}} \geq \sqrt{G} \,, 
\eeqn
and thus once again the possible precision of a position measurement is limited by
the Planck length.

A similar argument was made by Ng and van~Dam~\cite{Ng:1994zk}, who also pointed out
that with this thought experiment one can obtain a scaling for the uncertainty 
with the third root of the size of the detector. If one adds the position uncertainty
(\ref{deltasqrt}) from the non-vanishing commutator to the gravitational one, one
finds
\beqn
\Delta x \gtrsim \sqrt{\frac{R}{2M}} + G M \,. \label{deltasqrtplus}
\eeqn 
Optimizing this expression with respect to the mass that yields a minimal uncertainty, one finds $M \sim (R/l_{\mathrm{Pl}}^4)^{1/3}$ (up to
factors of order one) and, inserting this value of $M$ in (\ref{deltasqrtplus}), thus
\beqn
\Delta x \gtrsim \left( R l_{\mathrm{Pl}}^2 \right)^{\frac{1}{3}} \,. \label{Ng}
\eeqn
Since $R$ too should be larger than the Planck scale this is, of course, consistent with the
previously-found minimal uncertainty. 

Ng and van Dam further argue that this uncertainty induces a minimum error in measurements
of energy and momenta. By noting that the uncertainty $\Delta x$ of a length $R$ is indistinguishable
from an uncertainty of the metric components used to measure the length, $\Delta x^2 = R^2 \Delta g$, 
the inequality (\ref{Ng}) leads to
\beqn
\Delta g_{\mu\nu} \gtrsim \left( \frac{l_{\mathrm{Pl}}}{R} \right)^{\frac{2}{3}} \,.
\eeqn
But then again the metric couples to the stress-energy tensor $T_{\mu \nu}$, so this 
uncertainty for the metric further induces an uncertainty for the entries of $T_{\mu\nu}$
\beqn
( g_{\mu \nu} + \Delta g_{\mu \nu}) T^{\mu \nu} = g_{\mu \nu} ( T^{\mu \nu} + \Delta T^{\mu \nu} )\,. 
\eeqn 
Consider now using a test particle of momentum $p$ to probe the physics at scale $R$, 
thus $p \sim 1/R$. Then its uncertainty would be on the order of
\beqn
\Delta p \gtrsim p \left( \frac{l_{\mathrm{Pl}}}{R} \right)^{\frac{2}{3}} = p \left( \frac{p}{m_{\mathrm{Pl}}} \right)^{\frac{2}{3}} \,.
\eeqn

However, note that the scaling found by Ng and van Dam only follows if one
works with the masses that minimize the uncertainty (\ref{deltasqrtplus}). Then, even if one uses a 
detector of the approximate
extension of a cm, the corresponding mass of the `particle' we have to work with would be about a ton. 
With such a mass one has to worry about very different uncertainties. For particles with masses
below the Planck mass on the other hand, the size of the detector would have to be below the 
Planck length, which makes no sense since its extension too has to be subject to the minimal 
position uncertainty.

\subsubsection{Limits on the measurement of spacetime volumes}
\label{volumes}

The observant reader will have noticed that almost all of the above estimates have
explicitly or implicitly made use of spherical symmetry. The one exception is 
the argument by Adler and Santiago in Section~\ref{grhm} that employed cylindrical symmetry.  
However, it was also assumed there that the length and the radius of the cylinder
are of comparable size. 

In the general case, when the dimensions of the test particle in different directions
are very unequal, the Hoop conjecture \index{Hoop conjecture} does not forbid any
one direction to be smaller than the Schwarzschild radius to prevent
collapse of some matter distribution, as long as at least
one other direction is larger than the Schwarzschild radius. The question
then arises what limits that rely on black-hole formation can still
be derived in the general case.
\index{Black hole}

A heuristic motivation of the following argument can be found in~\cite{Doplicher:1994tu}, but
here we will follow the more detailed argument by Tomassini and Viaggiu~\cite{Tomassini:2011yu}.
In the absence of spherical symmetry, one may still use Penrose's isoperimetric-type conjecture, 
according to which the apparent horizon is always smaller than or equal to the
event horizon, which in turn is smaller than or equal to $16 \pi G^2 \omega^2$, where $\omega$
is as before the energy of the test particle. 

Then, without spherical symmetry the requirement that no black hole ruins our
ability to resolve short distances is weakened from the energy distribution having a
radius larger than the Schwarzschild radius, to the requirement that
 the area $A$, which encloses $\omega$ is large
enough to prevent Penrose's condition for horizon formation
\beqn
A \geq 16 \pi G^2 \omega^2 \,. \label{16}
\eeqn
The test particle interacts during a time $\Delta T$ that, by the normal uncertainty principle, 
is larger than $1/(2 \omega)$. 
Taking into account this uncertainty on the energy, one has
\beqn
A (\Delta T)^2 \geq 4 \pi G^2 \,. \label{this2}
\eeqn

Now we have to make some assumption for the geometry of the object, which will inevitably be a crude
estimate. While an exact bound will depend on the shape of the matter distribution, we will here
just be interested in obtaining a bound that depends on the three different spatial extensions, and
is qualitatively correct. To that end, we assume 
the mass distribution fits into some smallest box with side-lengths $\Delta x^1, \Delta x^2, \Delta x^3$, which
is similar to the limiting area
\beqn
A \sim \frac{\Delta x^1 \Delta x^2 + \Delta x^1 \Delta x^3 + \Delta x^2 \Delta x^3}{\alpha^2} \,,
\eeqn
where we added some constant $\alpha$ to take into account different possible geometries. A comparison with the 
spherical case, $\Delta x^i = 2 R$, fixes $\alpha^2 = 3/\pi$. With Eq.~(\ref{this2}) one obtains
\beqn
\left(\Delta t \right)^2 \left( \Delta x^1 \Delta x^2 + \Delta x^1 \Delta x^3 + \Delta x^2 \Delta x^3 \right) 
\geq 12 l_{\mathrm{p}}^4 \,.
\eeqn
Since 
\beqn
\left( \Delta x^1 + \Delta x^2 + \Delta x^3 \right)^2 \geq \Delta x^1 \Delta x^2 + \Delta x^1 \Delta x^3 + \Delta x^2 \Delta x^3 
\eeqn
one also has 
\beqn
\Delta t \left( \Delta x^1 + \Delta x^2 + \Delta x^3 \right) 
\label{unvolume}
\geq 12 l_{\mathrm{p}}^2 \,,
\eeqn
which confirms the limit obtained earlier by heuristic reasoning in~\cite{Doplicher:1994tu}.

Thus, as anticipated, taking into account that a black hole must not necessarily form if the spatial extension of a matter
distribution is smaller than the Schwarzschild radius in only one direction, the uncertainty we arrive at here depends on
the extension in all three directions, rather than applying separately to each of them. Here we have replaced $\omega$ by the 
inverse of $\Delta T$, rather than combining with Eq.~(\ref{usualdelta}), but this is just a matter of
presentation.

Since the bound on the volumes (\ref{unvolume}) follows from the bounds on spatial and temporal intervals we
found above, the relevant question here is not whether \ref{unvolumes} 
is fulfilled, but whether the bound $\Delta x \gtrsim l_{\mathrm{Pl}}$
can be violated~\cite{Hossenfelder:2012kk}. 

To address that question, note that the quantities $\Delta x_i$ in the above argument by Tomassini and Viaggiu 
differ from the
ones we derived bounds for in Sections~\ref{sec:heisenberg-microscope-newtonian}\,--\,\ref{xavier}.
Previously, the $\Delta x$
was the precision by which one can measure the position of a particle with help of the test particle. 
Here, the $\Delta x_i$ are the smallest possible extensions of the test particle (in the rest frame),
which with spherical symmetry would just be the Schwarzschild radius. The step in which one studies the motion of the measured particle that is induced by the gravitational 
field of the test particle is missing in this argument. Thus, while the above estimate correctly points out the relevance of
non-spherical symmetries, the argument does not support the conclusion that it is possible to
test spatial distances to arbitrary precision. 

The main obstacle to completion of this argument is that in the context of quantum field theory we
are eventually dealing with particles probing
particles. To avoid spherical symmetry, we would need different objects as probes, which would
require more information about the fundamental nature of matter. We will come back to this point in Section~\ref{branes}.

\subsection{String Theory}
\label{string}
\index{String Theory}

String theory is one of the leading candidates for a theory of quantum gravity. 
Many textbooks have been dedicated to the topic, and the interested reader can also find excellent resources online~\cite{Kiritsis:1997hj,Schwarz:2000ew,Mohaupt:2002py,Szabo:2002ca}. 
For the following we will not need many details. Most importantly, we need to know that a string is 
described by a 2-dimensional surface swept out in a higher-dimensional spacetime. The total
number of spatial dimensions that supersymmetric string theory requires for consistency is nine, i.e., there
are six spatial dimensions in addition to the three we are used to. In the following we will 
denote the total number of dimensions, both time and space-like, with $D$. In this Subsection,
Greek indices run from $0$ to $D$.

The two-dimensional surface swept out by the string in the $D$-dimensional spacetime
is referred to as the `worldsheet,' will be denoted by $X^\nu$, and will be parameterized by (dimensionless)
parameters $\sigma$ and
$\tau$, where $\tau$ is its time-like direction, and $\sigma$ runs conventionally from 0 to $2\pi$. A string has discrete excitations, and its
state can be expanded in a series of these excitations plus the motion of the center of mass. Due to conformal invariance, the worldsheet 
carries a complex structure and thus becomes a Riemann surface, whose complex coordinates we will denote with 
$z$ and $\overline z$. Scattering amplitudes in string theory are a sum over such surfaces. \index{Riemann surface}

In the following $l_{\mathrm{s}}$ is the string scale, and $\alpha' = l_{\mathrm{s}}^2$.
The string scale is related to the Planck scale by $l_{\mathrm{Pl}} = g_{\mathrm{s}}^{1/4} l_{\mathrm{s}}$, where $g_{\mathrm{s}}$ is the string coupling
constant. Contrary to what the name suggests, the string coupling constant is not constant, but depends on
the value of a scalar field known as the dilaton. 
\index{Planck scale}\index{String scale}

To avoid 
conflict with observation, the additional 
spatial dimensions of string theory have to be compactified. The compactification scale is usually 
thought to
be about the Planck length, and far below experimental accessibility. The possibility that
the extensions of the extra dimensions (or at least some of them) might be much larger than the Planck length and thus 
possibly experimentally accessible, has 
been studied in models with a large compactification volume and lowered Planck scale, see, e.g.,~\cite{Abel:2004rp}.\index{Planck scale!lowered} 
We will not discuss these models here, but mention in passing that they demonstrate the
possibility that the `true' higher-dimensional Planck mass is in fact much smaller than $m_{\mathrm{Pl}}$, and correspondingly
the `true' higher-dimensional Planck length, and with it the minimal length, much larger than $l_{\mathrm{Pl}}$. 
That such possibilities exist means, whether or not the model with extra dimensions are realized in nature,
that we should, in principle, consider the minimal length a free parameter that has to be constrained by
experiment. 

String theory is also one of the motivations to look into non-commutative geometries. Non-commutative geometry 
will be discussed
separately in Section~\ref{ncg}. A section on matrix models will be included in a future update.

\subsubsection{Generalized uncertainty}
\index{Generalized uncertainty principle} 

The following argument, put forward by Susskind~\cite{Susskind:1993ki,Susskind:1993aa}, 
will provide us with an insightful examination that illustrates how a string is different from a
point particle and what consequences this
difference has for our ability to resolve structures at shortest distances.
We consider a free string in light cone coordinates, $X_{\pm} = (X^0 \pm X^1)/\sqrt{2}$
with the parameterization $X_+ = 2 l_{\mathrm{s}}^2 P_+ \tau$, where $P_+$ is the momentum in the direction $X_+$ and
constant along the string. In the light-cone gauge, the string has no oscillations in the $X_+$ direction by
construction.

 The transverse dimensions are the remaining $X^i$ with
$i>1$. The normal
mode decomposition of the transverse coordinates has the form
\beqn
X^i(\sigma,\tau) = x^i(\sigma,\tau) + \mathrm{i} \sqrt{\frac{\alpha'}{2}} \sum_{n \neq 0} \left( \frac{\alpha^i_n}{n} e^{\mathrm{i}n(\tau+\sigma)} + \frac{\tilde{\alpha}^i_n}{n} e^{\mathrm{i}n (\tau -\sigma)} \right) , \label{stringexp}
\eeqn
where $x^i$ is the (transverse location of) the center of mass of the string. The coefficients $\alpha^i_n$ and $\tilde \alpha^i_n$ are
normalized to $[\alpha^i_n, \alpha^j_m] = [\tilde \alpha^i_n, \tilde \alpha^j_m] = - \mathrm{i} m \delta^{ij}\delta_{m,-n}$, and $[\tilde \alpha^i_n, \alpha^j_m] = 0$. Since the components $X^\nu$ are real, the coefficients have to fulfill the relations $(\alpha^i_n)^* = \alpha^i_{-n}$ and $(\tilde \alpha^i_n)^* = \tilde \alpha^i_{-n}$. 

We can then estimate the transverse size $\Delta X_\perp$ of the string by
\beqn
(\Delta X_\perp)^2 = \langle \sum_{i=2}^D (X^i - x^i)^2 \rangle \,,
\eeqn
which, in the ground state, yields an infinite sum
\beqn
(\Delta X_\perp)^2 \sim l_{\mathrm{s}}^2 \sum_n \frac{1}{n} \,.
\eeqn
This sum is logarithmically divergent because modes with arbitrarily high frequency are being summed over. To
get rid of this unphysical divergence, we note that testing the string with some energy $E$,
which corresponds to some resolution time $\Delta T = 1/E$, allows us to cut off \index{Cut-off}
modes with frequency $>1/\Delta T$ or mode number $n \sim l_{\mathrm{s}} E$. Then, for large $n$, the sum becomes approximately
\beqn
\Delta X^2_\perp \approx l_{\mathrm{s}}^2 \log ( l_{\mathrm{s}} E ) \,.
\eeqn
Thus, the transverse extension of the string grows
with the energy that the string is tested by, though only very slowly so. 

To determine the spread in the longitudinal direction $X_-$, one needs to know that in light-cone coordinates the constraint equations
on the string have the consequence that $X_-$ is related to the transverse directions so that it is given in
terms of the light-cone Virasoro generators
\beqn
X_-(\sigma,\tau) = x^-(\sigma,\tau) + \frac{\mathrm{i}}{P_+} \sum_{n \neq 0} \left( \frac{L_n}{n} e^{\mathrm{i}n(\tau+\sigma)} + \frac{\tilde{L}_n}{n} e^{ \mathrm{i}n (\tau - \sigma)} \right) ,
\eeqn
where now $L_n$ and $\tilde L_n$ fulfill the Virasoro algebra. Therefore, the longitudinal 
spread in the ground state gains a factor $\propto n^2$ over the transverse case, and  diverges as
\beqn
 (\Delta X_-)^2 \sim \frac{1}{P_+^2} \sum_n  n  \,.
\eeqn
Again, this result has an unphysical divergence, that we deal with the same way as before by taking into account
a finite resolution $\Delta T$, corresponding to the inverse of the energy by which the string is probed.  Then one finds
for large $n$ approximately
\beqn
 (\Delta X_-)^2  \approx \left( \frac{l_{\mathrm{s}}}{P_+} \right)^2 E^2 \,.
\eeqn
Thus, this heuristic argument suggests that the longitudinal spread of the string grows linearly with the energy at which it is
probed.

The above heuristic argument is supported by many rigorous calculations. 
That string scattering leads to a modification of the Heisenberg uncertainty relation has been shown in
several studies of string scattering at high energies performed in the late 1980s~\cite{Gross:1987ar,Veneziano:1989fc,Mende:1989wt}. 
Gross and Mende~\cite{Gross:1987ar} put forward a now well-known analysis of 
the classic solution for the trajectories of a string worldsheet describing a scattering 
event with external momenta $p_i^\nu$. 
In the lowest tree approximation they found for the extension of the string
\beqn
x^\nu(z,\overline z) \approx l_{\mathrm{s}}^2 \sum_i p_i^\nu \log |z - z_i | \,,
\eeqn
plus terms that are suppressed in energy relative to the first. Here, $z_i$ are the positions of the vertex operators on the 
Riemann surface corresponding to the asymptotic states with momenta $p_i^\nu$. Thus, as previously,  
the extension grows linearly with the energy. One also finds that the surface of the string grows with $E/N$, where $N$ is the genus of 
the expansion, and that the fixed angle scattering amplitude at high energies falls exponentially
with the square of the center-of-mass energy $s$ (times $l_{\mathrm{s}}^2$). 

One can interpret this spread of the string in terms of a GUP by
taking into account that at high energies the spread grows linearly with the energy. Together with the normal
uncertainty, one obtains
\beqn
\Delta x^\nu \Delta p^\nu \gtrsim 1 + l_{\mathrm{s}} E \,,
\eeqn
again the GUP that gives rise to a minimally-possible spatial resolution.

However, the exponential fall-off of the tree amplitude depends on the genus of the expansion, and is
dominated by the large $N$ contributions because these decrease slower. The Borel resummation of the
series has been calculated in~\cite{Mende:1989wt} and it was found that the tree level approximation
is valid only for an intermediate range of energies, and for $s l_{\mathrm{s}}^2 \gg g_{\mathrm{s}}^{-4/3}$ the
amplitude decreases much slower than the tree-level result would lead one to expect. 
Yoneya~\cite{Yoneya:2000bt} has furthermore argued that this
behavior does not properly take into account non-perturbative effects, and thus the generalized
uncertainty should not be regarded as generally valid in string theory. We will discuss this in Section~\ref{branes}.


It has been proposed that the resistance of the string to attempts to localize it plays a role in resolving
the black-hole information-loss paradox~\cite{Lowe:1995ac}. In fact, one can wonder if the high energy
behavior of the string acts against and eventually prevents the formation of black holes in elementary
particle collisions. It has been suggested in~\cite{Amati:1987uf,Amati:1987wq,Amati:1990xe} that string effects might become important at impact
parameters far greater than those required to form black holes, opening up the possibility that black holes might not
form.

The completely opposite point of view, that high energy scattering is ultimately entirely dominated by black-hole production,  
has also been put forward~\cite{Banks:1999gd,Giddings:2001bu}. Giddings and Thomas found an indication of how gravity prevents probes of distance 
shorter than the Planck scale~\cite{Giddings:2001bu} and discussed the `the end of short-distance physics'; Banks aptly named 
it `asymptotic darkness'~\cite{Banks:2003vp}. \index{Asymptotic darkness} 
A recent study of string scattering at high energies~\cite{Giddings:2007bw} 
found no evidence that the extendedness of the string interferes with black-hole formation. The subject of string scattering 
in the trans-Planckian regime is subject of ongoing research, see, e.g.,
\cite{Amati:2007ak,Ciafaloni:2009in,Giddings:2010pp} and references therein.

Let us also briefly mention that the spread of the string just discussed should not be confused with the length of the string. (For a 
schematic illustration see Figure \ref{stringlength}.) The length of a string in the transverse 
direction is
\beqn
L = \int \mathrm{d}\sigma \left( \partial_\sigma X^i \partial_\sigma X^i \right)^2  \,, 
\eeqn
where the sum is taken in the transverse direction, and has been
studied numerically in~\cite{Karliner:1988hd}. In this study, it has been shown that when one
 increases the cut-off on the modes, the string becomes 
space-filling, and fills space densely (i.e., it comes arbitrarily close to any point in space). 

\epubtkImage{string.png}{%
  \begin{figure}[ht]
    \centerline{\includegraphics[width=7.0cm]{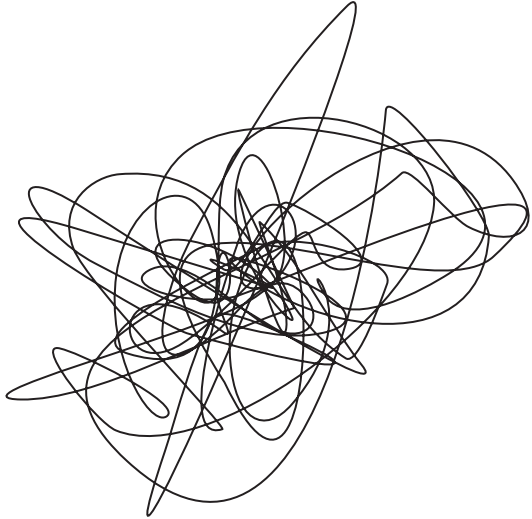}}
    \caption{The length of a string is not the same as its average
      extension. The lengths of strings in the groundstate were
      studied in~\cite{Karliner:1988hd}.}
    \label{stringlength}
\end{figure}}

\subsubsection{Spacetime uncertainty}

Yoneya~\cite{Yoneya:2000bt} argued that the GUP in string theory is not 
generally valid. To begin with, it is not clear whether the Borel resummation of the perturbative expansion leads 
to correct non-perturbative results. And, after the original works on the generalized uncertainty in string 
theory, it has become understood that string theory gives
rise to higher-dimensional membranes that are dynamical objects in their own right. These higher-dimensional
membranes significantly 
change the picture painted by high energy string scattering, as we will see in \ref{branes}.
However, even if the {\sc GUP} is not generally valid, there might be a different uncertainty principle that string 
theory conforms to, that is a spacetime uncertainty\index{Spacetime uncertainty} of the form
\beqn
\Delta X \Delta T \gtrsim l_{\mathrm{s}}^2 \,. \label{stu}
\eeqn

This spacetime uncertainty has been motivated by Yoneya to arise from conformal symmetry~\cite{Yoneya:1989ai,Yoneya:2000bt} as follows.
\index{Conformal invariance}

Suppose we are dealing with a Riemann surface with metric $\mathrm{d}s = \rho(z,\overline z) |\mathrm{d}z|$ that
parameterizes the string. In string theory, these surfaces appear in all path integrals and thus 
amplitudes, and they are thus of central importance for all possible processes. Let us denote with $\Omega$ a
finite region in that surface, and with $\Gamma$ the set of all curves in $\Omega$. The length of some
curve $\gamma \in \Gamma$ is then $L(\gamma, \rho) = \int_\gamma \rho |\mathrm{d}z|$. However, this length that
we are used to from differential geometry is not conformally invariant. To find a length that captures only 
the physically-relevant information, one can use a distance measure known as the `extremal length' $\lambda_{\Omega}$ 
\beqn
\lambda_\Omega(\Lambda) = \sup_\rho \frac{L(\Omega,\rho)^2}{A(\Omega, \rho)} \,,
\eeqn 
with
\beqn
L(\Gamma,\rho) = \inf_{\gamma \in \Lambda} L(\gamma,\rho) \,,\quad
A(\Omega,\rho) = \int_\Omega \rho^2 \, \mathrm{d}z \, \mathrm{d}\overline{z} \,.
\eeqn
The so-constructed length is dimensionless and conformally invariant. For simplicity, we assume that $\Omega$ is a generic polygon 
with four sides and four corners, with pairs of opposite sides named $\alpha,\alpha'$ and $\beta,\beta'$. Any more complicated
shape can be assembled from such polygons. Let 
$\Gamma$ be the set of all curves connecting $\alpha$ with $\alpha'$ and $\Gamma^*$ the set of all curves
connecting $\beta$ with $\beta'$. The extremal lengths $\lambda_\Omega(\Gamma)$ and $\lambda_\Omega(\Gamma^*)$ then
fulfill property~\cite{Yoneya:1989ai,Yoneya:2000bt}
\beqn
\lambda_\Omega(\Gamma^*) \lambda_\Omega(\Gamma) = 1 \,.
\eeqn

Conformal invariance allows us to deform the polygon, so instead of a general four-sided polygon, we can  
consider a rectangle in particular, where the Euclidean length of the sides $(\alpha, \alpha')$ will be 
named $a$ and that of sides $(\beta, \beta')$ will be named $b$. With a Minkowski metric, one of
these directions would be timelike and one spacelike. Then
the extremal lengths are~\cite{Yoneya:1989ai,Yoneya:2000bt}
\beqn
\lambda_\Omega(\Gamma^*) = \frac{b}{a} \,,\quad \lambda_\Omega(\Gamma) = \frac{a}{b} \,.
\eeqn
Armed with this length measure, let us consider the Euclidean path integral in the conformal gauge ($g_{\mu \nu} = \eta_{\mu \nu}$)
with the action
\beqn
\frac{1}{l_{\mathrm{s}}^2}\int_\Omega \mathrm{d}z \,
\mathrm{d}\overline{z} \, \partial_z X^\nu \partial_{\bar z} X^\nu \,.
\eeqn
(Equal indices are summed over). As before, $X$ are the 
target space coordinates of the string worldsheet. We now decompose the coordinate $z$ into its real and imaginary part 
$\sigma_1 = \mathrm{Re}(z),\sigma_2=\mathrm{Im}(z)$, and consider a rectangular piece of the surface with the boundary conditions
\beqn
X^\nu(0,\sigma_2) &=& X^\nu(a,\sigma_2) = \delta^{\nu 2} B \frac{\sigma_2}{b} \,, \nonumber\\
X^\nu(\sigma_1, 0) &=& X^\nu(\sigma_1,b) = \delta^{\nu 1} A \frac{ \sigma_1}{a} \,. 
\eeqn
If one integrates over the rectangular region, the action contains a factor $ab( (B/b)^2 + (A/a)^2)$ and the 
path integral thus contains a factor of the form 
\beqn
\exp\left( - \frac{1}{l_{\mathrm{s}}^2} \left( \frac{A^2}{\lambda(\Gamma)} + \frac{B^2}{\lambda(\Gamma^*)}\right) \right) .
\eeqn
Thus, the width of these contributions is given by the extremal length times the string scale, which quantifies the variance of $A$ and $B$ by
\beqn
\Delta A \sim l_{\mathrm{s}} \sqrt{\lambda(\Gamma)}\,,\quad \Delta B \sim l_{\mathrm{s}} \sqrt{\lambda(\Gamma^*)}\,.
\eeqn
In particular the product of both satisfies the condition
\beqn
\Delta A \Delta B \sim l_{\mathrm{s}}^2 \,.
\eeqn
Thus, probing short distances along the spatial and temporal directions simultaneously is not possible to
arbitrary precision, lending support to the existence of a spacetime uncertainty of the form (\ref{stu}). 
Yoneya notes~\cite{Yoneya:2000bt} that this argument cannot in this simple fashion 
be carried over to more complicated shapes. Thus, at present the spacetime uncertainty has the
status of a conjecture. However, the power of this argument rests in it only relying on
conformal invariance, which makes it plausible that, in contrast to the {\sc GUP}, it is universally and non-perturbatively valid.

\subsubsection{Taking into account Dp-Branes}
\index{Dirichlet-branes}
\label{branes}

The endpoints of open strings obey boundary conditions, either of the Neumann type or of the Dirichlet type or a mixture of both. For Dirichlet boundary conditions, the submanifold on which open strings end is called a Dirichlet brane, or Dp-brane for short, where p is an integer denoting the dimension
of the submanifold. A D0-brane is a point, sometimes called a D-particle; a D1-brane is a one-dimensional object, also called a D-string; and
so on, all the way up to D9-branes. 

These higher-dimensional objects that arise in
string theory have a dynamics in their own right, and have
given rise to a great many insights, especially with respect to dualities between different sectors of the theory, and the study of higher-dimensional black holes~\cite{Johnson:2000ch,Bachas:1998rg}.\index{Black hole!higher dimensional}

Dp-branes have a tension of 
$T_{p} = 1/(g_{\mathrm{s}} l_{\mathrm{s}}^{p+1})$; that is, in the weak coupling limit, they become very rigid. Thus, one might suspect D-particles to show evidence for structure on distances at least down to $l_{\mathrm{s}}g_{\mathrm{s}}$.

Taking into account the scattering of Dp-branes indeed changes the conclusions we
could draw from the earlier-discussed thought experiments. We have seen that this was already the case for strings, but we 
can expect that Dp-branes change the picture even more dramatically. At high energies, strings can convert energy into potential energy,
thereby increasing their extension and counteracting the attempt to probe small distances. Therefore, strings do not make
good candidates to probe small structures, and to probe the structures of Dp-branes, one would best scatter them off
each other. As Bachas put it~\cite{Bachas:1998rg}, the ``small dynamical scale of D-particles cannot be seen by using fundamental-string probes -- one cannot probe a
needle with a jelly pudding, only with a second needle!'' 

That with Dp-branes new scaling behaviors enter the physics of shortest distances has been pointed out by
Shenker~\cite{Shenker:1995xq}, and in particular the D-particle scattering has been studied in great detail by
Douglas et al.~\cite{Douglas:1996yp}. It was shown there
that indeed slow moving D-particles can probe distances below the (ten-dimensional) Planck scale and even below the string 
scale. For these D-particles, it has been found that structures exist down to $g_{\mathrm{s}}^{1/3}l_{\mathrm{s}}$.

To get a feeling for the scales involved here, let us first reconsider the scaling arguments on black-hole formation,
now in a higher-dimensional spacetime.
The Newtonian potential $\phi$ of a higher-dimensional point charge with energy $E$, or the perturbation of 
$g_{00} = 1 + 2\phi$, in $D$ dimensions, is qualitatively of the form
\beqn
\phi \sim \frac{G_D E}{(\Delta x)^{D-3}} \,,
\eeqn
where $\Delta x$ is the spatial
extension, and $G_D$ is the $D$-dimensional Newton's constant, related to the Planck length as $G_{D} = l_{\mathrm{Pl}}^{D-2}$. 
Thus, the horizon or the zero of $g_{00}$
is located at
\beqn
\Delta x \sim \left( G_D E \right)^{\frac{1}{D-3}} \,.
\eeqn
With $E \sim 1/\Delta t$, for some time by which we test the geometry, to prevent black-hole formation for $D=10$, one
thus has to require 
\beqn
(\Delta t) (\Delta x)^7 \gtrsim G_{10} = g^2_{\mathrm{s}} l^8_{\mathrm{s}} \,, \label{brane}
\eeqn
re-expressed in terms of string coupling and tension. We see that in the weak coupling limit, this lower bound can be small, 
in particular it can be much below the string scale. 

This relation between spatial and temporal resolution can now
be contrasted with the spacetime uncertainty (\ref{stu}), that sets the limits below which the classical
notion of spacetime ceases to make sense. Both of these limits are shown in Figure \ref{figureyoneya} for comparison. The
curves meet at
\beqn 
\Delta x_{\min} \sim l_{\mathrm{s}} g_{\mathrm{s}}^{1/3} \,,\quad
\Delta t_{\min} \sim l_{\mathrm{s}} g_{\mathrm{s}}^{-1/3} \,. \label{cross}
\eeqn
If we were to push our limits along the bound set by the spacetime uncertainty (red, solid line), then the best possible spatial resolution
we could reach lies at $\Delta x_{\min}$, beyond which black-hole production takes over.
Below the spacetime uncertainty limit, it would actually become meaningless to talk about black holes 
that resemble any classical object. 

\epubtkImage{stu3.png}{%
  \begin{figure}[ht]
    \centerline{\includegraphics[width=8.0cm]{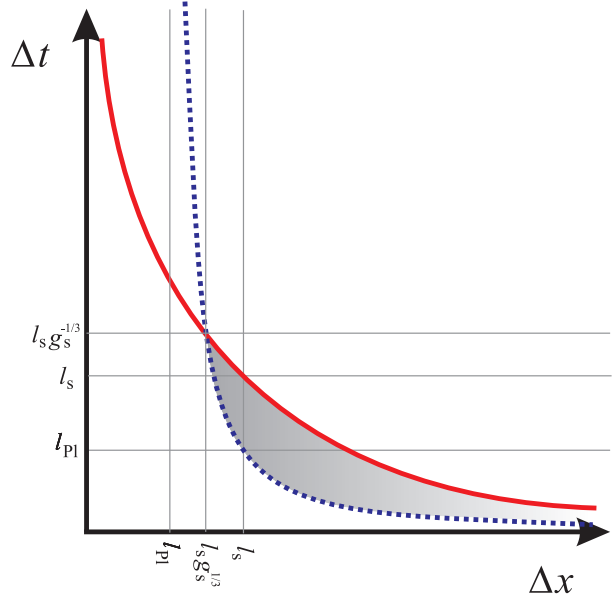}}
    \caption{Spacetime uncertainty (red, solid) vs uncertainty from
      spherical black holes (blue, dotted) in $D=10$ dimensions, for
      $g_{\mathrm{s}}<1$ (left) and $g_{\mathrm{s}}>1$
      (right). After~\cite{Yoneya:2000bt}, Figure~1. Below the bound from
      spacetime uncertainty yet above the black-hole bound that hides
      short-distance physics (shaded region), the concept of classical
      geometry becomes meaningless.}
    \label{figureyoneya}
\end{figure}}

At first sight, this argument seems to suffer from the same problem as the previously examined argument for
volumes in Section~\ref{volumes}. Rather than combining $\Delta t$ with $\Delta x$ to arrive at a weaker
bound than each alone would have to obey, one would have to show that in fact $\Delta x$ can become 
arbitrarily small. And, since the argument
from black-hole collapse in 10 dimensions is essentially the same as Mead's in 4 dimensions, just with a different $r$-dependence
of $\phi$, if one would consider point particles in 10 dimensions, one finds along the same 
line of reasoning as in Section~\ref{grhm}, that actually $\Delta t \gtrsim l_{\mathrm{Pl}}$ and $\Delta x \gtrsim l_{\mathrm{Pl}}$. 

However, here the situation is very different because fundamentally the objects we are dealing with
are not particles but strings, and the interaction between Dp-branes is mediated by strings stretched
between them. It is an inherently different behavior than what we can expect from the classical gravitational
attraction between point particles. 
At low string coupling, the coupling of gravity is weak and in this limit then, the backreaction 
of the branes on the background becomes negligible. For these reasons, the D-particles distort each other 
less than point particles in a quantum field theory would, and this is what allows one to use them 
to probe very short distances.

The following estimate from~\cite{Yoneya:2000bt} sheds light on the scales that we can test with D-particles
in particular. Suppose we use D-particles with velocity $v$ and mass $m_0=1/(l_{\mathrm{s}} g_{\mathrm{s}})$
to probe a distance of size $\Delta x$ in time $\Delta t$. Since $v \Delta t \sim \Delta x$, the uncertainty (\ref{brane}) gives
\beqn
(\Delta x )^8 \gtrsim v g_{\mathrm{s}}^2 l_{\mathrm{s}}^8 \,. \label{bh10}
\eeqn
thus, to probe very short distances one has to use slow D-particles. 

But if the D-particle is slow, then its wavefunction behaves like
that of a massive non-relativistic particle, so we have to take into account that the width spreads with time. For this, 
we can use the earlier-discussed bound Eq.~(\ref{deltasqrt})
\beqn
\Delta x_{\mathrm{spread}} \gtrsim \sqrt{\frac{\Delta t}{2m_0}} \,,
\eeqn
or
\beqn
\Delta x_{\mathrm{spread}} \gtrsim \frac{l_{\mathrm{s}} g_{\mathrm{s}}}{2 v} \,. \label{spread}
\eeqn
If we add the uncertainties (\ref{bh10}) and (\ref{spread}) and minimize the sum with respect to $v$, we find
that the spatial uncertainty is minimal for
\beqn
v \sim g_{\mathrm{s}}^{2/3} \,.
\eeqn
Thus, the total spatial uncertainty is bounded by
\beqn
\Delta x \gtrsim l_{\mathrm{s}} g_{\mathrm{s}}^{1/3} \,,
\eeqn
and with this one also has 
\beqn
\Delta t \gtrsim l_{\mathrm{s}} g_{\mathrm{s}}^{-1/3} \,,
\eeqn
which are the scales that we already identified in (\ref{cross}) to be those of the best possible resolution
compatible with the spacetime uncertainty. Thus, we see that the D-particles saturate the spacetime 
uncertainty bound and they can be used to test these short distances.
 
D-particle scattering has been studied in~\cite{Douglas:1996yp} by use of a quantum mechanical 
toy model in which the two particles are interacting by (unexcited) open strings stretched between them. The open strings 
create a linear potential between the branes. At moderate velocities, repeated collisions can take place, since the 
probability for all the open strings to annihilate between one collision and the next is small. At $v \sim g_{\mathrm{s}}^{2/3}$, 
the time between
collisions is on the order of $\Delta t \sim l_{\mathrm{s}} g^{-1/3}$, corresponding to a resonance of width 
$\Gamma \sim g_{\mathrm{s}}^{1/3}/l_{\mathrm{s}}$. By considering the conversion of kinetic energy into the potential of the strings, one
sees that the particles reach a maximal separation of $\Delta x \sim l_{\mathrm{s}}g_{\mathrm{s}}^{-1/3}$, realizing a test of
the scales found above.

Douglas et al.~\cite{Douglas:1996yp} offered a useful analogy of the involved scales to atomic physics; 
see Table~(\ref{douglastable}). The electron in a hydrogen atom moves with velocity determined by the
fine-structure constant $\alpha$, from which it follows the characteristic size of the atom. For the
D-particles, this corresponds to the maximal separation in the repeated collisions. The analogy 
may be carried further than that in that higher-order corrections
should lead to energy shifts.

\begin{table}
  \caption[Analogy between scales involved in \textit{D}-particle scattering
    and the hydrogen atom.]{Analogy between scales involved in
    \textit{D}-particle scattering and the hydrogen
    atom. After~\cite{Douglas:1996yp}.}
  \label{douglastable}
\centering
\renewcommand{\arraystretch}{1.3}
\begin{tabular}{ l l }
  \toprule
  \textbf{Electron}  & \textbf{\textit{D}-particle} \\
  \midrule
  mass $m_{\mathrm{e}}$  & mass $m_0 = 1/(l_{\mathrm{s}} g_{\mathrm{s}})$  \\
  Compton wavelength $1/m_{\mathrm{e}}$ & Compton wavelength $1/m_0 \sim g_{\mathrm{s}} l_{\mathrm{s}}$ \\
  velocity $\alpha$ & velocity $v = g_{\mathrm{s}}^{2/3}$ \\
  Bohr radius $\sim 1/(\alpha m_{\mathrm{e}})$ & size of resonance $1/(m_0 v) \sim g_{\mathrm{s}}^{1/3} l_{\mathrm{s}}$ \\
  energy levels $\sim \alpha^2 m_{\mathrm{e}}$ & resonance energy $\sim m_0 v^2 \sim g_{\mathrm{s}}^{1/3}/l_{\mathrm{s}}$ \\
  fine structure $\sim \alpha^4 m_{\mathrm{e}}$ & energy shifts $\sim m_0 v^4 \sim g_{\mathrm{s}}^{5/3}/l_{\mathrm{s}}$ \\
  \bottomrule
\end{tabular}
\end{table}

The possibility
to resolve such short distances with D-branes have been studied in many more calculations; 
for a summary, see, for example,~\cite{Bachas:1998rg} and references therein. For our purposes, this estimate of
scales will be sufficient. We take away that D-branes, should they exist, would allow us to probe distances
down to $\Delta x \sim g_{\mathrm{s}}^{1/3} l_{\mathrm{s}}$. 
 
\subsubsection{T-duality}
\index{T-duality}
\label{tdual}

In the presence of compactified spacelike dimensions, a string can acquire an entirely new property: It can wrap
around the compactified dimension. The number of times it wraps around, labeled by the integer $w$, is called 
the `winding-number.'
For simplicity, let us consider only one additional dimension, compactified on a radius $R$. Then, in the
direction of this coordinate, the string has to obey the boundary condition
\beqn
X^4(\tau, \sigma + 2 \pi) = X^4(\tau,\sigma) + 2 \pi w R \,. 
\eeqn 

The momentum in the direction of the additional coordinate is quantized in multiples of $1/R$, so the expansion (compare to Eq.~(\ref{stringexp})) reads
\beqn
X^4(\tau,\sigma) = x_0^4 + \frac{n l_{\mathrm{s}}^2}{R} \tau + w R \sigma + 
\mathrm{i} \frac{\alpha'}{2} \sum_{n \neq 0} \left( \frac{\alpha^i_n}{n} e^{\mathrm{i}n(\tau+\sigma)} + \frac{\tilde{\alpha}^i_n}{n} e^{\mathrm{i}n (\tau -\sigma)} \right) ,
\eeqn
where $x_0^4$ is some initial value. The momentum $P^4 = \partial_\tau x^4/(l_{\mathrm{s}}^2)$ is then
\beqn
P^4(\tau,\sigma) = \frac{n}{R} \tau + 
\frac{\mathrm{i}}{\sqrt{2} l_{\mathrm{s}}} 
\sum_{n \neq 0} \left( \alpha^i_n e^{\mathrm{i}n(\tau+\sigma)} + \tilde{\alpha}^i_n e^{\mathrm{i}n (\tau -\sigma)} \right) .
\eeqn

The total energy of the quantized string with excitation $n$ and winding number $w$ is formally divergent, due to the contribution of all the oscillator's zero point energies, and has to be renormalized. After renormalization, the energy is
\beqn
E^2 &=& \sum^3_{\mu =0} P^\mu P_{\mu} + m^2 \quad \mbox{with}\\
m^2 &=& \frac{n^2}{R^2} +  w^2 \frac{R^2}{\alpha'^2} + \frac{2}{\alpha'} \left( N + \tilde N - \frac{D-2}{12} \right) , \label{massex}
\eeqn
where $\mu$ runs over the non-compactified coordinates, and $N$ and $\tilde N$ are the levels of excitations of the left and right moving modes. Level
matching requires $nw + N - \tilde N = 0$. In addition to the normal contribution from the linear momentum, the string energy
thus has a geometrically-quantized contribution from the momentum into the extra dimension(s), labeled with $n$, an energy from the winding
(more winding stretches the string and thus needs energy), labeled with $w$, and a renormalized contribution from the Casimir energy. The 
important thing to note here is that this expression is invariant under the exchange
\beqn
R \leftrightarrow \frac{l_{\mathrm{s}}^2}{R} \,,\quad n \leftrightarrow w \,,
\eeqn
i.e., an exchange of winding modes with excitations leaves mass spectrum invariant. 

This symmetry is known as 
target-space duality, or T-duality for short. It carries over to multiples extra dimensions, and can be shown to hold not only for the free string but 
also during interactions. This means that for
the string a distance below the string scale $\sim l_{\mathrm{s}}$ is meaningless because it corresponds to a distance
larger than that; pictorially, a string that is highly excited also has enough energy
to stretch and wrap around the extra dimension. We have seen in Section~\ref{branes} that Dp-branes overcome
limitations of string scattering, but T-duality is a simple yet powerful
way to understand why the ability of strings to resolves short distances is limited.

This characteristic property of string theory has motivated a model that incorporates T-duality and compact extra dimensions into an effective 
path integral approach for a particle-like object that is described by the center-of-mass of the
string, yet with a modified Green's function, suggested in~\cite{Smailagic:2003hm,Fontanini:2005ik,Spallucci:2005bm}. 

In this approach it is assumed that the elementary constituents of matter are fundamentally strings 
that propagate in a higher dimensional spacetime with compactified additional dimensions, so that the strings
can have excitations and winding numbers. By taking into account the 
excitations and winding numbers, Fontanini et al.~\cite{Smailagic:2003hm,Fontanini:2005ik,Spallucci:2005bm} 
derive a modified Green's function for a
scalar field. In the resulting double sum over $n$ and $w$, the contribution from the $n=0$
and $w=0$ zero modes is dropped. Note that this discards all massless modes as one sees from Eq.~(\ref{massex}). 
As a result, the Green's function obtained in this way 
no longer has the usual contribution
\beqn
G(x,y) = - \frac{1}{(x-y)^2} \,.
\eeqn
Instead, one finds in momentum space
\beqn
G(p) = \sum_{N=0}^\infty \sum_{w,n=1}^\infty \frac{n l_0}{\sqrt{p^2 +
    m^2}} K_1 \left( n l_0 \sqrt{p^2 + m^2} \right) , \label{greensum}
\eeqn
where the mass term is given by Eq.~(\ref{massex}) and a function of $N,n$ and $w$. 
Here, $K_1$ is the modified Bessel function of the first kind, and $l_0 = 2\pi R$ is the compactification scale of the extra dimensions.
For $n=w=1$ and $N=0$, in the limit where $p^2 \gg m^2$ and the argument of $K_1$ is large compared to 1, $p^2 \gg 1/l^2_0$, the modified Bessel function can
be approximated by
\beqn
K_1 \to \frac{4 \pi^2}{\sqrt{p l_0}}  \exp \left( -l_0 \sqrt{p^2} \right),
\eeqn
and, in that limit, the term in the sum (\ref{greensum}) of the Green's function takes the form 
\beqn
\to \frac{\sqrt{l_0}}{p^{3/2}} \exp \left (-l_0 \sqrt{p^2} \right) .
\eeqn
Thus, each term of the modified Green's function falls off exponentially if the energies are large enough. 
The Fourier transform of this limit of the momentum space propagator is
\beqn
G(x,y) \approx \frac{1}{4 \pi^2} \frac{1}{(x-y)^2 + l_0^2} \,, \label{greens}
\eeqn
and one thus finds that the spacetime distance in the propagator acquires a finite correction term, which one can
interpret as a `zero point length', at least in the Euclidean case. 

It has been argued in~\cite{Smailagic:2003hm} that this ``captures the leading order correction from string theory''. 
This claim has not been supported by independent studies. However, this argument has been used as one of the motivations
for the model with path integral duality that we will discuss in Section~\ref{pathint}. The interesting thing to note here is
that the minimal length that appears in this model is not determined by the Planck length, but by the radius of
the compactified dimensions. It is worth emphasizing that this approach is manifestly Lorentz invariant.
\index{Zero-point length}

\subsection{Loop Quantum Gravity and Loop Quantum Cosmology} 
\label{loop}
\index{Loop Quantum Gravity}
\index{Loop Quantum Cosmology}
\index{LQG|see{Loop Quantum Gravity}}
\index{LQC|see{Loop Quantum Cosmology}}

Loop Quantum Gravity ({\sc LQG}) is a quantization of gravity by help of carefully constructed suitable
variables for quantization, variables that have become known as the Ashtekar variables~\cite{AA}. \index{Ashtekar variables}  
While {\sc LQG} theory still lacks experimental confirmation, during the last
two decades it has blossomed into an established research area. Here we will only roughly sketch
the main idea to see how it entails a minimal length scale. For technical details, the
interested reader is referred to the more specialized reviews~\cite{Ashtekar:2004eh,Thiemann:2006cf,Thiemann:2007zz,Mercuri:2010xz,Jorge}.

Since one wants to work with the Hamiltonian framework, one begins with the familiar 3+1 split of 
spacetime. That is, one assumes that spacetime has topology
${\mathbb{R}} \times \Sigma$, i.e., it can be sliced into
a set of spacelike 3-dimensional hypersurfaces. Then, the metric can be parameterized with the
lapse-function $N$ and the shift vector $N_i$
\beqn
\mathrm{d}s^2 = (N^2 - N_a N^a) \, \mathrm{d}t^2 - 2 N_a \, \mathrm{d}t \,
\mathrm{d}x - q_{ab} \, \mathrm{d}x^a \, \mathrm{d}x^b \,,
\eeqn
where $q_{ij}$ is the three metric on the slice. The three metric by itself does not suffice to completely 
describe the four dimensional spacetime. If one wants to stick with quantities that make sense on the
three dimensional surfaces, in order to prepare for quantization, one needs in addition 
the `extrinsic curvature' that describes how
the metric changes along the slicing
\beqn
K_{ab} = \frac{1}{2N} \left(\nabla_k N_i + \nabla_i N_k - \partial_t q_{ij} \right) , 
\eeqn
where $\nabla$ is the covariant three-derivative on the slice. So far, one is
used to that from general relativity. 

Next we introduce the triad or dreibein, $E_i^a$, which is a set of three vector fields
\beqn
q^{ab} = E^a_i E^b_j \delta^{ij} \,.
\eeqn
The triad converts the spatial indices $a,b$ (small, Latin, from the beginning of the alphabet)
to a locally-flat metric with indices $i,j$ (small, Latin, from the middle of the alphabet). The
densitized triad
\beqn
\tilde E_i^a = \sqrt{\det q} E_i^a,
\eeqn
is the first set of variables used for quantization. The other set of variables is an $\mathrm{su}(2)$ 
connection $A_a^i$, which is related to the connection on the manifold and the external curvature by
\beqn
A^i_a = \Gamma^i_a + \beta K^i_a \,,
\eeqn
where the $i$ is the internal index and $\Gamma^i_a = \Gamma_{ajk}\epsilon^{jki}$ is the spin-connection. 
The dimensionless constant $\beta$ is the `Barbero--Immirzi parameter'. Its value can be fixed by requiring the
black-hole entropy to match with the semi-classical case, and comes out to be of order one. 

From the triads one can reconstruct the internal metric, and from $A$ and the
triad, one can reconstruct the extrinsic curvature and thus one has a full description of spacetime. 
The reason for this somewhat cumbersome reformulation of general relativity is that these variables do not
only recast gravity as a gauge theory, but are also canonically conjugated 
in the classical theory
\beqn
\left\{ A^i_a(x), \tilde E^b_j(y) \right\} = \beta \delta_a^b \delta_j^i \delta^3(x-y) \,,
\eeqn 
which makes them good candidates for quantization. And so, under quantization one promotes $A$ and $E$ 
to operators $\hat A$ and $\hat E$ and replaces the 
Poisson bracket with commutators,
\beqn
\left[\hat A^j_b(x), \hat{\tilde E}^a_i(y) \right] = \mathrm{i} \beta \delta_a^b \delta_j^i \delta^3(x-y) \,.
\eeqn
The Lagrangian of general relativity can then be rewritten in terms of the new variables, 
and the constraint equations can be derived.  

In the so-quantized theory one can then work with different representations, like one works in quantum mechanics with
the coordinate or momentum representation, just more complicated. One such representation is the loop representation, an expansion
of a state in a basis of (traces of) holonomies around all possible closed loops. However, this basis
is overcomplete. A more suitable basis are spin networks $\psi_s$. Each such spin network is a graph with vertices and edges that carry 
labels of the respective $\mathrm{su}(2)$ representation. In this basis, the states of {\sc LQG} are then 
closed graphs, the edges of which 
are labeled by irreducible $\mathrm{su}(2)$ representations and the vertices by $\mathrm{su}(2)$ 
intertwiners. 

The details of this approach to quantum gravity are far outside the scope of this review; 
for our purposes we will just note that with this
quantization scheme, one can construct operators for areas and volumes, and with the expansion in the
spin-network basis $\psi_s$, one can calculate the eigenvalues of these operators, roughly as follows.

Given a two-surface $\Sigma$ that is
parameterized by two coordinates $x^1,x^2$ with the third coordinate $x^3 = 0$ on the surface, the area of the surface is
\beqn
A_\Sigma = \int_\Sigma \mathrm{d}x^1 \, \mathrm{d}x^2 \sqrt{\det q^{(2)}} \,,
\eeqn
where $\det q^{(2)} = q_{11}q_{22}-q_{12}^2$ is the metric determinant on the surface. In terms of
the triad, this can be written as
\beqn
A_\Sigma = \int_\Sigma \mathrm{d}x^1 \, \mathrm{d}x^2 \sqrt{\tilde E^3_i \tilde E^{3i}} \,.
\eeqn
This area can be promoted to an operator, essentially by making the triads operators, though to deal 
with the square root of a product of these operators one has to average the operators over smearing
functions and take the limit of these smearing functions to delta functions. One can then act with the so-constructed operator on the states of the spin network and obtain the eigenvalues
\beqn
\hat A_\Sigma \psi_s = 8 \pi l^2_{\mathrm{Pl}} \beta \sum_I \sqrt{j_I(j_I + 1)} \psi_s \,,
 \eeqn
where the sum is taken over all edges of the network that pierce the surface $\Sigma$, and $j_I$, a positive half-integer, are 
 the representation labels on the edge. This way, one finds that {\sc LQG} has a minimum area of 
\beqn
A_{\min} = 4 \pi \sqrt{3} \beta l^2_{\mathrm{Pl}} \,.
\eeqn 

A similar argument can be made for the volume operator, which also has a finite smallest-possible eigenvalue on the order
of the cube of the Planck length~\cite{Rovelli:1994ge,Thiemann:1996au,Ashtekar:1997fb}. These properties then lead 
to the following interpretation of the spin network:
the edges of the graph represent quanta of area with area $\sim l_{\mathrm{P}}^2 \sqrt{j(j+1)}$, and the vertices of the graph represent quanta of 3-volume. 

Loop Quantum Cosmology ({\sc LQC}) is a simplified version of LQG, developed
to study the time evolution of cosmological, i.e., highly-symmetric, models. The main simplification is that, rather than using the full quantized
theory of gravity and then studying models with suitable symmetries, one first reduces the symmetries and then quantizes
the few remaining degrees of freedom. 

For the quantization of the degrees of freedom one uses techniques similar to
those of the full theory. {\sc LQC} is thus not strictly speaking derived from {\sc LQG}, but an
approximation known as the `mini-superspace approximation.' For arguments why it is plausible
to expect that {\sc LQC} provides a reasonably good approximation and for a detailed treatment, the
reader is referred to~\cite{Ashtekar:2003hd,Ashtekar:2011ni,Bojowald:2008zzb,Bojowald:2007bg,MenaMarugan:2011me}. Here we will only pick out one
aspect that is particularly interesting for our theme of the minimal length.

In principle, one works in {\sc LQC} with operators for the triad and the connection, yet the semi-classical treatment
captures the most essential features and will be sufficient for our purposes. Let us first briefly recall the 
normal cosmological Friedmann--Robertson--Walker model coupled to scalar field $\phi$ in the new variables~\cite{Jorge}.
The ansatz for the metric is
\beqn
\mathrm{d}s^2 = - \mathrm{d}t^2 + a^2(t)\left( \mathrm{d}x^2 + \mathrm{d}y^2 + \mathrm{d}z^2 \right), \quad
\eeqn
and for the Ashtekar variables 
\beqn
A^i_a = c \delta^i_a \,, \quad \tilde E^a_i = p \delta^a_i \,.
\eeqn
The variable $p$ is dimensionless and related to the scale factor as $a^2 = |p|$, and $c$ has dimensions of energy. $c$ and $p$ are canonically 
conjugate and normalized so that the Poisson brackets are
\beqn
\left\{c,p\right\} = \frac{8\pi}{3} \beta \,.
\eeqn
The Hamiltonian constraint for gravity coupled to a (spatially homogeneous) pressureless scalar field with canonically conjugated 
variables $\phi,p_\phi$ is
\beqn
16 \pi G {\cal H} = - \frac{6}{\beta^2} c^2 |p|^{1/2} + 8 \pi G p^2_\phi |p|^{-3/2} = 0 \,. \label{frwham}
\eeqn
This yields
\beqn
\frac{c}{\beta} = 2 \sqrt{\frac{\pi}{3}} l_{\mathrm{Pl}} \frac{p_\phi}{|p|} \,. \label{cbeta}
\eeqn
Since $\phi$ itself does not appear in the Hamiltonian, the conjugated momentum $p_\phi$ is a constant of motion $\dot p_\phi = 0$, where
a dot denotes a derivative with respect to $t$. The equation of motion for $\phi$ is
\beqn
\dot \phi = \frac{p_\phi}{p^{3/2}} \,, \label{dotphi}
\eeqn
so we can identify 
\beqn
\rho_\phi = \frac{p_\phi^2}{2 |p|^3} 
\eeqn
as the energy density of the scalar field. With this, Equation (\ref{dotphi}) can be written in the more
familiar form
\beqn
\dot \rho_\phi = - \frac{3}{2} \frac{\dot p}{p^4} p_\phi^2 = -3 \frac{\dot a}{a} \rho_\phi \,. \label{dotrho}
\eeqn
The equation of motion for $p$ is
\beqn
\dot p = - \frac{8\pi}{3} \beta \frac{\partial {\cal H}}{\partial c} = 2 \frac{c}{\beta} |p|^{1/2} \,.
\eeqn
Inserting (\ref{cbeta}), this equation can be integrated to get
\beqn
p^{3/2} = 2 \sqrt{3 \pi} l_{\mathrm{Pl}} p_\phi t \,. \label{p32}
\eeqn
One can rewrite this equation by introducing the Hubble parameter $H = \dot a/a = \dot p/(2 p)$; then one finds
\beqn
H^2 = \frac{4 \pi}{3} l_{\mathrm{Pl}}^2 \frac{p_\phi^2}{p^3} = \frac{8 \pi}{3} G \rho_\phi \,,
\eeqn
which is the familiar first Friedmann equation. Together with the energy conservation (\ref{dotrho}) this fully 
determines the time evolution.

Now to find the Hamiltonian of {\sc LQC}, one considers an elementary cell that is repeated in all spatial directions because space is homogeneous. The holonomy around a loop is then just given by $\exp( \mathrm{i} \mu c )$,
where $c$ is as above the one degree of freedom in $A$, and $\mu$ is the edge length of the elementary
cell. We cannot shrink this length $\mu$ to zero because the area it encompasses has a minimum value. That
is the central feature of the loop quantization that one tries to capture in {\sc LQC}; $\mu$ 
has a smallest value on the order of $\mu_0 \sim l_{\mathrm{Pl}}$. Since one cannot shrink
the loop to zero, and thus cannot take the derivative of the holonomy with respect to $\mu$, one cannot use
this way to find an expression for $c$ in the so-quantized theory.

With that in mind, one can construct an effective Hamiltonian constraint from the classical Eq.~(\ref{frwham}) by
replacing  $c$ with $\sin(\mu_0 c)/\mu_0$ to capture the periodicity of the network due to the finite size of
the elementary loops. This replacement makes sense because the so-introduced operator can  be expressed and 
interpreted in terms of holonomies. (For this, one does not have to use the sinus function in particular; any 
almost-periodic
function would do~\cite{Ashtekar:2003hd}, but the sinus is the easiest to deal with.) This yields  
\beqn
16 \pi G H_{\mathrm{eff}} = - \frac{6}{\beta^2} |p|^{\frac{1}{2}} \frac{\sin^2 (\mu_0 c)}{\mu_0} + 8 \pi G \frac{1}{|p|^{\frac{3}{2}}} p^2_\phi \label{Heff} \,.
\eeqn
As before, the Hamiltonian constraint gives
\beqn
\frac{\sin (\mu_0 c)}{\mu_0 \beta} = 2 \sqrt{\frac{\pi}{3}} l_{\mathrm{Pl}} \frac{p_\phi}{|p|} \,. 
\eeqn
And then the equation of motion in the semiclassical limit is 
\beqn
\dot p = \{ p, H_{\mathrm{eff}} \} = - \frac{8 \pi}{3} \beta \frac{\partial H_{\mathrm{eff}}}{\partial c} = \frac{2 |p|^{\frac{1}{2}}}{\beta \mu_0} \sin (\mu_0 c) \cos (\mu_0 c) \,.
 \eeqn
With the previously found identification of $\rho_\phi$, we can bring this into a more familiar form
\beqn
{H}^2 = \frac{\dot p^2}{4 p^2} = \frac{8 \pi}{3} G \rho_\rho \left( 1 - \frac{\rho_\phi}{\rho_{\mathrm{c}}} \right) ,
\eeqn
where the critical density is
\beqn
\rho_{\mathrm{c}} = \frac{3}{8\pi G \beta^2 \mu_0^2 a} \,.
\eeqn
The Hubble rate thus goes to zero for a finite $a$, at
\beqn
a^2 = 4 \pi G p_\phi^2 \beta^2 \mu_0^2 \,,
\eeqn
at which point the time-evolution bounces without ever running into a singularity. The critical density at which this happens
depends on the value of $p_\phi$, which here has been a free constant. It has been argued in~\cite{Ashtekar:2006es}, that 
by a more careful treatment the parameter $\mu_0$ depends on the canonical variables and then the critical density can 
be identified to be similar to the Planck density. 

The semi-classical limit is clearly inappropriate when energy densities reach the Planckian regime, but the key feature of the bounce
and removal of the singularity survives in the quantized case~\cite{Bojowald:2001xe,Ashtekar:2011ni,Bojowald:2008zzb,Bojowald:2007bg}.
We take away from here that the canonical quantization of gravity leads to the existence of minimal areas and three-volumes,
and that there are strong indications for a Planckian bound on the maximally-possible value of energy density and curvature.

\subsection{Quantized Conformal Fluctuations}
\label{conformalqg}

The following argument for the existence of a minimal length scale has been put forward by 
Padmanabhan~\cite{Padmanabhan:1985jq, Padmanabhan:1986ny} in the context of conformally-quantized
gravity. That is, we consider fluctuations of the conformal factor only and quantize them. The metric
is of the form
\beqn
g_{\mu\nu}(x) = (1 + \phi(x))^2 \bar g_{\mu\nu}(x) \,,
\eeqn
and the action in terms of $\bar g$ reads
\beqn
S[\bar g, \phi] = \frac{1}{16 \pi G} \int \mathrm{d}^4 x \sqrt{-\bar g} \left( \bar R(1+\phi(x))^2 - 2 \Lambda (1 + \phi(x))^4 - 6 \partial^\nu \phi \partial_\nu \phi \right) . \label{confs}
\eeqn
In flat Minkowski background with $\bar g_{\nu \kappa} = \eta_{\nu \kappa}$ and in a vacuum state, we then want to address the
question what  is the expectation value of spacetime intervals
\beqn
\langle 0 | \mathrm{d}s^2 | 0 \rangle = \langle 0 | g_{\mu \nu} | 0
\rangle \, \mathrm{d}x^\mu \, \mathrm{d}x^\nu 
= (1+ \langle 0 | \phi(x)^2 | 0 \rangle ) \eta_{\mu\nu} \,
\mathrm{d}x^\mu \, \mathrm{d}x^\nu \,. 
\eeqn
Since the expectation value of $\phi(x)^2$ is divergent, instead of multiplying fields at the same point, one has to use covariant point-slitting to two points $x^\nu$ and $y^\nu = x^\nu + \mathrm{d}x^\nu$ 
and then take the limit of the two points approaching each other
\beqn
\langle 0 | \mathrm{d}s^2 | 0 \rangle = \lim_{\mathrm{d}x \to 0} (1+
\langle 0 | \phi(x)\phi(x + \mathrm{d}x) | 0 \rangle ) \eta_{\mu\nu}
\, \mathrm{d}x^\mu \, \mathrm{d}x^\nu \,.  
\eeqn

Now for a flat background, the action (\ref{confs}) has the same functional form as a massless scalar field (up to a sign), so we can tell
immediately what its Green's function looks like
\beqn
 \langle 0 | \phi(x)\phi(y) | 0 \rangle = \frac{l_{\mathrm{p}}^2}{4 \pi^2} \frac{1}{(x-y)^2} \,.
\eeqn
Thus, one can take the limit $\mathrm{d}x^\nu \to 0$ 
\beqn
\langle 0 | \mathrm{d}s^2 | 0 \rangle = \frac{l_{\mathrm{p}}^2}{4
  \pi^2}  \lim_{\mathrm{d}x \to 0} \frac{1}{(x-y)^2} \eta_{\mu\nu} \,
\mathrm{d}x^\mu \, \mathrm{d}x^\nu = \frac{l_{\mathrm{p}}^2}{4 \pi^2}
\,.  \label{conflimit}
\eeqn
The two-point function of the scalar fluctuation diverges and thereby counteracts the attempt to obtain a spacetime distance of length
zero; instead one has a finite length on the order of the Planck length.

This argument has recently been criticized by Cunliff in~\cite{Cunliff:2012zb} on the grounds that the 
conformal factor is not a dynamical degree of freedom in the pure Einstein--Hilbert gravity that was used in this
argument. However, while the classical
constraints fix the conformal fluctuations in terms of matter sources, for gravity coupled to
quantized matter this does not hold. Cunliff reexamined the argument, and found that the scaling behavior of
the Greens function at short distances then depends on the matter content; for normal matter 
content, the limit~(\ref{conflimit}) still goes to zero.

\subsection{Asymptotically Safe Gravity}
\label{asg}
\index{Asymptotically Safe Gravity}
\index{Renormalization}

String theory and LQG have in common the aim to provide
a fundamental theory for space and time different from general relativity; a theory 
based on strings or spin networks respectively. Asymptotically Safe
Gravity ({\sc ASG}), on the other hand, is an attempt to make sense of 
gravity as a quantum field theory by addressing the 
perturbative non-renormalizability of the Einstein--Hilbert action coupled to matter~\cite{tHooft:1974bx}. 

In {\sc ASG}, one considers general relativity merely as an effective theory
valid in the low energy regime that has to be suitably extended to high
energies in order for the theory to be renormalizable and make physical sense.
The Einstein--Hilbert action is then not  
the fundamental action that can be applied up to arbitrarily-high energy scales, but just
a low-energy approximation and its perturbative non-renormalizability need not worry us. 
What describes gravity at energies close by
and beyond the Planck scale (possibly in terms of non-metric 
degrees of freedom) is instead dictated by the non-perturbatively-defined 
renormalization flow of the theory. 

To see how that works, consider a generic Lagrangian of a local field
theory. The terms can be ordered by mass dimension and will come with,
generally dimensionful, coupling constants $g_i$. One redefines these
to dimensionless quantities $\tilde g_i = \lambda^{-d_i} g_i$, where $k$ is
an energy scale. It is a feature
of quantum field theory that the couplings will depend on the scale
at which one applies the theory; this is
described by the Renormalization Group (RG) 
flow of the theory. To make sense of the theory fundamentally, none of the dimensionless
couplings should diverge.  
\index{Renormalization group}
\index{RG|see{Renormalization group}} 

In more detail, one postulates that the RG flow of the theory, 
described by a vector-field in the infinite dimensional space of all possible functionals
of the metric, has a fixed point with finitely many ultra-violet (UV) attractive directions.
These attractive directions correspond to ``relevant'' operators (in perturbation
theory, those up to mass
dimension 4) and span the
tangent space to a finite-dimensional surface called the ``UV critical surface''.
The requirement that the theory holds up to arbitrarily-high energies then 
implies that the natural world must be described by an RG trajectory lying in this
surface, and originating (in the UV) from the immediate vicinity of the fixed point. 
If the surface has finite dimension $d$, then $d$ measurements performed
at some energy $\lambda$ are enough to determine all parameters, and then the remaining
(infinitely many) coordinates of the trajectory are a prediction of the theory, which can be 
tested against further experiments.

In ASG the fundamental gravitational
interaction is then considered asymptotically safe. This necessitates a modification of general relativity,
whose exact nature is so far unknown. Importantly, this scenario does not necessarily imply that the
fundamental degrees of freedom remain those of the metric at all energies. Also in {\sc ASG}, the metric itself might
turn out to be emergent from more fundamental degrees of freedom~\cite{Percacci:2010af}. 
Various independent works have
provided evidence that gravity is asymptotically safe, including studies of gravity in $2+\epsilon$ dimensions,
discrete lattice simulations, and continuum functional renormalization group methods. 
\index{Emergent gravity}

It is beyond the scope of this review to discuss how good this evidence for the asymptotic safety of gravity really is. The interested
reader is referred to reviews specifically dedicated to the topic, for example~\cite{Niedermaier:2006wt,Litim:2008tt,Percacci:2007sz}. 
For our purposes, in the following we will just assume that asymptotic safety is realized for general 
relativity.

To see qualitatively how gravity may become asymptotically safe, let $\lambda$ 
denote the RG scale. From a Wilsonian standpoint, we can refer to $\lambda$
as `the cutoff'. As is customary in lattice theory, we can take $\lambda$ as a
unit of mass and measure everything else in units of $\lambda$.
In particular, we define with
\beqn
\tilde G=G \lambda^2
\eeqn
the dimensionless number expressing Newton's constant in units of the cutoff.
(Here and in the rest of this subsection, a tilde indicates a dimensionless quantity.)
The statement that the theory has a fixed point means that $\tilde G$,
and all other similarly-defined dimensionless coupling constants, 
go to finite values when $\lambda \to\infty$.

The general behavior of the running of Newton's constant can be inferred already by
dimensional analysis, which suggests that the beta function of $1/G$ has the form 
\beqn
\lambda\frac{d}{d\lambda}\frac{1}{G}=\alpha \lambda^2 \,, \label{rgeq}
\eeqn
where $\alpha$ is some constant. This expectation is supported by a number of 
independent calculations, showing that the leading term in the beta function
has this behavior, with $\alpha>0$.
Then the beta function of $\tilde G$ takes the form
\beqn
\lambda\frac{d\tilde G}{d\lambda}=2\tilde G-\alpha \tilde G^2 \,.
\label{this}
\eeqn
This beta function has an IR attractive fixed point at $\tilde G=0$
and also a UV attractive nontrivial fixed point at $\tilde G_*=1/\alpha$.
The solution of the RG equation (\ref{rgeq}) is
\beqn
G(\lambda)^{-1} = G^{-1}_0 + \frac{\alpha}{2}  \lambda^2 \,, \label{gk}
\eeqn
where $G_0$ is Newton's constant in the low energy limit. Therefore, the Planck
length, $\sqrt{G}$, becomes energy dependent. \index{Planck scale!energy dependent} 

This 
running of Newton's constant is characterized by the
existence of two very different regimes:

\begin{itemize}
\item If $0<\tilde G\ll 1$ we are in the regime of sub-Planckian energies, and
the first term on the right side of Eq.~(\ref{this}) dominates. The solution of the flow equation is
\beqn
\tilde G(\lambda)=\tilde G_0\left(\frac{\lambda}{\lambda_0}\right)^2 ,
\eeqn
where $\lambda_0$ is some reference scale and $\tilde G_0=\tilde G(\lambda_0)$.
Thus, the dimensionless Newton's constant is linear in $\lambda^2$,
which implies that the dimensionful Newton's constant $G(\lambda)=G_0 = l_{\mathrm{Pl}}^2$ is constant.
This is the regime that we are all familiar with.

\item In the fixed point regime, on the other hand, the dimensionless Newton's constant 
$\tilde G=\tilde G_*$ is constant, 
which implies that the dimensionful Newton's constant runs according to its
canonical dimension, $G(\lambda)=\tilde G_*/\lambda^2$, in particular it goes to zero
for $\lambda \to \infty$. 
\end{itemize}

One naturally expects the threshold separating these two regimes to be near the Planck scale.
With the running of the RG scale, $\tilde G$ must 
go from its fixed point value at the Planck scale to very nearly zero
at macroscopic scales.

At first look it might seem like
{\sc ASG} does not contain a minimal length scale because there is no limit to the energy
by which structures can be tested. In addition, towards the fixed point regime, 
the gravitational interaction becomes weaker, and with it weakening the argument from
thought experiments in Section~\ref{grhm}, which relied on the distortion caused by
the gravitational attraction of the test particle. It has, in fact, been argued~\cite{Basu:2010nf,Falls:2010he} that in ASG the formation of a black-hole horizon must not necessarily occur, and we
recall that the formation of a horizon was the main spoiler for increasing the
resolution in the earlier-discussed thought experiments.  

However, to get the right picture one has to identify physically-meaningful quantities and a procedure to measure them, which leads to the
following  general argument for
the occurrence of a minimal length in ASG~\cite{Calmet:2010tx,Percacci:2010af}. 

Energies have to be measured in some
unit system, otherwise they are physically meaningless. To assign meaning to
the limit of $\lambda \to \infty$ itself, $\lambda$ too has to be expressed in
some unit of energy, for example as $\lambda \sqrt{G}$, and that unit 
in return has to be defined by some measurement process. In general, the unit 
itself will depend on the scale that is probed in any one particular
experiment. The physically-meaningful energy that we can probe distances
with in some interaction thus will generally not go to $\infty$ with $\lambda$. 
In fact, since $\sqrt{G} \to 1/\lambda$, an energy measured in units of
$\sqrt{G}$ will be bounded by the Planck energy; it will go to one in units
of the Planck energy. 

One may think that one could just use some system of units other than Planck
units to circumvent the conclusion, but if one takes any other dimensionful coupling
as a unit, one will arrive at the same conclusion if the theory is asymptotically safe.
And if it is not, then it is not a fundamental theory that will break down at
some finite value of energy and not allow us to take the limit $\lambda \to \infty$.
As Percacci and Vacca pointed out in~\cite{Percacci:2010af}, it is essentially a tautology
that an asymptotically-safe theory comes with this upper bound when measured in
appropriate units.  

A related argument was offered by Reuter and Schwindt~\cite{Reuter:2005bb} who 
carefully distinguish measurements of distances or momenta with a fixed metric 
from measurements with the physically-relevant metric that solves the equations 
of motion with the  couplings evaluated at the scale $\lambda$ that is being 
probed in the measurement.
In this case, the dependence on $\lambda$ naturally can be moved into the metric. Though
they have studied a special subclass of (Euclidian) manifolds, their finding
that the metric components go like $1/\lambda^2$ is interesting and possibly of
more general significance. 

The way such a $1/\lambda^2$-dependence of the metric on the scale $\lambda$ at which
it is tested leads to a finite resolution is as follows. Consider a scattering process
with in and outgoing particles in a space, which, in the infinite distance from the
scattering region, is flat. In this limit,  to good precision spacetime has the metric 
$g_{\kappa \nu}(\lambda \to 0) =\eta_{\kappa \nu}$.  Therefore, we define the
momenta of the in- and outgoing particles, as well as their sums and differences,
and from them as usual the Lorentz-invariant Mandelstam variables, to be of the form 
$s = \eta_{\kappa \nu}p^\kappa p^\nu$. However, since the metric depends on the
scale that is being tested, the physically-relevant quantities in the collision region 
have to be evaluated with the metric $g_{\kappa \nu}(\sqrt{s}) =
\eta_{\kappa \nu} m^2_{\mathrm{Pl}}/s $. With that one 
finds that the effective Mandelstam variables, and thus also the momentum transfer in the
collision region, actually go to $g_{\kappa \nu}(\sqrt{s})p^\nu p^\mu = m_{\mathrm{Pl}}^2$,
and are bounded by the Planck scale. 

This behavior can be further illuminated by considering in more detail the scattering process in
an asymptotically-flat spacetime~\cite{Percacci:2010af}. The dynamics of this process
is described by some Wilsonian effective action with a suitable momentum scale $\lambda$. 
This action already takes into account the effective contributions of loops with momenta
above the scale $\lambda$, so one may evaluate scattering at tree level in the
effective action to gain insight into the scale-dependence. In particular, we will
consider the scattering of two particles, scalars or fermions, by exchange of a graviton.

Since we want to unravel the effects of ASG, we assume 
the existence of a fixed point, which enters the cross sections of the scattering
by virtual graviton exchange through the running of Newton's constant. The
tree-level amplitude contains a factor $1/m_{\mathrm{Pl}}$ for each vertex. In the $s$-channel, 
the
squared amplitude for the scattering of two scalars is
\beqn
|M_{\mathrm{s}}^2| = \frac{1}{4 m_{\mathrm{Pl}}} \frac{t^2 u^2}{s^2} \,,
\eeqn
and for fermions
\beqn
|M_{\mathrm{f}}^2| = \frac{1}{128 m_{\mathrm{Pl}}} \frac{t^4-6t^3u +18 t^2 u^2 - 6 tu^3 +u^4}{s^2} \,.
\eeqn
As one expects, the cross sections scale with the fourth power of energy over the Planck mass. In
particular, if the Planck mass was a constant, the perturbative expansion would break down at
energies comparable to the Planck mass. However, we now take into account that in {\sc ASG} the
Planck mass becomes energy dependent. For the annihilation process in the $s$-channel, it
is $\sqrt{s}$, the total energy in the center-of-mass system, that encodes what 
scale can be probed. Thus, we replace $m_{\mathrm{Pl}}$ with $1/\sqrt{G(s)}$. One proceeds similarly
for the other channels.

From the above amplitudes the total cross section is found to be~\cite{Percacci:2010af}
\beqn
\sigma_{\mathrm{s}} = \frac{s G(s)}{1920 \pi} \,, \quad \sigma_{\mathrm{f}} = \frac{s G(s)}{5120 \pi} \,,
\eeqn
for the scalars and fermions respectively. Using the running of the gravitational coupling
constant (\ref{gk}), one sees that the cross section has a maximum at $s = 2 G_0/\alpha$ and goes to
zero when the center-of-mass energy goes to infinity. For illustration, the cross section for the scalar scattering
is depicted in Figure~\ref{2} for the case with a constant
Planck mass in contrast to the case where the Planck mass is energy dependent. 

\epubtkImage{fig_roberto.png}{%
  \begin{figure}[ht]
    \centerline{\includegraphics[width=9.0cm]{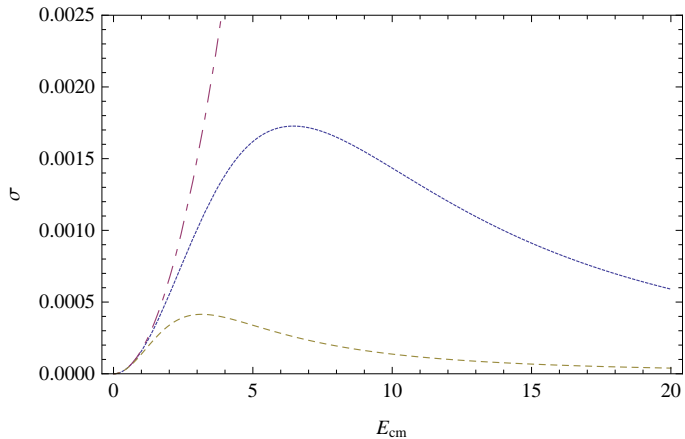}}
    \caption{Cross section for scattering of two scalar particles by
      graviton exchange with and without running Planck mass, in units
      of the low-energy Planck mass $1/\sqrt{G_0}$. The dot-dashed
      (purple) line depicts the case without asymptotic safety; the
      continuous (blue) and dashed (grey) line take into account the
      running of the Planck mass, for two different values of the
      fixed point, $1/\sqrt{\tilde G*}=0.024$ and $0.1$
      respectively. Figure from~\cite{Percacci:2010af}; reproduced with
      permission from IOP.}
    \label{2}
\end{figure}}

If we follow our earlier argument and
use units of the running Planck mass, then the cross section as well as the physically-relevant 
energy, in terms of the asymptotic quantities $G (\sqrt{s}) \sqrt{s}/G_0$,
become constant at the Planck scale. These indications for the existence of a minimal length scale 
in {\sc ASG} are  intriguing, in particular because the dependence of the cross section on the energy
offers a clean way to define a minimal length scale from observable quantities, for
example through the (square root of the) cross section at its maximum value. 

However, it is not 
obvious how the above argument should be extended to interactions in which no graviton exchange takes
place. It has been argued on general grounds in~\cite{Calmet:2010tx}, that even in these cases the dependence 
of the background on the energy of the exchange particle reduces the momentum transfer so that the
interaction would not probe distances below the Planck length and cross sections would 
stagnate once the fixed-point regime has been reached, but the details require more study.
Recently, in~\cite{Anber:2011ut} it has been argued that it is difficult to universally define
the running of the gravitational coupling because of the multitude of kinematic factors present at higher order.
In the simple example that we discussed here, the dependence of $G$ on the $\sqrt{s}$ seems like a reasonable
guess, but a cautionary note that this argument might not be possible to generalize is in order.

\subsection{Non-commutative geometry}
\label{ncg} \index{Non-commutative geometry}

Non-commutative geometry is both a modification of quantum mechanics and quantum field theory that arises within 
certain approaches towards quantum gravity, and a class of theories i n its own right. Thus, it 
could rightfully claim a place both in this section with motivations for a minimal length scale, and in Section~\ref{models} 
with applications. We will discuss the general idea of 
non-commutative geometries in the motivation because
there is a large amount of excellent literature that covers the applications and 
phenomenology of non-commutative geometry. Thus, our treatment here will be very brief.
For details, the interested reader is referred to~\cite{Douglas:2001ba,Hinchliffe:2002km}
and the many references therein. 

String theory and M-theory are among the motivations to look at non-commutative geometries 
(see, e.g., the nice summary in~\cite{Douglas:2001ba}, section VII) and there have been indications that LQG may give rise to a certain type
of non-commutative geometry known as $\kappa$-Poincar\'e. This approach has been very fruitful
and will be discussed in more detail later
in Section~\ref{models}. 

The basic ingredient to non-commutative geometry is that, upon quantization, spacetime
coordinates $x^\nu$ are associated to Hermitian operators $\hat x^\nu$ that are non-commuting.
The simplest way to do this is of the form
\beqn
[\hat x^\nu, \hat x^\mu] = \mathrm{i} \theta^{\mu \nu} \,.
\eeqn
The real-valued, antisymmetric two-tensor $\theta_{\mu \nu}$ of dimension length squared is the deformation parameter 
in this modification of quantum theory, known as the Poisson tensor.\index{Poisson tensor} In the limit $\theta_{\mu \nu} \to 0$ one
obtains ordinary spacetime. In this type of non-commutative geometry, the Poisson tensor is not a dynamical field and
defines a preferred frame and thereby breaks Lorentz invariance.
\index{Lorentz invariance}

The deformation parameter enters here much like $\hbar$ in the commutation\index{Commutation relations}
relation between
position and momentum; its physical interpretation is that of a smallest observable area in the $\mu\nu$-plane. 
 The above commutation relation leads to a
minimal uncertainty among spacial coordinates of the form
\beqn
\Delta x_\mu \Delta x_\nu \gtrsim \frac{1}{2} |\theta^{\mu \nu} | .
\eeqn
One expects the non-zero entries of $\theta_{\mu\nu}$ to be on the order of about the square of the Planck length,
though strictly speaking they are free parameters that have to be constrained by experiment. 

Quantization under the assumption of a non-commutative geometry can be extended from
the coordinates themselves to the algebra of functions $f(x)$ by using Weyl quantization. What one
looks for is a procedure $W$ that assigns to each element $f(x)$ in the algebra of functions ${\cal A}$ a 
Hermitian operator $\hat f = W(f)$ in the algebra of operators $\hat {\cal A}$. One does that 
by choosing a suitable basis for elements of each algebra and then identifies them with each other. The most 
common choice%
\epubtkFootnote{An example of a different choice of basis can be
  found in~\cite{Wohlgenannt:2006dx}.}
is to use a Fourier decomposition of the function $f(x)$
\beqn
\tilde f(k) = \frac{1}{(2 \pi)^4} \int \mathrm{d}^4 x \, e^{-\mathrm{i} k_\nu x^\nu} f(x) \,, \label{hin}
\eeqn
and then doing the inverse transform with the non-commutative operators $\hat x^\nu$
\beqn
\hat f = W(f) = \frac{1}{(2 \pi)^4} \int \mathrm{d}^4 k \, e^{-\mathrm{i} k_\nu \hat x^\nu} \tilde f(k) \,. \label{her}
\eeqn

One can extend this isomorphism between the vector spaces to an algebra isomorphism by constructing
a new product, denoted $\star$, that respects the map $W$, \index{Star product}
\beqn
W(f \star g)(x) = W(f) \cdot W(g) = \hat f \cdot \hat g \,,  \label{starproduct}
\eeqn
for $f,g \in {\cal A}$ and $\hat f, \hat g \in \hat {\cal A}$. From Eqs.~(\ref{her}) and (\ref{starproduct}) one
finds the explicit expression
\beqn
W(f \star g) = \frac{1}{(2 \pi)^4} \int \mathrm{d}^4 k \, \mathrm{d}^4
p \, e^{\mathrm{i} k_\nu \hat x^\nu} \, e^{\mathrm{i} p_\nu \hat
  x^\nu} \, \tilde f(k) \tilde g(p) \,.
\eeqn
With the Campbell--Baker--Hausdorff formula
\beqn
e^Ae^B = e^{A + B + \frac{1}{2} [A,B] + \frac{1}{12}[A,[A,B]] - \frac{1}{12}[B,[A,B]] - \frac{1}{24} [B,[A,[A,B]]] + \dots} 
\eeqn
one has
\beqn
e^{\mathrm{i} k_\nu \hat x^\nu} 
e^{\mathrm{i} p_\nu \hat x^\nu} = 
e^{\mathrm{i} (k_\nu + p_\nu) \hat x^\nu - \frac{\mathrm{i}}{2} k_\nu \theta^{\nu\kappa} p_\kappa}\,,
\eeqn
and thus
\beqn
W(f \star g) = \frac{1}{(2 \pi)^4} \int \mathrm{d}^4 \, k \mathrm{d}^4
p \, e^{\mathrm{i} (k_\nu + p_\nu) \hat x^\nu - \frac{\mathrm{i}}{2} k_\nu \theta^{\nu\kappa} p_\kappa}
\tilde f(k) \tilde g(p) \,.
\eeqn
This map can be inverted to
\beqn
f(x) \star g(x) = \int \frac{\mathrm{d}^4 p}{(2 \pi)^4 }
\frac{\mathrm{d}^4 k}{(2 \pi)^4 } \, \tilde f(k) \tilde g (p)
e^{- \frac{\mathrm{i}}{2} k_{\kappa} \theta^{\kappa \nu} p_\nu} e^{- \mathrm{i} ( k_{\kappa} + p_\kappa)x^\kappa}   \label{fstarg} \,.
\eeqn
If one rewrites the $\theta$-dependent factor into a differential operator acting on the plane-wave--basis, one can also express this in the form
\beqn
f(x) \star g(x) = \exp \left( \frac{\mathrm{i}}{2} \frac{\partial}{\partial x^\nu} \theta^{\nu\kappa} \frac{\partial}{\partial x^\kappa} \right)
f(x) g(y) \Big|_{x\to y} \,,
\eeqn
which is known as the Moyal--Weyl product~\cite{Moyal:1949sk}.

The star product is a particularly useful way to handle non-commutative geometries, because one can continue to
work with ordinary functions, one just has to keep in mind that they obey a modified product rule in the algebra. 
With that, one can build non-commutative quantum field theories by replacing normal products of fields in the
Lagrangian with the star products.

To gain some insight into the way this product modifies the physics, it is useful to compute the star product with 
a delta function. For that, we rewrite Eq.~(\ref{fstarg}) as
\beqn
f(x) \star g(x) &=& \int \frac{\mathrm{d}^4 p}{(2 \pi)^4 }
\mathrm{d}^4 y \, f(x+\frac{1}{2} \theta k) g (x+ y)
 e^{- \mathrm{i} k_{\kappa} y^\kappa}  \nonumber \\
&=&
\frac{1}{\pi^4 \left| \det \theta \right|} \int \mathrm{d}^4 z \, \mathrm{d}^4 y\, f(x+z) g(x+ y) e^{-2\mathrm{i} z^\nu \theta^{-1}_{\nu \kappa} y^\kappa}\,. 
\eeqn
And so, one finds the star product with a delta function to be
\beqn
\delta(x) \star g(x) = 
\frac{1}{\pi^4 \left|\det\theta \right|} \int \mathrm{d}^4 y \, e^{2 \mathrm{i} x^\nu \theta^{-1}_{\nu \kappa} y^\kappa} g(y) \,.
\eeqn
In contrast to the normal product of functions, this describes a highly non-local operation. This non-locality, which
is a characteristic property of the star product, is the most relevant feature of non-commutative geometry.

It is clear that the non-vanishing commutator by itself already introduces some notion of fundamentally-finite resolution,
but there is another way to see how a minimal length comes into play in non-commutative geometry. To see that, we look at a 
Gaussian centered around zero. Gaussian distributions are of interest not only because they are widely used field 
configurations, but, for example, also because they may describe solitonic solutions in a potential~\cite{Gopakumar:2000zd}. 

For simplicity, we will consider only two spatial dimensions and spatial commutativity, 
so then we have
\beqn
[\hat x^i, \hat x^j] = \mathrm{i} \theta \epsilon^{ij}\,,
\eeqn
where $i,j \in \{1,2 \}$, $\epsilon^{ij}$ is the totally antisymmetric tensor, and $\theta$ is the one remaining
free parameter in the Poisson tensor. This is a greatly simplified scenario, but it
will suffice here. 

A normalized Gaussian in position space centered around zero with covariance $\sigma$ 
\beqn
\Psi_{\sigma}(x) = \frac{1}{\pi \sigma} \exp \left( - \frac{x^2}{\sigma^2}\right) \quad
\eeqn
has the Fourier transform
\beqn
\tilde \Psi_{\sigma}(k) = \int d^2 x e^{ik x} \Psi_{\sigma}(x) = \exp \left( -  \pi^2 k^2 \sigma^2 \right) .
\eeqn
We can then work out the star product for two Gaussians with two different spreads $\sigma_1$ and $\sigma_2$ to
\beqn
\tilde \Psi_{\sigma_1} \star \tilde \Psi_{\sigma_2} (k) &=& \int d^2 k \Psi_{\sigma_1}(k) \Psi_{\sigma_2}(p-k) 
\exp \left( \frac{\mathrm{i}}{2} k^i \epsilon_{ij}p^j\right) \nonumber \\
&=& \frac{\pi}{(4 \sigma^2_1 + \sigma^2_2)^2} \exp   \left( - \frac{p^2 \sigma^2_{12}}{4}\right) ,
\eeqn
where
\beqn
\sigma^2_{12} = \frac{\sigma_1^2 \sigma_2^2 + \theta^2}{\sigma_1^2 + \sigma_2^2} \,. \label{sigma12}
\eeqn
Back in position space this yields
\beqn
\Psi_{\sigma_1} \star \tilde \Psi_{\sigma_2} (x) &=& 
\frac{1}{\pi \sigma_{12} (4 \sigma^2_1 + \sigma^2_2)^2} \exp   \left( - \frac{x^2}{\sigma^2_{12}} \right) .
\eeqn
Thus, if we multiply two Gaussians with $\sigma_1, \sigma_2 < \theta$, the width of the product $\sigma_{12}$ is
larger than $\theta$. In fact, if we insert $\sigma_1 = \sigma_2 = \sigma_{12} = \sigma$ in Eq.~(\ref{sigma12}) and
solve for $\sigma$, we see that a Gaussian with width $\sigma = \theta$ squares to itself. Thus, since Gaussians
with smaller width than $\theta$ have the effect to spread, rather than to focus, the product, one can think
of the Gaussian with width $\theta$ as having a minimum effective size. 

In non-commutative quantum mechanics, even in more than one dimension,
Gaussians with this property constitute solutions to polynomial potentials with a mass 
term (for example for a cubic potential this would be of the form $V(\phi) = m^2 \phi \star \phi + a_2 \phi \star \phi \star \phi \star \phi$)~\cite{Gopakumar:2000zd}, because they square to themselves, and so only higher powers continue to reproduce the original function. 
\index{Non-commutative soliton}

\subsection{Miscellaneous}
\label{motmisc}

Besides the candidate theories for quantum gravity so far discussed, there are also 
discrete approaches, reviewed, for example, in~\cite{Loll:1998aj}. 
For these approaches, no general statement can be made with respect to the notion of a minimal length scale. Though one has 
lattice parameters that play the role of regulators, the goal is to eventually let the lattice spacing go 
to zero, leaving open the question of whether observables in this limit allow an arbitrarily good resolution 
of structures or whether the resolution remains bounded. One example of a discrete approach, where 
a minimal length appears, is the lattice approach by Greensite~\cite{Greensite:1990jm} (discussed also in Garay~\cite{Garay:1994en}), in which the 
minimal length scale appears for much the same reason as it appears in the case of 
quantized conformal metric fluctuations discussed in Section~\ref{conformalqg}. Even if the lattice spacing does not go to zero,
it has been argued on general grounds in~\cite{Bojowald:2011jd} that discreteness does not
necessarily imply a lower bound on the resolution of spatial distances. \index{Discrete spacetime}

One discrete approach in which a minimal length scale makes itself noticeable in yet another
way are Causal Sets~\cite{Sorkin:2003bx}. In this approach, one considers as fundamental 
the causal structure of spacetime, as realized by a partially-ordered, locally-finite set of points. 
This set, represented by a discrete sprinkling of points, replaces the smooth background manifold of 
general relativity. The ``Hauptvermutung'' (main conjecture) of the Causal Sets approach is that 
a causal set uniquely determines the macroscopic (coarse-grained) spacetime manifold. In
full generality, this conjecture is so far unproven, though it has been proven in
a limiting case~\cite{Bombelli:1989mu}. Intriguingly, the causal sets approach to a discrete spacetime 
can preserve Lorentz invariance.  This can be achieved by using not a regular but a random sprinkling 
of points; there is thus no meaningful lattice parameter in the ordinary sense. It has been shown in~\cite{Bombelli:2006nm}, 
that a Poisson process fulfills the desired property. This sprinkling has a finite density, \index{Causal sets}
which is in principle a parameter, but is usually assumed to be on the order of the Planckian density.

Another broad class of approaches to quantum gravity that we have so far not mentioned  \index{Emergent gravity}
are emergent gravity scenarios, reviewed in~\cite{Barcelo:2005fc,Sindoni:2011ej}. Also in these
cases, there is no general statement that can be made about the existence of a minimal length scale. 
Since gravity is considered to
be emergent (or induced), there has to enter some energy scale at which the true fundamental,
non-gravitational, degrees of freedom make themselves noticeable. Yet, from this alone we do not know whether this 
also prevents a resolution of structures. In fact, in the absence of
anything resembling spacetime, this might not even be a meaningful question to ask. 

Giddings and Lippert~\cite{Giddings:2001pt,Giddings:2004ud,Giddings:2006vu} have proposed that the gravitational obstruction to test 
short distance probes should be translated into a fundamental limitation in quantum gravity distinct from the GUP.
Instead of a modification of the uncertainty principle, fundamental limitations should arise due to strong gravitational (or other) dynamics,
 because the concept of locality is only approximate, giving rise to a  `locality bound' beyond which the notion of locality ceases to be meaningful. 
When the locality bound is violated, the usual
 field theory description of matter no longer accurately describes the quantum state and one loses the rationale for the usual Fock space description 
of the states; instead, one would have to deal with states able to describe a quantum black hole, whose full and proper quantum description is presently
unknown.
\index{Locality bound}

Finally, we should mention an interesting recent approach by Dvali et al.\ that takes very seriously the
previously-found bounds on the resolution of structures by black-hole formation~\cite{Dvali:2010bf} and is partly
related to the locality bound. However, rather than identifying a regime where quantum field theory breaks
down and asking what quantum theory of gravity would 
allow one to consistently deal with
strong curvature regimes, in Dvali et al.'s approach of `classicalization', super-Planckian degrees of freedom cannot 
exist. On these grounds, it has been argued that classical gravity is in this sense UV-complete exactly because an arbitrarily
good resolution of structures is physically
impossible~\cite{Dvali:2010ue}.
\index{Classicalization}


\subsection{Summary of Motivations}

In this section we have seen that there are many indications, from thought experiments as well
as from different approaches to quantum gravity, that lead us to believe in a fundamental limit to
the resolution of structure. But we have also seen that these limits appear in different forms. 

The most commonly known form is a lower bound on spatial and temporal resolutions given by the Planck
length, often realized by means of a GUP, in which the spatial
uncertainty increases with the increase of the energy used to probe the structures. Such an uncertainty
has been found in string theory, but we have also seen that this uncertainty does not seem to
hold in string theory in general. Instead, in this particular approach to quantum gravity, it is more
generally a 
spacetime uncertainty that 
still seems to hold. One also has to keep in mind here that this bound is given by the string scale, which
may differ from the Planck scale. LQG and the simplified model for LQC give rise to bounds on the eigenvalues of the area and volume operator, and limit
the curvature in the early universe to a Planckian value. 

Thus, due to these different types of bounds, it is somewhat misleading to speak of a `minimal length,'
since in many cases a bound on the length itself does not exist, but only on the powers of
spatio-temporal distances. Therefore, it is preferable to speak more generally of a `minimal length
scale,' and leave open the question of how this scale enters into the measurement of physical
quantities.

\newpage

\section{Models and Applications}
\label{models}

In this section we will investigate some models that have been developed 
to deal with a minimal length or, more often,
a maximum energy scale. The models discussed in the following are not in themselves 
approaches to a fundamental description of spacetime like the ones previously discussed 
that lead us to seriously
consider a finite resolution of structures. Instead, the models discussed in
this section
are attempts to incorporate the notion of a minimal length into the standard model of
particle physics and/or general relativity
by means of a modification of quantum mechanics, quantum field theory and Poincar\'e symmetry. 
These models are meant to provide an effective description of the possible
effects of a minimal length with the intention to make contact with
phenomenology and thereby ideally constrain the possible modifications.%
\epubtkFootnote{The word `effective' should here not be read as a
  technical term.}
 
As mentioned previously, the non-commutative geometries discussed in Section~\ref{ncg} 
could also have rightfully claimed a place in this section on models and applications.

Before we turn towards the implementation, let us spend
some words on the interpretation because the construction of a suitable model 
depends on the physical picture one aims to realize.

\subsection{Interpretation of a minimal length scale}

It is the central premise of this review that there exists a minimal length scale
that plays a fundamental role in the laws of nature. In the discussion in Section~\ref{disc} we 
will consider the possibility that this premise is not fulfilled, but for now we
try to incorporate a minimal length scale into the physical description of the
world. There are then still different ways to think about a minimal length scale
or a maximum energy scale. 

One perspective that has been put forward in the literature~\cite{AmelinoCamelia:2000ge,KowalskiGlikman:2001gp,AmelinoCamelia:2002wr,Magueijo:2002am} is that for the 
 Planck mass to be observer independent it should be invariant under Lorentz boosts.
Since normal Lorentz boosts do not allow this, the observer independence of the
Planck mass is taken as a motivation to modify special relativity to what has become
known as `Deformed Special Relativity' (DSR). (See also Section~\ref{dsr}).
In brief, this modification of special relativity allows one to perform a Lorentz boost
on momentum space in such a way that an energy of Planck mass remains invariant. 
\index{Observer independence}
\index{Lorentz invariance}
\index{Deformed Special Relativity} 
\index{DSR|see{Deformed Special Relativity}}
 
While that is a plausible motivation to look into such departures from special
relativity, one has to keep in mind that just because a quantity is of dimension length (mass)
it must not necessarily transform under Lorentz boosts as a spatial or time-like
component of a spacetime (or momentum) four vector.  A  constant of dimension length can be invariant 
under normal Lorentz boosts, for example, if it is a spacetime (proper) distance. This interpretation 
is an essential ingredient to Padmanabhan's path integral approach (see Section~\ref{pathint}).\index{Path-integral duality}
Another example is the actual
mass of a particle, which is invariant under Lorentz boosts by merit of being 
a scalar. Thinking back to our historical introduction, we recall that the 
coupling constant of Fermi's theory for the weak interaction is proportional 
to the inverse of the $W$-mass and therefore observer independent in the sense
that it does not depend on the rest frame in which we determine it -- and that without the
need to modify special relativity. 

Note also that coupling constants do depend 
on the energy with which structures are probed, as discussed in Section~\ref{asg}
on ASG\index{Asymptotically Safe Gravity}\index{ASG|see{Asymptotically Safe Gravity}}. 
There we have yet another interpretation for a
minimal length scale, that being the energy range (in terms of normally Lorentz-invariant
Mandelstam variables for the asymptotically in/out-going states) where the running Planck
mass comes into the
fixed-point regime. The center-of-mass energy corresponding to the turning point of
the total cross section in graviton scattering, for example, makes it a clean and
observer-independent definition of an energy scale, beyond which there is no
more new structure to be found. 

One should also keep in mind that the outcome of a Lorentz boost is not an observable per se.
To actually determine a distance in some reference frame one has to perform a
measurement. Thus, for the observer independence of a minimal length, it is sufficient if 
there is no operational procedure that allows one to resolve structures to a precision better
than the Planck length. It has been argued in~\cite{Hossenfelder:2006cw} that this does
not necessitate a modification of Lorentz boosts for the momenta of free particles; it is
sufficient if the interactions of particles do not allow one to resolve structures beyond
the Planck scale. There are different ways this could be realized, for example, by an
off-shell modification of the propagator that prevents arbitrarily-high momentum transfer.
As previously discussed, there are some indications that {\sc ASG} might realize such a feature
and it can, in a restricted sense, be interpreted as a version of {\sc DSR}~\cite{Calmet:2010tx}.
\index{Planck scale}

An entirely different possibility that we mentioned in Section~\ref{motmisc}, has 
recently been put forward in~\cite{Dvali:2010bf,Dvali:2010ue}, where it was argued that it is exactly because of the
universality of black-hole production that the Planck length already plays a fundamental role in classical 
gravity and there is no need to complete the theory in the high energy range. \index{Classicalization}

\subsection{Modified commutation relations}
\index{Commutation relations}
\label{mcr}

The most widely pursued approach to model the effects of a minimal length scale in quantum mechanics
and quantum field theory is to reproduce the GUP\index{Generalized uncertainty principle} 
\index{GUP|see{Generalized Uncertainty Principle}} starting from a modified commutation relation for position
and momentum operators. This modification may or may not come with a modification also of the commutators of
these operators with each other, which would mean that the geometry in position or momentum space becomes
non-commuting. 

The modified commutation relations imply not only a {\sc GUP},
but also a modified dispersion relation and a modified Poincar\'e-symmetry in momentum
and/or position space. The literature on the subject is vast, but the picture is still incomplete and under construction.
\index{Modified dispersion relation}

\subsubsection{Recovering the minimal length from modified commutation relations}
\label{simpleexample}

To see the general idea, let us start with a simple example that shows the relation of modified commutation relations to the minimal length scale. Consider variables $\mathbf{k} = (\vec k, \omega)$, where $\vec k$ is the three vector, components of which will be labeled with small Latin indices, and $\mathbf{x} = (\vec x, t) $. Under quantization, the associated operators obey the standard commutation relations
\beqn
\left[x^\nu,  x^\kappa\right] &=& 0 \,, \quad [ x^\nu, k_\kappa ] = \mathrm{i} \delta^\nu_\kappa \quad , \quad  
\left[k_\nu, k_\kappa\right] = 0  \,. \label{normalcomm}
\eeqn
Now we define a new quantity $\mathbf{p} = (\vec p, E) = f(\mathbf{k}) $, where $f$ is an injective function, so that $f^{-1}(\mathbf{p}) = \mathbf{k}$ is well defined.
We will also use the notation $p_\mu = h^\alpha_{\;\mu}(\mathbf{k}) k_\alpha$ with the inverse $k_\alpha = h_\alpha^{\; \mu}(\mathbf{p}) p_\mu$.  
For the variables $\mathbf{x}$ and $\mathbf{p}$ one then has the commutation relations
\beqn
\left[ x^\nu,  x^\kappa\right] &=& 0 \,, \quad \left[ x^\nu,  p_\kappa\right] = \mathrm{i} \frac{\partial f_\kappa}{\partial k_\nu} 
\,,\quad \left[p_\nu,  p_\kappa\right] = 0 \,. \label{pxother}
\eeqn
If one now looks at the uncertainty relation between $x^i$ and $p_i$, one finds
\beqn
\Delta x_i \Delta p_i \geq \frac{1}{2} \langle \frac{\partial f_i}{\partial k^i} \rangle \,. \label{ggup}
\eeqn
To be concrete, let us insert some function, for example a generic expansion of the form
$\vec p \approx \vec k (1 + \alpha k^2/m_{\mathrm{Pl}}^2)$ plus higher orders in $k/m_{\mathrm{Pl}}$, so that the
inverse relation is $\vec k \approx \vec p ( 1 - \alpha p^2/m_{\mathrm{Pl}}^2 )$. Here, $\alpha$ is some
dimensionless parameter. One then has
\beqn
\frac{\partial f_i}{\partial k^j} \approx \delta_{ij} \left(1 + \alpha \frac{p^2}{m_{\mathrm{Pl}}^2} \right) + 2 \alpha \frac{p_i p_j}{m_{\mathrm{Pl}}^2} \,.
\eeqn
Since this
function is convex, we can rewrite the expectation value in~(\ref{ggup}) to 
  \beqn
\Delta x_i \Delta p_i \geq \frac{1}{2} \left( 1 + \alpha \frac{\langle p^2 \rangle }{m_{\mathrm{Pl}}^2} + 2 
\alpha \frac{\langle p^2_i \rangle }{ m_{\mathrm{Pl}}^2}   \right) .
\eeqn
And inserting the expression for the variance $\langle A^2 \rangle - \langle A \rangle ^2 = \Delta A^2$, one obtains
  \beqn
\Delta x_i \Delta p_i &\geq& \frac{1}{2} \left( 1 + \alpha \frac{\Delta p^2 + \langle p \rangle^2}{m_{\mathrm{Pl}}^2} + 2 \alpha  
\frac{\Delta p_i^2 + \langle p_i \rangle^2}{ m_{\mathrm{Pl}}^2} \right) \nonumber\\
&\geq& 
\frac{1}{2} \left( 1 + 3 \alpha \frac{\Delta p_i^2}{m_{\mathrm{Pl}}^2} \right) ,
\eeqn
or 
\beqn
\Delta x_i &\geq& 
\frac{1}{2} \left( \frac{1}{\Delta p_i} + 3 \alpha \frac{\Delta p_i}{m_{\mathrm{Pl}}^2} \right) . \label{gupkp}
\eeqn
Thus, we have reproduced the GUP that we found in the thought experiments
in Section~\ref{thought} with a minimal possible uncertainty for the position~\cite{Kempf:1993bq,AmelinoCamelia:1999pm,Maggiore:1993kv,Maggiore:1993zu,AmelinoCamelia:1999wk}. 

However, though the nomenclature here is deliberately suggestive, one has to be careful with interpreting this
finding. What the inequality (\ref{gupkp}) tells us is that we cannot measure the position to arbitrary precision if we
do it by varying the uncertainty in $\mathbf{p}$. That has an operational meaning only if we assign to $\mathbf{p}$
the meaning of a physical momentum, in particular it should be a Hermitian operator. 
To distinguish between the physical quantity $\mathbf{p}$, and $\mathbf{k}$ that fulfills the
canonical commutation relations, the $\mathbf{k}$ is sometimes referred to as the `pseudo-momentum' or, because it
is conjugated to $\mathbf{x}$, as `the wave vector.' One can then
physically interpret the non-linear relation between $\mathbf{p}$ and $\mathbf{k}$ as an 
energy dependence of Planck's constant~\cite{Hossenfelder:2006rr}. \index{Pseudo-momentum}\index{Planck's constant}

To further clarify this, let us turn towards the question of Lorentz invariance. If we do not make 
statements in addition to the commutation relations, we do not know anything about the transformation behavior of the quantities.
For all we know, they could have an arbitrary transformation behavior and 
Lorentz invariance could just be broken. If we require Lorentz invariance to be preserved, this opens the
question of how it is preserved, what is the geometry of its phase space, and what is its Poisson structure. 
And, most importantly, how do we identify physically-meaningful coordinates on that space?\index{Phase space}

At the time of writing,
no agreed upon picture has emerged. Normally, phase space is the cotangential bundle of the spacetime
manifold. One might generalize this to a bundle of curved momentum spaces, an idea that dates back
at least to Max Born in 1938~\cite{Born}. In a more radical recent
approach, the `principle of relative locality'~\cite{Smolin:2010xa,Smolin:2010mx,Jacob:2010vr,AmelinoCamelia:2010qv,AmelinoCamelia:2011bm,AmelinoCamelia:2011yi}
phase space is instead considered to be the cotangential bundle of momentum space. 
\index{Principle of relative locality}

To close this gap in our example, let us add some more structure and assume that the phase space is a 
trivial bundle ${\cal S} = {\cal M} \otimes {\cal P}$, where ${\cal M}$ is spacetime and ${\cal P}$ is momentum space.
Elements of this space have the form $(\mathbf{x},\mathbf{p})$. If we add that the quantity $\mathbf{p}$ is a coordinate on ${\cal P}$
and transforms according to normal 
Lorentz transformations under a change of inertial frames, and $\mathbf{k}$ is another coordinate system, then we know right away how $\mathbf{k}$ transforms because
 we can express it 
by use of the function $f$. If we do a Lorentz transformation
on $\mathbf{p}$, so that ${\bf p'} = \Lambda \mathbf{p}$, then we have
\beqn
\mathbf{k}' = f({\bf p'}) = f(\Lambda \mathbf{p}) = f(\Lambda f^{-1}(\mathbf{k})) \,,  \label{finite}
\eeqn
which we can use to construct the modified Lorentz transformation as $\mathbf{k}' = \widetilde{\Lambda}(\mathbf{k})$.
In particular, one can chose $f$ in such a way that it maps an infinite value of $\mathbf{p}$ (in either the spatial or temporal
entries, or both) to a finite value of $\mathbf{k}$ at the Planck energy. The so-constructed Lorentz transformation
on $\mathbf{p}$ will then keep the Planck scale invariant, importantly without introducing any preferred frame. This
is the basic idea of deformations of special relativity, some explicit examples of which we will meet in
Section~\ref{dsr}. 

If one assumes that $\mathbf{p}$ transforms as a normal Lorentz vector, one has
\beqn
[J_{\kappa\nu},p_\mu] = \mathrm{i}\left( p_\nu \eta_{\kappa \mu}  - p_\kappa \eta_{\nu\mu} \right ) .
\eeqn
Since this commutator commutes with $\mathbf{p}$, one readily finds
\beqn
[J_{\kappa\nu},k_\mu] = \mathrm{i}\left( p_\nu \eta_{\kappa \alpha}  - p_\kappa \eta_{\nu \alpha} \right) \frac{\partial k_\mu}{\partial p_\alpha}  \,,
\label{Jk}
\eeqn
which gives us the infinitesimal version of (\ref{finite}) by help of the usual expansion
\beqn
k_{'\nu} = k_\nu - \frac{\mathrm{i}}{2} \omega^{\alpha \beta} [ J_{\alpha \beta}, k_\nu ] + {\cal O}(\omega^2) \,,
\eeqn
where $\omega^{\alpha \beta}$ are the group parameters of $\Lambda$.  

Now that we know how $\mathbf{k}$ transforms, we still need to add information for how the coordinates on ${\cal P}$ are
supposed to be matched to those on ${\cal M}$. One requirement that we
can use to select a basis on ${\cal M}$ along with that on ${\cal P}$ under a Lorentz transformation is
that the canonical form of the commutation relations should remain preserved. 
With this requirement, one then finds for the infinitesimal transformation of $\mathbf{x}$~\cite{Panes:2011cd}
\beqn
[J_{\kappa\nu}, x^\mu] = \mathrm{i} x^\alpha \frac{\partial}{\partial k_\mu} \left(p_\kappa \eta_{\beta \nu}\frac{\partial k_\alpha}{\partial p_\beta} 
- p_\nu \eta_{\beta \kappa}\frac{\partial k_\alpha}{\partial p_\kappa}\right)  \label{Jx} \quad
\eeqn
and the finite transformation
\beqn
{x'}^\nu = \frac{\partial p'_\alpha}{\partial k'_\mu} \Lambda^\nu_{\; \alpha} \frac{\partial k_\beta}{\partial p_\nu} x^\kappa  \,.
\eeqn 
One finds the latter also directly by noting that this transformation behavior is required to keep the symplectic
form $w = dx^\alpha \wedge d k_\alpha$ canonical.\index{Symplectic form}

A word of caution is in order here because the innocent looking indices on these quantities do now implicitly stand for
different transformation behaviors under Lorentz transformations. One can amend this possible confusion by a more complicated notation, but this is for
practical purposes usually unnecessary, as long as one keeps in mind that the index itself does not tell the
transformation behavior under Lorentz transformations. In particular, the derivative $\partial k_\alpha/\partial p_\beta$ has a mixed transformation
behavior and, in a Taylor-series expansion, this and higher derivatives yield factors that convert the normal to
the modified transformation behavior.

So far this might have seemed like a rewriting, so it is important to stress the following: just writing down the
commutation relation leaves the structure under-determined. To completely specify the 
model, one needs to make an additional assumption about how a Lorentz transformation is defined, 
how the coordinates in position space ought to be chosen along with those in momentum space 
under a Lorentz transformation and, most importantly, what the metric on the curved momentum space (and possibly spacetime) is.

Needless to say, the way we have fixed the transformation behavior in this simple example is 
not the only way to do it. Another widely used choice is to require that 
$\mathbf{k}$ obeys the usual transformation behavior, yet interpret $\mathbf{p}$ as the physical momentum, or $\mathbf{x}$ as `pseudo-coordinates'
(though this is not a word that has been used in the literature). We will meet an example of this in Section~\ref{snyderbasis}. 
This variety is the main reason why the literature on the subject of modified commutation relations is
confusing.

At this point it should be mentioned that a function $f$ that maps an infinite value of $p$ to an asymptotically-finite value of $f(p)$ cannot be 
a polynomial of finite order. Its Taylor series expansion necessarily needs to have an infinite number of terms. 
Now if $k(p)$ becomes constant for large $p$, then $\partial k/\partial p$ goes to zero and $\partial p/\partial k$
increases without bound, which is why the uncertainty (\ref{ggup}) increases for large $p$. Depending on the choice
of $f$, this might be the case for the spatial or temporal components or both. If one wants to capture the regularizing
properties of the minimal length, then a perturbative expansion in powers of $E/m_{\mathrm{Pl}}$ will not work in the
high energy limit. In addition, such expansions generically add the complication that any
truncation of the series produces for the dispersion relation a polynomial of finite order, which will have 
additional zeros, necessitating
additional initial values~\cite{Hossenfelder:2007re}. This can be prevented by not truncating the series,
but this adds other complications, discussed in Section~\ref{qftml}.

Since
$\partial p/\partial k$ is a function of $\mathbf{p}$, the position operator $\mathrm{i} (\partial p_\alpha/\partial k_\nu) \partial/\partial p_\alpha$,  is not Hermitian if $\mathbf{p}$ is Hermitian. This is unfortunate for a quantity that is supposed to be a physical observable, but we have to keep in mind that the
operator itself is not an observable anyway. To obtain an observable, we have to take an expectation value. To ensure that the
expectation value produces meaningful results, we change its evaluation so that the condition $\langle x \Psi | \Phi \rangle = \langle \Psi | x \Phi \rangle$,
which in particular guarantees that expectation values are real, is still fulfilled. That is, we want the operator to be symmetric, rather than
Hermitian.%
\epubtkFootnote{According to the Hellinger--Toeplitz theorem, an
  everywhere-defined symmetric operator on a Hilbert space is
  necessarily bounded. Since some operators in quantum mechanics are
  unbounded, one is required to deal with wave functions that are not
  square integrable. The same consideration applies here.}
To that end, one changes the measure in momentum space to~\cite{Kempf:1993bq,Kempf:1994su}
\beqn
\mathrm{d}^3 p \to \mathrm{d}^3 k = \Big| \frac{\partial k}{\partial
  p} \Big| \mathrm{d}^3 p \,, \label{measuremomentum}
\eeqn
which will exactly cancel the non-Hermitian factor in $\hat x = \mathrm{i} \partial/(\partial k)$, because now
\beqn
\langle \Psi | x \Phi \rangle &=& \mathrm{i} \int \mathrm{d}^3 p \Big| \frac{\partial k}{\partial p} \Big| \Psi^* \frac{\partial}{\partial k}  \Phi  \nonumber \\
&=& \mathrm{i} \int \mathrm{d}^3 k \Psi^* \frac{\partial}{\partial k}
\Phi  = \mathrm{i} \int \mathrm{d}^3 k (\frac{\partial}{\partial k}  \Psi^*) \Phi \nonumber \\
&=& \langle  x  \Psi  | \Phi \rangle \,.
\eeqn
Note that this does not work without the additional factor because then the integration measure does not fit to the derivative,
which is a consequence of the modified commutation relations. We will see in the next section that there is another way
to think of this modified measure.

The mass-shell relation, $g^{\mu \nu}(\mathbf{p}) p_\mu p_\nu - m^2 =0$, is invariant under normal Lorentz transformations acting on $\mathbf{p}$, and thus $g^{\mu \nu}(\mathbf{k})p_\mu(\mathbf{k}) p_\nu(\mathbf{k})^2 - m^2 = 0$ is invariant under the modified Lorentz transformations 
acting on $\mathbf{k}$. The clumsy notation here has stressed that the metric on momentum
space will generally be a non-linear function of the coordinates. The mass-shell relation will yield the Hamiltonian
constraint of the theory. 

There is a subtlety here; since $\mathbf{x}$ is not Hermitian, we can't use the representation if this operator. 
The way this can be addressed depends on the model. One can in many cases just work the momentum representation.
In our example, we would note that the operator $\bar x^\nu = (\partial k_\nu /\partial p_\mu ) x^\mu$ is Hermitian, and use
its representation. In this representation, the Hamiltonian constraint becomes a higher-order operator,
and thus delivers a modification of the dispersion relation. However, the interpretation of the dispersion
then hinges on the interpretation of the coordinates. Depending on the suitable identification of position space
coordinates and the function $f$, the speed of massless particles in this model may thus become a function of the momentum 
four-vector $\mathbf{p}$. It should be noted however that this is not the case for all 
choices of $f$~\cite{Hossenfelder:2005ed}.

Many of the applications of this model that we
will meet later only use the first or second-order expansion of $f$. While these are sufficient for some interesting
consequences of the {\sc GUP}, the
Planck energy is then generically not an asymptotic and invariant value. If the complete function $f$ is considered,
it is usually referred to as an `all order GUP.'\index{Generalized uncertainty principle!to all orders} The
expansion to first or second order is often 
helpful because it allows one to parameterize the possible modifications by only a few dimensionless
quantities. 

To restrict the form of modifications possible in this approach, sometimes the Jacobi identities\index{Jacobi identities} 
are drawn upon. 
It is true that the Jacobi identities restrict the possible commutation
relations, but seen as we did here starting from the standard commutation relations, this statement is
somewhat misleading. The Jacobi identities are, as the name says, identities. They say more about the 
properties of the binary operation they represent than about the quantities this operation acts on. 
They are trivially  fulfilled for the commutators of all new variables $f(k)$ one can define (coordinates one can choose) if the old ones fulfilled the 
identities. However, if one does not start from such a function, one can draw upon the Jacobi identities as a consistency
check. 

A requirement that does put a restriction on the possible form of the commutation relation is that 
of rotational invariance. Assuming that $f_i(\vec k) = k_i h(k)$, where $k = |\vec k|$, one has
\beqn
\frac{\partial f_i}{\partial k_j} = \delta_i^j h(k) + k_i \frac{\partial h (k)}{\partial k_j} \,.
\eeqn 
The expansion to 3rd order in $k$ is
\beqn
 h(k) = 1 + \alpha k + \beta k^2 + {\cal{O}}(k^3) \,,\quad \frac{\partial h}{\partial k_j} = \alpha \frac{k_j}{k} + 2 \beta k_j + {\cal{O}}(k^2) \,. 
\eeqn
We denote the inverse of $f_i(\vec k)$ with $f_i^{-1}(\vec p) = p_i \tilde h(p)$. An expansion of $\tilde h(p)$ and comparison
of coefficients yields to third order
\beqn
\tilde h(p) = 1 - \alpha p - (\beta -2\alpha^2) p^2 \,. \label{hinverse}
\eeqn
With this, one has then
\beqn
\left[x_i, p_j \right] &=& \delta_{ij} + \alpha \left( k \delta_{ij} + \frac{k_i k_j}{k} \right) + \beta \left( k^2 \delta_{ij} +2 k_j k_i \right) +   {\cal{O}}(k^3)\nonumber \\
&=& \delta_{ij} + \alpha \left( p \delta_{ij} + \frac{p_i p_j}{p} \right) + (\beta - \alpha^2) p^2 \delta_{ij} + (2 \beta - \alpha^2) p_i p_j +   {\cal{O}}(p^3)\,. \label{param}
\eeqn
For the dimensions to match, the constant $\alpha$ must have a dimension of length and $\beta$ a dimension of length squared. One
would expect this length to be of the order Planck length and play the role of the fundamental length. 

It is often assumed that $\beta = \alpha^2$, 
but it should be noted that
this does not follow from the above. In particular, $\alpha$ may be zero and the 
modification be even in $k$, starting only at second order. Note that an expansion of the commutator in the form
$[x_i,p_j] = \mathrm{i} \delta_{ij}(1 + \beta p^2)$ does not fulfill the above requirement.

To summarize this section, we have seen that a GUP that gives rise to a
minimal length scale can be realized by modifying the canonical commutation relations. We have seen
that this modification alone does not completely specify the physical picture, we have in addition to 
fix the transformation behavior under Lorentz transformations and the metric on momentum space. In Section~\ref{snyderbasis} we will see how this can be done.

\subsubsection{The Snyder basis}
\label{snyderbasis}

As mentioned in the previous example, $\mathbf{p}$ is not canonically conjugate to $\mathbf{x}$,
and the wave vector $\mathbf{k}$, which is canonically conjugate, is the quantity that transforms under the modified Lorentz transformations. 
But that is not necessarily the case for models with modified commutation relations, as we will see in this section.

Let us start again from the normal commutation relations (\ref{normalcomm}) and now define new position coordinates $\mathbf{X}$ by $X_\nu = x_\nu - x^\alpha k_\alpha k_\nu/ m_{\mathrm{Pl}}^2$, as discussed in~\cite{Girelli:2007sz}. (The Planck mass $m_{\mathrm{Pl}}$ could enter here with an additional
dimensionless factor that one would expect to be of order one, if one describes a modification that has its origin in quantum 
gravitational effects. In the following we will not carry around
such an additional factor. It can easily be inserted at any stage just by replacing $m_{\mathrm{Pl}}$ with $\alpha m_{\mathrm{Pl}}$.) In addition, we
now require that $\mathbf{k}$ transforms under the normal Lorentz transformations. With this
replacement, the $\mathbf{k}$'s are then still commutative as usual and the remaining commutation relations have the form
\beqn
\left[ X_\mu, X_\nu \right] = - \frac{1}{m_{\mathrm{Pl}}} J_{\mu \nu} \,, \quad \left[ X_\mu, k_\nu \right] = i \left( \eta_{\mu\nu} - \frac{k_\mu k_\nu}{m_{\mathrm{Pl}}^2} \right) , \label{Snydercomm}
\eeqn
where we recognize
\beqn
J_{\mu \nu} = x_\mu k_\nu - x_\nu k_\mu = X_\mu k_\nu - X_\nu k_\mu \,,
\eeqn
as the generators of Lorentz transformations.  This reproduces the commutation relations of Snyder's original proposal~\cite{Snyder}.

The commutator between $\mathbf{X}$ and $\mathbf{k}$ leads to a {\sc GUP} by taking the expectation value in the same way as previously, though the
reason here is a different one: If it is the $X_\nu$'s that are representing physically-meaningful positions in 
spacetime, then it is their non-commutativity that spoils the resolution of structures at the Planck scale. Note that the
transformation from $\mathbf{x}$ to $\mathbf{X}$ is not canonical exactly for the reason that it does change the commutation
relations. 

In a commonly-used notation, $J_{i0} = N_i$ are the generators of
boosts and $\{ J_{23}, J_{31}, J_{12} \}$ are the generators of the rotations $\{M_1, M_2, M_3\}$
that fulfill the normal Lorentz algebra
\beqn
\left[N_i, N_j \right] &=& - i \epsilon_{ijk} N_k \,, \quad \left[M_i, M_j \right] = i \epsilon_{ijk} N_k,\nonumber\\
\left[M_i, N_j \right] &=& i \epsilon_{ijk} N_k \,. \label{normallorentz}
\eeqn
Since we have not done anything to the transformation of the momentum $\mathbf{k}$, in the $\mathbf{X},\mathbf{k}$ phase-space coordinates one also has
\beqn
\left[M_i, k_j\right] &=& i \epsilon_{ijk} k_k \,,\quad \left[M_i, k_0\right] = 0, \nonumber\\
\left[N_i, k_j\right] &=& i \delta_{ij} k_0 \,,\quad \left[N_i, k_0\right] = i k_i \,. \label{normallorentz2} 
\eeqn

There are two notable things here. First, as in Section~\ref{snyderbasis},
 the Lorentz algebra remains entirely unmodified. Second, the $\mathbf{X}$ by construction 
transforms covariantly under normal Lorentz transformations if the $\mathbf{x}$ and $\mathbf{k}$ do. However, we see 
that there is exactly one $\mathbf{x}$ for which $\mathbf{X}$
does not depend on $\mathbf{k}$, and that is $\mathbf{x} = 0$. If we perform a translation by use of the generator $\mathbf{k}$; the
coordinate $\mathbf{x}$ will be shifted to some value $\mathbf{x}' = \mathbf{x}+{\bf a}$. 
Alternatively, one may try to take a different generator for
translations than $\mathbf{k}$, the obvious choice is the operator canonically conjugated to $X_\nu$
\beqn
\frac{\partial}{\partial X_\nu} = \frac{\partial}{\partial x_\alpha}\frac{\partial x_\alpha}{\partial X_\nu} \,.
\eeqn
If one contracts $X_\nu = x_\nu + x^\alpha k_\alpha k_\nu /m_{\mathrm{Pl}}^2$ with $k^\nu$, one can invert $\mathbf{X}(\mathbf{x})$ to
\beqn
x_\nu = X_\nu - \frac{X_\alpha k^\alpha}{m_{\mathrm{Pl}}^2 + k_\kappa k^\kappa} k_\nu \,. 
\eeqn
Then one finds the translation operator
\beqn
\frac{\partial}{\partial X_\nu} = k^\nu \left( \frac{1}{1 - k^\alpha k_\alpha/ m_{\mathrm{Pl}}^2}\right) .
\eeqn
Therefore, it has been argued~\cite{Mignemi:2008kn} that one should understand this type of theory 
as a modification of translation invariance rather than a modification of Lorentz symmetry. However, this
depends on which variables are assigned physical meaning, which is a question that is still under discussion.

We should at this point look at Snyder's original motivation for it is richer than just the
commutation relations of position, momenta and generators and adds to it in an important way. Snyder originally considered a 5-dimensional
flat space of momenta, in which he looked at a hypersurface with de~Sitter geometry. The full metric
has the line element
\beqn
\mathrm{d}s^2 = \eta^{AB} \, \mathrm{d}\eta_A \, \mathrm{d}\eta_B \,,
\eeqn
where the coordinates $\eta_A$ have dimensions of energy and capital Latin indices run from 0 to 4.
This flat space is invariant under the action of the group $SO(4,1)$, which has a total of 10 generators. In that
5-dimensional space, consider a 4-dimensional hyperboloid defined by
\beqn
- m_p^2 = \eta^{AB} \eta_A \eta_B = \eta^{\mu \nu} \eta_\mu \eta_\nu - \eta_4^2 \,.
\eeqn
This hypersurface is invariant under the $SO(3,1)$ subgroup of $SO(4,1)$. It describes a de~Sitter space and can be parameterized by 
four coordinates. Snyder
chooses the projective coordinates $k_\nu = m_{\mathrm{Pl}} \eta_\nu/\eta_4$. (These coordinates are nowadays rarely used to
parameterize de~Sitter space, as the fifth coordinate of the embedding space $\eta_4$ is not constant on
the hyperboloid.) The remaining four generators of 
$SO(4,1)$ are then identified with the coordinates
\beqn
J_{4\nu} = X_\nu = i \left( \frac{\eta_4}{m_{\mathrm{Pl}}} \frac{\partial}{\partial \eta_\nu} +  \frac{\eta_\nu}{m_{\mathrm{Pl}}} \frac{\partial}{\partial \eta_4} \right) .
\eeqn
From this one obtains the same commutation relations (\ref{Snydercomm}), (\ref{normallorentz}), and (\ref{normallorentz2}) as above~\cite{Blaut:2003wg}.

However, the Snyder approach contains additional information: We know that the commutation
relations seen previously can be obtained by a variable substitution from the normal ones. In
addition, we know that the momentum space is curved. It has a de~Sitter \index{De Sitter space} geometry, a non-trivial curvature tensor and curvature scalar
$12/m_{\mathrm{Pl}}^2$. It has the corresponding parallel transport and a volume measure. In these coordinates, 
the line element has the form
\beqn
\mathrm{d}s^2 = \frac{\eta^{\mu \nu} \, \mathrm{d}k_\mu \, \mathrm{d}k_\nu}{1 - \eta^{\alpha \kappa} k_\alpha k_\kappa/m_{\mathrm{Pl}}^2} \,.
\eeqn
Thus, we see how the previously found need to adjust the measure in momentum space (\ref{measuremomentum}) arises here naturally
from the geometry in momentum space. The mass-shell condition is \index{Momentum space curvature}
\beqn
m^2 = \frac{\eta^{\mu \nu} k_\mu k_\nu}{1 - \eta^{\alpha \kappa} k_\alpha k_\kappa/m_{\mathrm{Pl}}^2} \,. \label{snydershell}
\eeqn
We note that on-shell this amounts to a redefinition of the rest mass. 

The $(\mathbf{X},\mathbf{k})$ coordinates on phase space have become known as the Snyder basis.\index{Snyder basis}

\subsubsection{The choice of basis in phase space}
\label{otherbasis}
\index{Phase space}

The coordinates $\boldsymbol{\eta}$ that Snyder chose to parameterize the hyperbolic 4-dimensional submanifold are not
unique. There are infinitely many sets of coordinates we can choose on this space; most of them will be non-linear
combinations of each other. Such non-linear redefinitions of momenta will change the commutation relations between
position and momentum variables. More generally, the question
that arises here is what coordinates on phase space should be chosen, since we have seen in the
previous Section~\ref{snyderbasis} that a change of coordinates in phase space that mixes position and momentum operators
creates non-commutativity. For example, one could use a transformation that mixes $\mathbf{p}$ and $\mathbf{x}$ to
absorb the unusual factor in the $[\mathbf{x},\mathbf{p}]$ commutator in (\ref{pxother}) at the expense of creating 
a non-commutative momentum space. 

Besides the above-discussed coordinate systems $(\mathbf{x},\mathbf{k})$, $(\mathbf{x}, \mathbf{p})$, $(\mathbf{X}, \boldsymbol{\eta})$, and $(\mathbf{X},\mathbf{k})$, there 
are various other choices of coordinates that can be found in the literature. 
One choice that is very common are coordinates $\tilde{\mathbf{x}}$ that are related to the Snyder
position variables~\cite{KowalskiGlikman:2003we} via
\beqn
\tilde x_0 = X_0  \,, \quad \tilde x_i = X_i + \frac{N_i}{m_{\mathrm{p}}} \,. \label{xX}
\eeqn
This leads to the commutation relations
\beqn
\left[ \tilde x_0, \tilde x_i \right] = - \mathrm{i} \frac{x_i}{m_{\mathrm{Pl}}} \,, \quad \left[ \tilde x_i, \tilde x_j \right] = 0 \,.
\eeqn
The non-commutative spacetime described by these coordinates has become known as $\kappa$-Minkowski spacetime.
The name derives from the common nomenclature in which the constant $m_{\mathrm{Pl}}$ (that, as we have warned previously, might differ from
the actual Planck mass by a factor of order one) is $\kappa$. \index{$\kappa$-Minkowski} \index{Non-commutative geometry}

Another choice of coordinates that can be found in the literature~\cite{Girelli:2005dc,Smolin:2010xa} is obtained by the transformation 
\beqn
{\chi}_0 = x_0 + x_i k^i/m_{\mathrm{Pl}} \,, \quad {\chi}_i = x_i  \label{chix}
\eeqn
on the normal coordinates $x_\nu$. This leads to the commutation relations
\beqn
[\chi_0, \chi_i] &=& \mathrm{i} \frac{\chi_i}{m_{\mathrm{Pl}}} \,,\quad [\chi_i, {\chi_j}]  = 0 \,, \quad [k_i, {\chi_j}]  = \mathrm{i} \delta_{ij}, \nonumber\\
\left[ \chi_0, k_0 \right] &=& \mathrm{i} \,,\quad  [\chi_0, k_i] =  - \mathrm{i} \frac{k_i}{m_{\mathrm{Pl}}} \,, \quad [k_0, \chi_i] = \mathrm{i} \,.
\eeqn
This is the full $\kappa$-Minkowski phase space~\cite{Lukierski:1992dt}, which is noteworthy because it was shown by Kowalski-Glikman and 
Nowak~\cite{KowalskiGlikman:2003we} that the geometric approach to phase space is equivalent
to the algebraic approach that has been pursued by deforming the Poincar\'e-algebra (the algebra of generators of Poincar\'e transformations, i.e., boosts, rotations and momenta) to a Hopf algebra~\cite{Majid:1994cy}\index{Hopf algebra} with deformation parameter $\kappa$,
the $\kappa$-Poincar\'e Hopf algebra, giving rise to the above $\kappa$-Minkowski phase space.
\index{$\kappa$-Poincar\'e}
 
A Hopf algebra generally consists of two algebras that are related by a dual structure and associated product rules that have to
fulfill certain compatibility conditions. Here, the dual
to the $\kappa$-Poincar\'e algebra is the $\kappa$-Poincar\'e group, whose elements are identified as Lorentz transformations
and position variables. The additional structure that we found in the geometric approach to be the curvature
of momentum space is, in the algebraic approach, expressed in the co-products
and antipodes of the Hopf algebra. As in the geometrical approach, there is an ambiguity in the choice of
coordinates in phase space.

\index{Bicrossproduct}
In addition to the various choices of position space coordinates, one can also use different coordinates
on momentum space, by choosing different parameterizations of the hypersurface than that of Snyder.
One such parameterization is using coordinates $\pi_\nu$, that are related to the Snyder basis by
\beqn
\eta_0 &=& - m_{\mathrm{Pl}} \sinh \left( \frac{\pi_0}{m_{\mathrm{Pl}}}\right) - \frac{\vec \pi^2}{2 m_{\mathrm{Pl}}} \exp \left( \frac{\pi_0}{m_{\mathrm{Pl}}}\right), \nonumber\\
\eta_i &=& - \pi_i \exp \left( \frac{\pi_0}{m_{\mathrm{Pl}}}\right), \nonumber \\
\eta_4 &=& - m_{\mathrm{Pl}}  \cosh \left( \frac{\pi_0}{m_{\mathrm{Pl}}}\right) - \frac{\vec \pi^2}{2 m_{\mathrm{Pl}}} \exp \left( \frac{\pi_0}{m_{\mathrm{Pl}}}\right) .
\label{bicross}
\eeqn
(Recall that $\eta_4$ is not constant on the hypersurface.) The $\pi_\nu$'s are the bicrossproduct basis of the
Hopf algebra~\cite{Majid:1994cy}, and they make a natural choice for the algebraic approach. With the $\kappa$-Minkowski coordinates $\tilde x_\mu$, one then has the commutators~\cite{KowalskiGlikman:2003we} 
\beqn
[\pi_0 , \tilde x_0] &=& \mathrm{i} \,, \quad [\pi_i,\tilde x_0 ] = - \mathrm{i} \frac{\pi_i}{m_{\mathrm{Pl}}} , \nonumber \\
\left[\pi_i, \tilde x_j \right] &=& - \mathrm{i} \delta_{ij} \,, \quad [ \pi_0, \tilde x_i] = 0 \,.
\eeqn

Another choice of coordinates on momentum space is the
Magueijo--Smolin basis ${\cal P}_\mu$, which is related to the Snyder coordinates by
\beqn
k_\mu = \frac{{\cal P}_\mu}{1- {\cal P}_0/m_{\mathrm{Pl}}} \,. \label{magsmo}
\eeqn
From the transformation behavior of the $\mathbf{k}$ (\ref{normallorentz2}), one can work out the transformation
behavior of the other coordinates, 
and reexpress the mass-shell condition \ref{snydershell} in the new sets of coordinates. 

Since there are infinitely many other choices of coordinates, listing them all is beyond the scope of 
this review.  So long as one can identify a new
set of coordinates by a coordinate transformation from other coordinates, the commutation relations
will fulfill the Jacobi identities automatically. Thus, these coordinate systems are consistent
choices. One can also, starting from the transformation of
the Snyder coordinates, derive the transformation behavior under Lorentz transformation for all 
other sets of coordinates. For the above examples the transformation behavior can be found
in~\cite{Girelli:2005dc,KowalskiGlikman:2003we} 

In summary, we have seen here that there are many different choices of coordinates on phase
space that lead to modified commutation relations. We have met some oft used examples
and seen that the most relevant information is in the geometry of momentum space. Whether
there are particular choices of coordinates on phase space that lend themselves to easy interpretations
and are thus natural in some sense is presently an open question. 
\index{Jacobi identities}

\subsubsection{Multi-particle states}
\label{soccer}

One important consequence of the modified Lorentz symmetry that has been left out in our discussion so far
is the additivity of momenta, which becomes relevant when considering
interactions. 

In the example from Section~\ref{simpleexample}, the function $f$ has
to be non-linear to allow a maximum value of (some components of) $\mathbf{k}$ to remain Lorentz invariant, and consequently the Lorentz transformations
$\widetilde{\Lambda}$ are non-linear functions of $\mathbf{k}$. But that means that the transformation of a sum of 
pseudo-momenta $\mathbf{k}_1 + \mathbf{k}_2$ is not the same as the sum of the transformations:
\beqn
\widetilde{\Lambda}(\mathbf{k}_1 + \mathbf{k}_2) \neq \widetilde{\Lambda}(\mathbf{k}_1) + \widetilde{\Lambda}(\mathbf{k}_2) \,. 
\eeqn
Now in the case discussed in Section~\ref{simpleexample}, it is the $\mathbf{p}$ that is the physical momentum that is conserved, 
and since it transforms under normal Lorentz transformations its conservation is independent of the rest frame.

However, if one has chosen the $\mathbf{p}$ rather than the $\mathbf{k}$
to obey the normal Lorentz transformation, as was the case in Sections~\ref{snyderbasis} and \ref{otherbasis}, then this 
equation looks exactly the other way round
\beqn
\widetilde{\Lambda}(\mathbf{p}_1 + \mathbf{p}_2) \neq \widetilde{\Lambda}(\mathbf{p}_1) + \widetilde{\Lambda}(\mathbf{p}_2) \,. 
\eeqn
And now one has a problem, because the momentum $\mathbf{p}$ should be conserved in interactions, and 
the above sum is supposed to be conserved in an interaction in one rest frame, it
would not be conserved generally in all rest frames. (For the free particle, both $\mathbf{p}$ and $\mathbf{k}$
are conserved since the one is a function of the other.)

The solution to this problem is to define a new,
non-linear, addition law $\oplus$ that has the property that it remains invariant and that can be rightfully
interpreted as a conserved quantity. This
is straightforward to do if we once again use the quantities $\mathbf{k}$ that in this case by 
assumption transform under the normal Lorentz transformation. To each momentum we have an associated 
pseudo-momentum $\mathbf{k}_1 = f^{-1}(\mathbf{p}_1)$,
 $\mathbf{k}_2 = f^{-1}(\mathbf{p}_2)$. The sum $\mathbf{k}_1 + \mathbf{k}_2$ is invariant under normal 
Lorentz transformations, and so we construct the sum of the $\mathbf{p}'s$ as
\beqn
\mathbf{p}_1 \oplus \mathbf{p}_2 = f(\mathbf{k}_1 + \mathbf{k}_2) =   f(f^{-1}(\mathbf{p}_1) + f^{-1}(\mathbf{p}_2)) \,.
\eeqn
It is worthwhile to note that this modified addition law can also be found from the algebraic approach; it
is the bicrossproduct of the $\kappa$-Poincar\'e algebra~\cite{KowalskiGlikman:2003we}.
\index{Bicrossproduct}

This new definition for a sum is now observer independent by construction, but we have created a new problem. 
If the function $f$ (or some of its components) has a maximum of the Planck mass, then the sum of
momenta will never exceed this maximal energy. The Planck mass is a large energy as far as
particle physics is concerned, but in everyday units it is about $10^{-5}$ gram, a
value that is easily exceeded by some large molecules. This problem of reproducing a
sensible multi-particle limit when one chooses the physical momentum to transform under modified
Lorentz transformations has become known as the `soccer-ball problem.'\index{Soccer-ball problem}

The soccer-ball problem is sometimes formulated in a somewhat different form. If one makes an expansion 
of the function $f$ to include the first correction terms in $p/m_{\mathrm{Pl}}$, and from that derives
the sum $\oplus$, then it remains to be shown that the correction terms stay small if one
calculates sums over a large number of momenta, whose ordinary sum describes macroscopic
objects like, for example, a soccer ball. One expects that the sum then has approximately the form
$\mathbf{p}_1 \oplus \mathbf{p}_2 \approx \mathbf{p}_1 + \mathbf{p}_2 + \mathbf{p}_1 \mathbf{p}_2 \Gamma/m_{\mathrm{Pl}}$,
where $\Gamma$ are some coefficients of order one. If one iterates this sum for $N$ terms, the
normal sum grows with $N$ but the additional term with $\sim N^2$, so that it will eventually
become problematic. 

Note that this problem is primarily about sums of momenta, and not even necessarily about 
bound states. If one does not symmetrize the new
addition rule, the result may also depend on the order in which momenta are added. 
This means in particular the sum of two momenta can depend on a third term that may
describe a completely unrelated (and arbitrarily far away) part of the universe, which
has been dubbed the `spectator problem'~\cite{KowalskiGlikman:2004qa,Girelli:2004ue}. \index{Spectator problem}

There have been various attempts to address the problem, but so far none has been
generally accepted. For example, it has been suggested that with the addition of $N$ particles,
the Planck scale that appears in the Lorentz transformation, as well as in the modified
addition law, should be rescaled to $m_{\mathrm{Pl}} N$~\cite{Magueijo:2002am,Judes:2002bw,Magueijo:2006qd}. 
It is presently difficult to see how this ad-hoc solution would follow from the theory. Alternatively, it has been suggested
that the scaling of modifications should go with the density~\cite{Hossenfelder:2007fy,Olmo:2011sw} rather than with
the total momentum or energy respectively. While the energy of macroscopic objects is larger
than that of microscopic ones, the energy density decreases instead. This seems
a natural solution to the issue but would necessitate a completely different ansatz
to implement. A noteworthy recent result is~\cite{AmelinoCamelia:2011uk,Hossenfelder:2012vk}, where it has
been shown how the problem can be alleviated in a certain model, such that the nonlinear term
in the sum does not go with $N^2$ but with $N^{3/2}$. 

One should also note that this problem does not occur in the case in which the modification 
is present only off-shell, which seems to be suggested in some interpretations. Then, if one
identifies the momenta of particles as those of the asymptotically free states, the 
addition of their momenta is linear as usual. For the same reason, the problem also does not
appear in the interpretation of such modifications of conservation laws as being caused by a 
running Planck's constant, put forward~\cite{Percacci:2010af,Calmet:2010tx}, and discussed in Section~\ref{asg}. 
In this case, the relevant energy is the momentum transfer, and for bound states this remains small 
if the total mass increases.

So we have seen that demanding the physical momentum
rather than the pseudo-momentum to transform under modified Lorentz transformations leads to
the soccer ball and the spectator problem. This is a disadvantage of this choice. On the other
hand, this choice has the advantage that it has a geometric base,
which is missing for the case discussed in Section~\ref{simpleexample}.  

\subsubsection{Open problems}

We have, in Section~\ref{snyderbasis} and \ref{otherbasis},
seen different examples for modified commutation relations with
a curved momentum space. But
commutation relations alone don't make for physics. To derive physical meaning, one has to 
define the dynamics of the system and its observables. 

This raises the question of which principles to use for the 
dynamics and how to construct observables. While there are several approaches to this, some of which 
we will meet in the following, there exists to date no agreed-upon framework by which to derive observables,
and therefore the question of whether there is a physical reason to prefer one basis over another is open.

One finds some statements in the literature that a different choice of coordinates leads to
a different physics, but this statement is somewhat misleading. One should more precisely say that
a choice of coordinates and the corresponding commutation relations do not in and of themselves 
determine the
physics. For that, one has to specify not only the geometry of the phase space, which is not 
contained merely in the commutation relations, but also a unique procedure to arrive at 
equations of motion. 

Given the complete geometry (or, equivalently, the operations on the Hopf algebra), the 
Hamiltonian in some basis can be identified as (a function of) the Casimir operator of the
Lorentz group.%
\epubtkFootnote{The Lorentz group has a second Casimir operator, which
  is the length of the Pauli--Lubanski pseudovector. It can be
  identified by it being a function of the angular momentum operator.}
It can be expressed in any basis one wishes by substitution. 
However, if the transformation between one basis of phase space and the other is not canonical, then
transforming the Hamiltonian by substitution will not preserve the Hamiltonian equations. 
In particular, $\partial H/\partial p$ and $[H,x]$ will generically not yield the same result, 
thus the notion of velocity requires careful interpretation, especially when the coordinates
in position space are in addition non-commuting. It has been argued by Smolin in~\cite{Smolin:2010mx}
that commuting coordinates are the sensible choice. The construction of observables with
non-commuting coordinates has been worked towards, e.g., in~\cite{Smolin:2010xa,Smolin:2010mx,Jacob:2010vr,AmelinoCamelia:2010qv,AmelinoCamelia:2011bm,AmelinoCamelia:2011yi,Rychkov:2003ar,Tezuka:2003vt,Girelli:2004xy}.

The one modification that all of these approaches have in common is a non-trivial measure in 
momentum space that in the geometric approach results from the volume element of the now curved space. 
But this raises the question of what determines the geometry. Ideally one would like an axiomatic approach that allows one to
derive the geometry from an underlying principle, and then everything else from the geometry. One
would, in the general case, if dynamical matter distributions are present, not expect the structure of momentum space to be entirely
fixed. A step towards a dynamical momentum space has been made in~\cite{Chang:2010ir}, but clearly the
topic requires more investigation.

Another open problem with this class of models is
the type of non-locality that arises. If the Planck length acts as a minimal length,
there clearly has to be some non-locality. However, it has been shown that for those types of
models in which the speed of light becomes energy dependent, the non-locality becomes 
macroscopically large. Serious conceptual problems arising from
this were pointed out in~\cite{AmelinoCamelia:2002vy,Schutzhold:2003yp,Hossenfelder:2006rr}, 
and shown to be incompatible with observation in~\cite{Hossenfelder:2009mu,Hossenfelder:2010tm}. 
\index{Non-locality}

A very recent development to
address this problem is to abandon an absolute notion of locality, and instead settle
for a relative one. This `principle of relative locality' 
\cite{Smolin:2010xa,Smolin:2010mx,Jacob:2010vr,AmelinoCamelia:2010qv,AmelinoCamelia:2011bm,AmelinoCamelia:2011yi,Carmona:2011wc}
is a promising development. It remains to be seen how it mitigates the problem of non-local
particle interactions. For some discussion, see~\cite{Hossenfelder:2010yp,Hossenfelder:2010jn,Hossenfelder:2010xr}.
It should be stressed that this problem does not occur if the
speed of light remains constant for free particles. 
\index{Principle of relative locality}

\subsection{Quantum mechanics with a minimal length scale}

So we have seen that modified commutation
relations necessarily go together with a GUP, a modified measure 
in momentum space and a modified Lorentz symmetry. These models may or may not give rise to
a modified on-shell dispersion relation and thus an energy-dependent
speed of light~\cite{Hossenfelder:2005ed}, but the modified
commutation relations and the generalized uncertainty cannot be
treated consistently without taking care of the momentum space integration and the transformation behavior. 
 
The literature on the subject of quantum mechanics with a minimal length scale is partly confusing 
because many models use only some of the previously-discussed ingredients and do not 
subscribe to all of the modifications, or at least they are not explicitly stated. Some
differ in the interpretation of the quantities; notoriously there is the question of what is
a physically-meaningful definition of velocity and what is the observable momentum. 

Thus, the topic of a minimal length scale has thus given rise to many related approaches that run under
the names `modified commutation relation,' `generalized uncertainty,' `deformed special relativity,'
`minimal length deformed quantum mechanics,' etc.\ and are based on only some features of
the modified phase space discussed previously. It is not clear in all cases whether this is
consistently possible or what justifies a particular interpretation. For example, one may argue
that in the non-relativistic limit, a modified transformation
behavior under boosts, that would only become relevant at relativistic energies, is irrelevant. However, one has to keep in mind that the non-linear
transformation behavior of momenta results in a non-linear addition law, which becomes problematic for the
treatment of multi-particle states. Thus, even in the non-relativistic case, one has to be careful
if one deals with a large number of particles. 

The lack of clean, agreed upon, axiomatic approach
has inevitably given rise to occasional criticism. It has been argued in~\cite{Ahluwalia:2002wf}, for
example, that the
deformations of special relativity are operationally indistinguishable from special relativity. Such
misunderstandings are bound to arise if the model is underspecified. Maybe, the easiest way to see that the minimal length
modified quantum theory is not equivalent to the unmodified case is to keep in mind that the
momentum space is curved: There is no coordinate
transformation that will make the curvature of momentum space go away. The non-trivial metric
will also produce an infinite series of higher-order derivatives in the Hamiltonian constraint, a reflection of the non-locality
that the existence of a minimal length scale implies.


In the following, we will not advocate one particular approach, but just report what results 
are presently available. Depending on which quantities are raised to physical importance, 
the resulting model can have very different properties. The speed of light might be energy dependent~\cite{AmelinoCamelia:1999pm}
or not~\cite{Tamaki:2001ck,Daszkiewicz:2003yr},  there might be an upper limit to energies and/or momenta, or not,
addition laws and thresholds might be modified, or not, coordinates might be non-commuting, or not, there 
might be non-localities or not, the modification might only be present off-shell, or not. This is
why the physical meaning of different bases in phase space is a problem in need of being addressed
in order to arrive at more stringent predictions.

\subsubsection{Maximal localization states}
\index{Maximal localization states}

The most basic information about the minimal length modified quantum mechanics is in the position operator
itself. While, in the momentum representation, there exist eigenvectors of the position operator
that correspond to arbitrarily--sharply-peaked wave functions, 
these do not describe physically-possible configurations. It has been shown in~\cite{Kempf:1994su} that
the sharply-peaked wavefunctions with spread below the minimal position uncertainty carry an infinite energy,
and thus do not represent a physically-meaningful basis.
Instead, one can construct quasi-localized states that are as sharply focused as physically possible. These
states are then no longer exactly orthogonal. In~\cite{Kempf:1994su}, the maximal localization states have
been constructed in one spatial dimension for a 2nd-order expansion of the {\sc GUP}.  

\subsubsection{The Schr{\"o}dinger equation with potential}
\label{schroedinger}

The most straight-forward modification of quantum mechanics that one can construct with the modified commutation
relations is leaving the Hamiltonian unmodified. For the harmonic oscillator for example, 
one then has the familiar expression
\beqn
H = \frac{\mathbf{p}^2}{2m} + m \omega^2 \frac{\mathbf{x}^2}{2} \,.
\eeqn
However, due to the modified commutation relations, if one inserts the operators, the resulting differential
equation becomes higher order. In one dimension, for example, in the momentum space representation, one would have
to second order
$\hat x = \mathrm{i} (1 + l^2_{\mathrm{Pl}} p^2) \partial/\partial p$  and thus for the stationary equation
\beqn
\frac{\partial^2}{\partial p^2} \psi(p) + \frac{2 l_{\mathrm{Pl}}}{1+ l_{\mathrm{Pl}}^2 p^2} \frac{\partial}{\partial p} \Psi(p) + 
\frac{2E/(2m \omega^2) - p^2/(m\omega)^2}{(1+l_{\mathrm{Pl}}^2 p^2 )^2} \psi(p) = 0 \,.
\eeqn

The same procedure can be applied to other types of potentials in the Schr{\"o}dinger equation, and in principle this can be done not only in the
small-momentum expansion, but to all orders. In this fashion, in the leading-order approximation, 
the harmonic oscillator in one dimension has been
studied in~\cite{Kempf:1994su,Hossenfelder:2003jz,Ali:2011fa,Ghosh:2011ze}, the harmonic oscillator in 
arbitrary dimensions in~\cite{Chang:2011jj,Kempf:1996fz,Chang:2001kn,Dadic:2002qn}, the energy levels of the hydrogen atom in 
\cite{Hossenfelder:2003jz,Brau:1999uv,Stetskohydrogen,Bouaziz:2010he,Pedram:2012ub}, the particle in a box in~\cite{Ali:2009zq}, Landau levels and
the tunneling current in~\cite{Ali:2011fa,Das:2008kaa,Das:2009hs}, the uniform gravitational potential in~\cite{Nozari:2010qy,Chang:2011jj},
the inverse square potential in~\cite{Bouaziz:2007gs,Bouaziz:2010hc}, 
neutrino oscillations in~\cite{Sprenger:2010dg},
reflection and transmission coefficients of a potential step and potential barrier in~\cite{Ali:2011fa,Das:2009hs},
the Klein paradox in~\cite{Ghosh:2012cv}, 
and corrections to the gyromagnetic moment of the muon in~\cite{Harbach:2003qz,Das:2011tq}. Note that
these leading order expansions do not all use the same form of the {\sc GUP}. 

In order to obtain the effects of the minimal length on the transition rate of ultra cold neutrons in gravitational
spectrometers, Pedram et al.\ calculated
the quantization of the energy spectrum of a particle in a linear gravitational field in the {\sc GUP}
leading-order approximation~\cite{Pedram:2011xj} and to all orders~\cite{Pedram:2012pt}. The
harmonic oscillator in one dimension with an asymptotic GUP has been considered in~\cite{Pedram:2011aa,Pedram:2011gw}.

While not, strictly speaking, falling into the realm of quantum mechanics, let us also mention here
the Casimir effect, which has been studied in~\cite{Harbach:2005yu,Bachmann:2005km,Nouicer:2005dp,Frassino:2011aa,Dorsch:2011qf},
and Casimir--Polder intermolecular forces, which have been looked at in~\cite{Panella:2007kd}.

All these calculations do, in principle, cause corrections to results obtained in standard quantum
mechanics. As one expects, the correction terms are unobservably small
 if one assumes the minimal length scale to be on the order of the Planck length. However, as argued previously,
since we have no good explanation as to why the Planck length as the scale at which quantum gravity should
become important is so small, the minimal length should, in principle, be regarded as a free parameter 
and then be bound by
experiment. A compilation of bounds from the above calculations is presently not available and unfortunately 
no useful comparison is possible due to the different parameterizations and assumptions used. One
can hope that this might improve in the future if a more standardized approach becomes established,
for example, using the parameterization (\ref{param}). 

\subsubsection{The Klein--Gordon and Dirac equation}

The Klein Gordon equation can be obtained directly from the invariant $\mathbf{p}(\mathbf{k})^2 - m^2 =0$. 
The Dirac equation can be constructed using the same prescription that lead to the Schr{\"o}dinger equation,
except that, to make sure Lorentz invariance is preserved, one should first bring it into a suitable form
\beqn
( \gamma^\nu p_\nu - m) \Psi = 0 \,,  
\eeqn
and then replace $p_\nu$ with its operator as discussed in Section~\ref{simpleexample}. In the position representation, this will generally 
produce higher-order derivatives not only in the spatial, but also in the temporal, components. In
order to obtain the Hamiltonian that generates the time evolution, one then has to invert the temporal part.

The Dirac equation with modified commutation relations has been discussed in~\cite{Hossenfelder:2003jz,Kober:2010sj}. The Klein--Gordon equation and the Dirac particle in a rectangular and spherical box 
has been examined in~\cite{Das:2010zf}.

\subsection{Quantum field theory with a minimal length scale}
\label{qftml}

One can construct a quantum field theory along the lines of the quantum mechanical treatment, starting 
with the modified commutation relations. If the
position-space coordinates are non-commuting with a constant Poisson tensor, this leads to the territory of non-commutative
quantum field theory for which the reader is referred to the literature specialized on that
topic, for example~\cite{Douglas:2001ba,Hinchliffe:2002km} and the many references therein. 

Quantum field theory with the $\kappa$-Poincar\'e algebra on the non-commuting $\kappa$-Minkowski
spacetime coordinates has been pioneered in~\cite{Kempf:1994qp,Kempf:1996ss,Kempf:1996nk}. In~\cite{Kempf:1996mv}
it has been shown that introducing the minimal length uncertainty principle into quantum field theory works as a regulator in the
ultraviolet, at least for $\phi^4$ theory. Recently, there has 
been a lot of progress on the way towards field quantization~\cite{Arzano:2010jw}, by developing the Moyal--Weyl product, the
Fock space~\cite{Camacho:2003dm,Arzano:2007ef}, and the conserved Noether charges~\cite{Freidel:2007hk,Agostini:2006nc,Arzano:2007gr}.
The case of scalar field theory has been investigated in~\cite{Arzano:2009ci,Meljanac:2010ps}. A different
second-order modification of the commutation relation has been investigated for the spinor and
Klein--Gordon field in~\cite{Moayedi:2011ur} and~\cite{Moayedi:2010vp} respectively. \cite{Carmona:2009ra} studied the situation in which the Hamiltonian remains
unmodified and only the equal time commutation relations are modified.
There are,
as yet, not many applications in the literature that investigate modifications of the standard model of
particle physics, but one can expect these to follow soon.

Parallel to this has been the development of quantum field theory in the case where coordinates are
commuting, the physical momentum transforms under the normal Lorentz transformation, and the speed
of light is constant~\cite{Hossenfelder:2003jz}.
This approach has the advantage of being easier to interpret, yet has the disadvantage of delivering more
conservative predictions. In this approach, the modifications one is left with are the modified measure in
momentum space and the higher-order derivatives that one obtains from the metric in momentum space. 
The main difficulty in this
approach is that, when one takes into account gauge invariance, one does not only obtain an infinite
series of higher-derivate corrections to the propagator, one also obtains an infinite number of 
interaction terms. Whether these models are unitary is an open question.

In order to preserve the super-Planckian limit that is necessary to capture the presence
of the UV-regulating properties, it has been suggested in~\cite{Hossenfelder:2003jz} that one
expand the Lagrangian in terms of $g E/m_{\mathrm{Pl}}$, where $g$ is the coupling constant 
and $E$ is the energy scale. This means that the corrections to the propagator (which
do not contain any $g$) are kept entirely, but one has only the first vertex of the infinite
series of interaction terms. One can then explore the interesting energy range $>m_{\mathrm{Pl}}$ until $m_{\mathrm{Pl}}/g$. 
The virtue of this expansion, despite its limited range of applicability, is that, by
not truncating the power series of the propagator, one does not introduce additional
poles. The expansion in terms of
$E/m_{\mathrm{Pl}}$ was looked at in~\cite{Kober:2010sj}. 

In this modified quantum field theory, in~\cite{Hossenfelder:2004up} the running gauge couplings, possibly with
additional compactified spatial dimensions, were investigated. In~\cite{Kober:2011dn}, 
the electro-weak gauge interaction with minimal length was studied. In~\cite{Mimasu:2011sa},
the top quark phenomenology in the case with a lowered Planck scale was studied, and in
\cite{Hossenfelder:2004ze,Cavaglia:2004jw} it has been argued that if the Planck scale is indeed lowered, then
its role as a minimal length would decrease the production of black holes. 

One recurring
theme in these models is the suppression of phase space at high energies~\cite{Hossenfelder:2003jz}, which is
a direct consequence of the modified measure in momentum space. This has also been found,
for the same reason, in the $\kappa$-Poincar\'e approach~\cite{VilelaMendes:2004up}.
Another noteworthy feature of these quantum field theories with a minimal length is that the 
commutator between the fields $\phi(x)$ and their
canonical conjugate $\pi(y)$ are not equal to a delta function~\cite{Hossenfelder:2007fy,Maziashvili:2011dx},
which is an expression of the non-locality that the higher-order derivatives bring in. \index{Non-locality}

In~\cite{Kober:2011uj} it has furthermore been suggested that one apply this modification of the quantization
procedure to quantum cosmology, which is a promising idea that might allow one to make contact with
phenomenology.

 \subsection{Deformed Special Relativity}
\label{dsr}\index{Deformed Special Relativity}

Deformed special relativity (DSR) is concerned with the departure from Lorentz symmetry 
that results from the postulate that the Planck energy transforms like a (component of a)
momentum four vector and remains an invariant, maximal energy scale. 
While the modified commutation relations
necessarily give rise to some version of {\sc DSR}, one can 
also try to extract information from the deformation of
the Lorentz symmetry, or the addition law, directly. This gives rise to what Amelino-Camelia
has dubbed `test theories'~\cite{AmelinoCamelia:2004ht}: simplified and reduced versions of the quantum theory
with a minimal length. Working with these test theories has the advantage that one can make contact with phenomenology
without working out the -- still not very well understood -- second quantization and interaction. 
It has the disadvantage that it makes the ambiguity in identifying physically-meaningful 
observables worse.

The literature on the topic is vast, and we can not cover it in totality here. For
more details on the {\sc DSR} phenomenology, the reader is referred to~\cite{AmelinoCamelia:2008qg,AmelinoCamelia:2010pd}.
We will just mention the most relevant properties of these types of models here. 

As we have
seen earlier, a non-linear relation $\mathbf{p}(\mathbf{k})$, where $\mathbf{k}$ transforms under a normal Lorentz transformation,
generates the deformed Lorentz transformation for $\mathbf{p}$ by Eq.~(\ref{finite}). Note that in {\sc DSR} it is the
non-linearly transforming $\mathbf{p}$ that is considered the physical momentum, while the $\mathbf{k}$ that
transforms under the normal Lorentz transformation is considered the pseudo-momentum. 
There is an infinite number of such functions, and thus there is an infinite number of
ways to deform special relativity. 

We have already met the common choices in the literature; they constitute bases
in $\kappa$-Minkowski phase space, for example, the coordinates (\ref{magsmo}) proposed by Magueijo and Smolin in~\cite{Magueijo:2001cr}. 
With this relation, a boost in the $z$-direction takes the form
\beqn
{\cal P}'_0 &=& \frac{\gamma ({\cal P}_0 - v {\cal P}_z)}{1+ (\gamma-1){\cal P}_0/m_{\mathrm{Pl}} -  \gamma v {\cal P}_z/m_{\mathrm{Pl}}}\\
{\cal P}'_z &=& \frac{\gamma ({\cal P}_z - v {\cal P}_0)}{1+ (\gamma-1){\cal P}_0/m_{\mathrm{Pl}} - \gamma v {\cal P}_z/m_{\mathrm{Pl}}} \,.
\eeqn
This example, which has become known as {\sc DSR2}, is particularly illustrative because this deformed Lorentz boost transforms 
$(m_{\mathrm{Pl}},m_{\mathrm{Pl}}) \to (m_{\mathrm{Pl}},m_{\mathrm{Pl}})$, and thus keeps the Planck energy invariant. 
Note that since $k_0/k = {\cal P}_0/{\cal P}$, a so defined speed of light remains constant in this case.

Another example that has entered the literature under the name DSR1~\cite{Bruno:2001mw} makes use of the
bicrossproduct basis (\ref{bicross}). Then, 
the dispersion relation for massless particles takes the form
\beqn
\cosh(\pi_0/m_{\mathrm{Pl}}) = \frac{1}{2} \frac{\vec \pi^2}{m_{\mathrm{Pl}}^2} e^{\pi_0/m_{\mathrm{Pl}}} . \label{dsr1}
\eeqn
The relation between the momenta and the pseudo-momenta and their inverse has been worked out in~\cite{Judes:2002bw}.
In this example, the speed of massless particles that one derives from the dispersion relation depends
on the energy of the particle. This effect may be observable in high-frequency light reaching Earth from
distant sources, for example from $\gamma$-ray bursts. This interesting prediction is covered in more detail in~\cite{AmelinoCamelia:1999pm,AmelinoCamelia:2008qg,AmelinoCamelia:2010pd}.
\index{Bicrossproduct}

As discussed in Section~\ref{soccer}, the addition law in this type of model has
to be modified in order to obtain Lorentz-invariant conserved sums of momenta. This gives rise to the soccer-ball
problem and can lead to changes in thresholds of particle interactions~\cite{AmelinoCamelia:2008qg,AmelinoCamelia:2010pd,Carmona:2010ze}.
It had originally been argued that this would shift the {\sc GZK} cut-off~\cite{AmelinoCamelia:2002gv}, but
this argument has meanwhile been revised. However, in the `test theory' one does not actually have a
description of the particle interaction, so whether or not the kinematical considerations would be
realized is unclear. 

In these two {\sc DSR} theories, it is usually assumed that the position variables conjugated 
to $\mathbf{k}$ are not commutative~\cite{Daszkiewicz:2003yr}, thereby delivering a particular realization of
$\kappa$-Minkowski space.  
In the {\sc DSR1} model, the speed of massless particles that one derives from
the above dispersion relation (\ref{dsr1}) is energy dependent. However, the interpretation of that
speed hinges on the meaning of the conjugated position-space coordinates, which is why it has
also been argued that the physically-meaningful speed is actually constant 
\cite{Tamaki:2001ck,Daszkiewicz:2003yr}. Without an identification of observable positions, it
is then also difficult to say whether this type of model actually realizes a minimal length.
One can expect that recent work on the principle of relative locality 
\cite{Smolin:2010xa,Smolin:2010mx,Jacob:2010vr,AmelinoCamelia:2010qv,AmelinoCamelia:2011bm,AmelinoCamelia:2011yi,Carmona:2011wc}
will shed light on this question. A forthcoming review~\cite{Jerzy} will be especially dedicated to
the development of relative locality.
\index{$\kappa$-Minkowski}\index{Principle of relative locality}

\subsection{Composite systems and statistical mechanics}

As mentioned previously, a satisfactory treatment of multi-particle states in those
models in which the free particles' 
momenta are bound by a maximal energy scale is still lacking. Nevertheless, approaches to the
description of composite systems or many particle states have been made, 
based on the modified commutation
relations either without subscribing to the deformed Lorentz transformations, and thereby generically
breaking Lorentz invariance, or by employing an ad hoc solution by rescaling the bound on the energy with
the number of constituents. While these approaches are promising in so far that modified statistical
mechanics at Planckian energies would allow one to use the early universe as a laboratory, they should
be regarded with some caution, because the connection to the single particle description with
deformed Lorentz symmetry is missing, and the case in which Lorentz symmetry is broken is strongly
constrained already~\cite{Kostelecky:2008ts}.

That having been said, the statistical mechanics from the $\kappa$-Poincar\'e algebra was investigated
in general in~\cite{KowalskiGlikman:2001ct,Fityo:2008zz}. In~\cite{Quesne:2009vc} 
corrections to the effective Hamiltonian of macroscopic bodies have been studied, and 
in~\cite{Pikovski:2011zk} observational consequences of modified commutation relations for a massive oscillator have been considered. 
In
\cite{KalyanaRama:2001xd} statistical mechanics with a generalized uncertainty and possible
applications for cosmology have been looked at. The
partition functions of minimal-length quantized statistical mechanics have been derived in~\cite{Pedram:2011aa}, in
\cite{Ali:2011ap} the consequences of the GUP on the Liouville theorem were investigated,
and in~\cite{Chang:2001bm} the modification of the density of states and the arising
consequences for black-hole thermodynamics were studied. In~\cite{Nozari:2006gg}, one finds 
 the effects of the {\sc GUP} on the thermodynamics of ultra-relativistic particles in the early universe,
and relativistic thermodynamics in~\cite{Das:2009qb}. \cite{Wang:2011iv} studied the 
equation of state for ultra-relativistic Fermi gases in compact stars, the ideal gas
was addressed in~\cite{Chandra:2011nj} and photon gas
thermodynamics in~\cite{Zhang:2011ms}. 

\subsection{Path-integral duality}
\label{pathint}
\index{Path-integral duality}
\index{Zero-point length}

In Sections~\ref{tdual} and \ref{conformalqg} we have discussed two motivations for limits
of spacetime distances that manifest themselves in the Green's function. While one may 
question how convincing these motivations are,  
the idea is interesting and may be considered as a model on its own right. 
Such a modification that realizes a finite `zero point length' of spacetime
intervals had been suggested by Padmanabhan~\cite{Padmanabhan:1986ny, Padmanabhan:1987au, Padmanabhan:1996ap, Padmanabhan:1998yya} as a way to effectively take into account
metric fluctuations below the Planck scale (the motivation from string theory discussed in Section~\ref{tdual} was added after
the original proposal). This model has the merit of not requiring a modification of Lorentz invariance.

The starting point is to
rewrite the Feynman propagator $G_F(x,y)$ as a sum over all paths $\gamma$ connecting $x$ and $y$
\beqn
G_F(x,y) = \sum_{\gamma} e^{-m D_{\gamma}(x,y)} = \int d\tau e^{-m \tau} K(x,y,\tau) \,, \label{GF}
\eeqn
where $D_\gamma (x,y)$ is the proper length of $\gamma$, and
$m$ is a constant of dimension mass. Note that the length of the path depends on the background metric, which is why one
expects it to be subject to quantum gravitational fluctuations. The path integral kernel is
\beqn
K(x,y,\tau) = \int {\cal D} x \exp \left( -\frac{m}{4} \int_0^\tau \mathrm{d}\tau' g_{\mu\nu} \dot x^\mu \dot x^\nu \right) ,
\eeqn
where a dot indicates a derivative with respect to $\tau'$.
The relevant difference between the middle and right expressions in (\ref{GF}) is that $D(x,y)$ has a square root in it. 
The equivalence has been shown using a Euclidean lattice approach in~\cite{Padmanabhan:1998yya}. Once
the propagator is brought into that form, one can apply Padmanabhan's postulated ``principle of duality'' according
to which the weight for each path should be invariant under the transformation $D_\gamma(x,y) \to l^2_{\mathrm{Pl}}/D_\gamma(x,y)$.
This changes the propagator~(\ref{GF}) to
\beqn
\widetilde G_F(x,y) = \sum_{\gamma} \exp\left[ -m \left( D_\gamma(x,y) + \frac{l_{\mathrm{Pl}}^2}{D_\gamma(x,y)} \right) \right] . \label{GFtilde}
\eeqn
Interestingly enough, it can be shown~\cite{Padmanabhan:1998yya} that with this modification in the Schwinger representation, the path integral 
kernel remains unmodified, and one merely obtains a change of the weight 
\beqn
\widetilde G_F(x,y) = \int \mathrm{d}\tau \exp\left[ - m \left( \tau + \frac{l_{\mathrm{Pl}}^2}{\tau} \right) \right] K(x,y,\tau) \,. 
\eeqn
When one makes the Fourier transformation of this expression, the propagator in momentum space takes the form
\beqn
\widetilde G_F(p) &=& \frac{2l_{\mathrm{Pl}}}{\sqrt{p^2 + m^2}} K_1( 2l_{\mathrm{Pl}} \sqrt{p^2 + m^2}) \,,
\eeqn
where $K_1$ is the modified Bessel function of the first kind. This expression has the limiting values (compare to Eq.~(\ref{greensum}))
\beqn
\widetilde G_F(p) \to \left\{ 
\begin{array}{lll} 
\displaystyle \frac{1}{p^2 + m^2}  & {\mbox{for}} & \sqrt{p^2 + m^2} \ll m_{\mathrm{Pl}} \\
\displaystyle \frac{\exp(-2 l_{\mathrm{Pl}} \sqrt{p^2 + m^2})}{\sqrt{2 l_{\mathrm{Pl}}} (p^2 + m^2)^{3/4}} & {\mbox{for}} & m_{\mathrm{Pl}} \ll \sqrt{p^2 + m^2} 
\end{array}\right. \,.
\eeqn 

This postulated duality of the path integral thus suppresses the super-Planckian contributions to amplitudes. As mentioned in Section~\ref{tdual}, in
position space, the Feynman propagator differs from the ordinary one by the shift $(x-y)^2 \to (x-y)^2 + 2l_{\mathrm{Pl}}$. (This idea is so 
different not from that of March~\cite{March}, who in 1936
proposed to replace ordinary spacetime distances with a modified distance $\mathrm{d}\tilde{s}= ds - \rho$. Though at that time,
the `minimal length' $\rho$ was supposed to be of about the size of the atomic nucleus. March's interpretation was that
when the newly defined distance between two points vanishes, the points become indistinguishable.)

Some applications for this model for {\sc QED}, for example the Casimir effect, have been worked out in~\cite{Srinivasan:1997rs,Shankaranarayanan:2000sp}, 
and consequences for inflation and cosmological models have been looked at in~\cite{Sriramkumar:2006qt,Kothawala:2009fa}. For a recent
criticism see~\cite{Campo:2010ce}.

\subsection{Direct applications of the uncertainty principle}
\index{Generalized uncertainty principle}

Maybe the most direct way to look for effects of the minimal length is to start from the
GUP itself. This procedure is limited in its applicability
because there are only so many insights one can gain from an inequality for variances of
operators. However, cases that can be studied this way are everything that can be concluded from 
modifications of
the Bekenstein argument, and with it corrections to the black-hole entropy that one obtains 
taking into account the modification of the uncertainty principle and 
modified dispersion relations.
\index{Black hole}
\index{Black hole!remnant}

Most interestingly, in~\cite{Medved:2004yu,AmelinoCamelia:2004fk} it has been argued that comparing
the corrections to the black-hole entropy obtained from the {\sc GUP} to the corrections obtained in 
string theory and LQG may be used to restrict the functional form of the {\sc GUP}. 

It has also been argued that taking into account the GUP may give rise to black-hole 
remnants~\cite{Adler:2001vs}, a possibility that has been explored in many follow-up works, e.g.,~\cite{Chen:2002tu,Nozari:2005ah,Xiang:2009yq}. 
Corrections to the thermodynamical properties of a Schwarzschild black hole have been looked at  
in~\cite{AmelinoCamelia:2005ik,Zhao:2006xf,Dehghani:2011zzb,Majumder:2011xg,Majumder:2011bv}, the Reissner--Nordstr{\"o}m black hole has been considered in~\cite{Yoon:2007aj}, 
and black holes in anti-de~Sitter space in~\cite{Setare:2004sr,Setare:2005sj,Bolen:2004sq}. Black-hole thermodynamics with a 
{\sc GUP} has been studied in~\cite{Li:2002xb,Myung:2006qr,Bina:2010ir,Kim:2007hf}, the thermodynamics of 
Kerr--Newman black holes in~\cite{Xiang:2009yq}, and the entropy of a
charged black hole in $f(R)$ gravity in~\cite{Said:2011dg}. In~\cite{Carr:2011pr} the consequences of the {\sc GUP}
for self-dual black holes found in the mini-superspace approximation of LQC have been analyzed.
\index{Hawking-radiation}

The thermodynamics of anti-de Sitter space has been looked at in~\cite{Vakili:2008tt}, and the dynamics of the
Taub cosmological model with {\sc GUP} in~\cite{Battisti:2007zg}. The thermodynamics of Friedmann--Robertson--Walker
in four-dimensional spacetimes with {\sc GUP} can be found in~\cite{Battisti:2007jd,Majumder:2011eg}, and with
additional dimensions in~\cite{Sefiedgar:2010we}.
The relations of the {\sc GUP} to holography in 
extra dimensions have been considered in~\cite{Scardigli:2003kr}, the effects of {\sc GUP} on perfect 
fluids in cosmology in~\cite{Majumder:2011ad}, and the entropy of the bulk scalar field in the Randall--Sundrum 
model with {\sc GUP} in~\cite{Kim:2006rx}. In~\cite{Garattini:2011aa} it has been suggested that there is a relationship between
black-hole entropy and the cosmological constant. The relations of the {\sc GUP} to Verlinde's entropic gravity 
have been discussed in~\cite{Nozari:2011gj}
\index{Black hole}

\subsection{Miscellaneous}

While most of the work on modified uncertainty relations has  focused on the {\sc GUP}, 
the consequences of the spacetime uncertainty that arises in string theory for the spectrum of cosmological
perturbations have been studied in~\cite{Brandenberger:2002nq}. In~\cite{Raghavan:2009dk} it has been
proposed that it might be possible to test Planck scale modifications of the energy-time uncertainty relation by monitoring tritium decay. 
It should also be mentioned that
the classical mechanics of $\kappa$-Poincar\'e has been worked out in~\cite{Lukierski:1993wx}, and the
kinematics of a classical free relativistic particle with deformed phase space in~\cite{Girelli:2005dc,Ghosh:2006cb,Arzano:2010kz}.
The effects of such a deformed phase space on scalar field cosmology have been investigated in~\cite{PerezPayan:2011sq}.

In~\cite{Kempf:1999xt} an interesting consequence of the
minimal length was studied, the implication of a finite bandwidth for physical fields. Making this connection
allows one to then use theorems from classical information theory, such as Shannon's sampling
theorem. It was shown in~\cite{Kempf:1999xt} that fields on a space with minimum
length uncertainty can be reconstructed everywhere if known only on a
discrete set of points  (any set of points), if these points are, on average, spaced densely enough. These continuous 
fields then have a literally finite information density. In~\cite{Kempf:2003qu}, it was shown that this information-theoretic meaning of the minimal length generalizes naturally to curved spacetime and in~\cite{Kempf:2010rx} it was
then argued that for this reason spacetime
would be simultaneously continuous and discrete in the same way that
information can be.

A model for spacetime foam in terms of non-local
interactions as a description
for quantum gravitational effects, which serves as an origin of a minimal length scale, has been put forward 
in~\cite{Garay:1998wk,Garay:1998as}. This model is
interesting because it ties together three avenues towards a phenomenology of quantum gravity: the minimal
length scale, decoherence from spacetime foam, and non-locality.

Finally, we mention that a minimum time and length uncertainty in rainbow gravity has been
found in~\cite{Galan:2004st,Galan:2005ju,Galan:2006by}.
\index{Finite information density}
\index{Spacetime uncertainty}
\index{Non-locality}

\newpage

\section{Discussion}
\label{disc}

After the explicit examples in Sections~\ref{motivations} and \ref{models},
 here we will collect some general considerations. 

One noteworthy remark for models with a minimal length scale is that discreteness seems neither necessary nor sufficient 
for the existence of a minimal length scale. String theory is an example that documents that discreteness is not
necessary for a limit to the resolution of structures, and~\cite{Bojowald:2011jd} offered
example in which discreteness does not put a finite limit on the resolution of spatial distances (though the
physical interpretation, or the observability of these quantities requires more study). 

We have also seen that the minimal length scale is not necessarily the Planck length. 
In string theory, it is naturally
the string scale that comes into play, or a product of the string coupling and the string
scale if one takes into account D-branes. Also in {\sc ASG}, or emergent gravity
scenarios, the Planck mass might just appear as a coupling constant in some effective
limit, while fundamentally some other constant is relevant. We usually talk
about the Planck mass because we know of no higher energy scale that is relevant to
the physics we know, so it is the obvious candidate, but not
necessarily the right one.
\index{String scale}

\subsection{Interrelations}

The previously-discussed theories and models are related in various ways. We had already
mentioned that the path-integral duality (Section~\ref{pathint}) is possibly related to T-duality (Section~\ref{tdual}) 
or conformal fluctuations in quantum gravity (Section~\ref{conformalqg}), and that string theory is one of 
the reasons to study non-commutative geometries. 
In addition to this, it has also been argued that the coherent-state approach to non-commutative geometries 
represents another model for minimal length modified quantum mechanics~\cite{Sprenger:2012uc}. The
physics of black holes in light of the coherent state approach has been reviewed in~\cite{Nicolini:2008aj}.
\index{Non-commutative geometry}

{\sc DSR} has been motivated by {\sc LQG}, though no rigorous derivation exists to
date. However, there are non-rigorous arguments that {\sc DSR} may emerge from a semiclassical limit 
of quantum gravity theories in the form of an effective field theory with  an energy dependent metric~\cite{AmelinoCamelia:2003xp},
or that {\sc DSR} (in form of a $\kappa$-Poincar\'e algebra) may result from a version of path integral 
quantization~\cite{KowalskiGlikman:2008fj}. In addition, it has been shown that
 in 2+1 dimensional gravity coupled to matter, the gravitational degrees of freedom can be integrated out, leaving 
an effective field theory for the matter, which is a quantum field theory on $\kappa$-Minkowski spacetime, realizing
a particular version of {\sc DSR}~\cite{Freidel:2003sp}. Recently, it has also been suggested that
{\sc DSR} could arise via {\sc LQC}~\cite{Bojowald:2009ey}. 
\index{Loop Quantum Cosmology}
\index{Loop Quantum Gravity}
\index{Deformed Special Relativity}
\index{$\kappa$-Minkowski}

As already mentioned, it has been argued in~\cite{Calmet:2010tx} that {\sc ASG} may give rise to
{\sc DSR} if one carefully identifies the momentum and the pseudo-momentum. In~\cite{Girelli:2006sc} how the running of the Planck's mass can give rise to a modified dispersion relation was studied.  
\index{Modified dispersion relation}

\subsection{Observable consequences}

The most relevant aspect of any model is to make contact with phenomenology. We have mentioned a
few phenomenological consequences that are currently under study, but for completeness we 
summarize them here. 

To begin with, experimental evidence that speaks for any one of the approaches to quantum
gravity discussed in Section~\ref{motivations} will also shed light on the nature of a
fundamental length scale. Currently, the most promising areas
to look for such evidence are cosmology (in particular the polarization of the cosmic 
microwave background) and miscellaneous signatures of Lorentz-invariance violation. 
The general experimental possibilities to make headway on a theory of quantum gravity
have been reviewed in~\cite{AmelinoCamelia:2008qg, Hossenfelder:2010zj}. One notable
recent development, which is especially interesting for the question of a minimal length
scale, is the possibility that direct evidence for the discrete nature of
spacetime may be found in the emission spectra of primordial black holes, if such black holes exist and
can be observed~\cite{Barrau:2011md}.

Signatures directly related to the minimal length proposal are a transplanckian cut-off
that would make itself noticeable in the cosmic microwave background in the way that the spectrum of fluctuations would not be
exactly scale invariant~\cite{Martin:2000xs, Brandenberger:2002nq}. Imprints from
scalar and tensor perturbations have been studied
in~\cite{Ashoorioon:2004vm, Ashoorioon:2004wd, Ashoorioon:2005ep}, and
in~\cite{Chialva:2011iz, Chialva:2011hc} it has been argued in that observable consequences 
arise at the level of the {\sc CMB} bispectrum. Deformations of special relativity 
can lead to an energy-dependent dispersion, which might be an observable effect for photons
reaching Earth from $\gamma$-ray bursts at high
redshift~\cite{AmelinoCamelia:1999pm, AmelinoCamelia:2008qg, AmelinoCamelia:2010pd}. Minimal length
deformations do, in principle, give rise to computable correction terms to a large number
of quantum mechanical phenomena (see Section~\ref{schroedinger}). This allows one to put bounds on the
parameters of the model. These bounds are presently many orders of magnitude
away from the regime where one would naturally expect quantum gravitational effects.
While it is therefore unlikely that evidence for a minimal length can be found in these
experiments, it should be kept in mind that we do not strictly speaking know that the
minimal length scale is identical to the Planck scale and not lower, and scientific care
demands that every new range of parameter space be scrutinized.

Recently, it was proposed that a massive quantum mechanical oscillator might allow one to test
Planck-scale physics~\cite{Pikovski:2011zk} in a parameter range close to the Planck
scale. This proposal should
be regarded with caution because the deformations for composite systems used therein do
not actually follow from the ones that were motivated by our considerations in Section~\ref{motivations},
because the massive oscillator represents a multi-particle state.
If one takes into account the ad-hoc solutions to the soccer-ball problem, that are necessary for consistency 
of the theory when considering multi-particle states (see Section~\ref{soccer}), then the expected effect is suppressed 
by a mass many orders of magnitude above the Planck mass. Thus, it is unlikely that the proposed experiment 
will be sensitive  to Planck-scale physics.

It is clearly desireable to be able to study composite systems and ensembles, which
would allow us to make use of recent advances in quantum optics and data from the
early high-density era of the universe. Thus, solving the soccer-ball problem is of central
relevance for making contact between these models and phenomenology.

\subsection{Is it possible that there is no minimal length?}

\begin{quote}
\textit{
``The last function of reason is to recognize that there are an infinity of things which surpass it.''}

-- Blaise Pascal
\end{quote}
After having summarized all the motivations for the existence of a minimal length scale, we have to
take care that our desire for harmony does not have us neglecting evidence to the contrary. 

We have already discussed that there are
various possibilities for a minimal length scale to make itself noticeable, and this does
not necessarily mean that it appears as a lower bound on the spatial resolution. We could
instead merely have a bound on products of spatial and temporal extensions. So in this sense
there might not be a minimal length, just a minimal length scale. Therefore,
we answer the question posed in this section's title in the affirmative. Let us
then ask if it is possible that there is no minimal length scale.

The case for a minimal length scale seems clear in
string theory and LQG, but it is less clear in emergent gravity
scenarios. If gravity is emergent, and the Planck
 mass appears merely as a coupling constant in the effective limit, this raises the
question, of there is some way in which the fundamental theory
cannot have a limiting value at all. \index{Emergent gravity} 

In {\sc ASG}, the arguments we have reviewed in Section~\ref{asg} are suggestive but not entirely conclusive. 
The supporting evidence that we discussed comes from
graviton scattering, and from a study of a particular type of Euclidean quantum spacetimes.%
\epubtkFootnote{Though it has meanwhile been shown that the fixed
  point behavior can be found also in the Lorentzian
  case~\cite{Manrique:2011jc}.}
Notwithstanding the question of whether general relativity actually has a
(physically-meaningful) fixed point, the evidence for a minimal length
is counterintuitive even in {\sc ASG}, because gravity becomes weaker
at high energies, so, naively, one would expect its distorting effects to also become weaker. 

As Mead carefully pointed out in his article investigating the
Heisenberg microscope with gravity:
\begin{quote}
``We have also neglected the effect of quantum fluctuations in the
  gravitational field. However, these would be expected to provide an
  additional source of uncertainty, not remove those already
  present. Hence, inclusion of this effect would, if anything,
  strengthen the result.'' (\cite{Mead}, p.~B855)
\end{quote}
That is correct, one might add, unless gravity itself weakens and counteracts the effect of the quantum fluctuations.  In fact, 
in~\cite{Basu:2010nf} the validity of the Hoop conjecture in a thought experiment testing short-distance structures
has been re-examined in the context of {\sc ASG}. It was found that the
running of the Planck mass avoids the necessity of forming a trapped surface at the scale of the experiment. However,
it was also found that still no information about the local physics can be transmitted to an observer in the asymptotic
distance.  \index{Asymptotically Safe Gravity} \index{Hoop conjecture}

As previously mentioned, there is also no obvious reason for the existence of a 
minimal length scale in discrete approaches where the lattice spacing is taken to zero~\cite{Ambjorn:2010rx}. 
To study the question, one needs to investigate the behavior of suitably constructed
observables in this limit. We also note the central role of the Hoop conjecture for
our arguments, and that it is, for general configurations, an unproven conjecture.
\index{Discrete spacetime}

These questions are presently very much under
discussion; we mention them to show that the case is not as settled as it might have
seemed from Sections~\ref{motivations} and \ref{models}.

\newpage

\section{Summary}

We have seen in this review that there are many motivations for the existence of a minimal
length scale. Various thought experiments suggest there are limits to how well we can
resolve structures. String theory and LQG, presently the two most widely pursued
approaches to quantum gravity, both bring with them a minimal length scale, if in very
different realizations. It has been argued that a minimal length scale also exists in
the scenario of ASG, and that non-commutative geometries have a
minimal length scale built in already. 

With that extensive motivation, many models have been proposed that aim at incorporating
a minimal length scale into the quantum field theories of the standard
model, rather than waiting for a theory of quantum gravity to be developed and eventually
connected to the standard model. We have discussed some of these approaches, and
also identified some key open problems. While a lot of work has been done directly studying the implications of modified dispersion relations, deformations of special relativity and a GUP, the underlying framework is not
yet entirely understood. Most importantly, there is the question of how to construct
physically-meaningful observables. One possibility to address this and some
other open questions is to develop an axiomatic approach based on the
geometry of phase space. 

Exploring the consequences of 
a minimal length scale is one of the best motivated avenues to make contact with
the phenomenology of quantum gravity, and to gain insights about the fundamental
structure of space and time.

\newpage

\section{Acknowledgements}

I thank Xavier Calmet, Florian Girelli, Luis Garay, Steven Giddings, Arun Gupta, Amit Hagar, Achim Kempf, Jerzy Kowalski-Glikman, Boris Panes, Roberto Percacci, Stefan Scherer, Lee Smolin, 
Giorgio Torrieri and Phil Warnell for helpful feedback, and Renate Loll for suggesting this review -- it was
only by working on it that I realized an up-to-date comprehensive summary has been overdue.  I am also very grateful for Luis Garay's 1994
review~\cite{Garay:1994en}, which was an invaluable starting point. 

I want to apologize to all those whose work was not treated in the depth it deserved. I have tried to make up for brevity with
an extensive literature list. I hope the reader finds this compilation of work on the minimal length in quantum gravity 
scenarios useful.

\clearpage
\phantomsection
\addcontentsline{toc}{section}{Index}
\printindex

\newpage

\bibliography{refs}

\end{document}